\documentclass[11pt]{article}
\usepackage{amsfonts,amsmath,amssymb}
\usepackage{color,bbold}
\usepackage{verbatim}
\usepackage{pslatex}
\usepackage[section]{placeins} 
\usepackage{float}
\usepackage{subfigure}
\usepackage{graphicx,wrapfig,hyperref}
\usepackage[small,bf]{caption}
\numberwithin{equation}{section}
\begin{document}

\title{{\bf A magnetic instability of \\ the non-Abelian Sakai-Sugimoto model} }
\author{Nele Callebaut\thanks{ncalleba.callebaut@ugent.be}\,\,,
David Dudal\thanks{david.dudal@ugent.be}\\
{\small \it  Ghent University, Department of Physics and Astronomy, Krijgslaan 281-S9, 9000 Gent, Belgium}\normalsize}
\date{}
\maketitle

\begin{abstract}
In this follow-up paper of \cite{Callebaut:2011ab} we further discuss the occurrence of a magnetically induced tachyonic instability of the rho meson in the two-flavour Sakai-Sugimoto model, uplifting two remaining approximations in the previous paper. That is, firstly, the magnetically induced splitting of the branes is now taken into account,  evaluating without approximations the symmetrized trace which enters in the non-Abelian Dirac-Born-Infeld (DBI) action. This leads to an extra mass generating effect for the charged heavy-light rho meson through a holographic Higgs mechanism. Secondly, we compare the results  in the approximation to second order in the field strength to the results using the full DBI-action. Both improvements cause an increase of the critical magnetic field for the onset of rho meson condensation. In addition, the stability in the scalar sector in the presence of the magnetic field is discussed.
\end{abstract}

\tableofcontents

\section{Introduction}
The Sakai-Sugimoto model \cite{Sakai:2004cn,Sakai:2005yt} is one of the most used holographic QCD-models to study effective low-energy effects of a QCD-like theory at strong coupling.
Its main merits are the incorporation of spontaneous chiral symmetry breaking, closely related to the description of confinement in the model, and the fact that previously constructed effective low-energy QCD models (such as the Skyrme model for pions, the hidden local symmetry approach for the coupling of pions and rho mesons, vector meson dominance for the pion formfactor, etc.) drop out automatically.

In this paper we further investigate the stability of the two-flavour Sakai-Sugimoto model in the presence of a magnetic field, and this in the confinement phase. We will find stability in the scalar and an instability in the charged vector sector.
Previous stability analyses of the Sakai-Sugimoto model (SSM) have mainly focused on the case of a background chemical potential. In particular Chern-Simons-induced instabilities to spatially modulated phases have received quite some attention recently
\cite{Ooguri:2010xs,Fukushima:2013zga,deBoer:2012ij,Bayona:2011ab,BallonBayona:2012wx,Domokos:2013kha}. Earlier works in this context include \cite{Chuang:2010ku}, and \cite{Bergman:2007wp} on the $(T,\mu,B)$ phase diagram in the Sakai-Sugimoto model.
More relevant for our current purposes is the DBI-induced instability in the presence of an isospin  chemical potential studied in \cite{Aharony:2007uu},
where a tachyonic instability of the rho meson and ensuing rho meson condensation was described.
We will encounter a somewhat similar phenomenon here, but as a result of the presence of a background magnetic field $B$ and zero chemical potential.

The papers referred to above which include magnetic fields, use the original antipodal SSM in which the flavour branes are positioned $B$-independently at opposite points on the supersymmetry-breaking circle of the background.
We will focus on the more general non-antipodal embedding of flavour branes, in which case the embedding does depend on the magnetic field, corresponding to chiral magnetic catalysis in the dual field theory \cite{Bergman:2008sg}.

The stability of the embedding of the flavour branes has been checked in \cite{Sakai:2004cn} for the antipodal case, and in \cite{Ghoroku:2009iv} for the non-antipodal case. We extend this analysis to the non-antipodal, $B$-dependent embedding,  finding what we referred to as `stability in the scalar sector' earlier.

We believe we are also the first to consider multiple non-antipodal embedded flavour branes $N_f>1$ that couple to the external magnetic field with different electric charges,  modeling differently charged up- and down-quarks. Taking this complication into account will create a magnetically induced splitting of the flavour branes, interpreted as explicit breaking of the $U(N_f)$ chiral symmetry to a product of Abelian $U(1)$ chiral symmetries, which makes the evaluation of the symmetrized trace in the action significantly more cumbersome.

In the end, we find a holographic description of the
instability towards rho meson condensation in the presence of a very strong magnetic field, first discussed in phenomenological QCD-models in \cite{Chernodub:2010qx, Chernodub:2011mc}.
This is one of the many effects studied recently in the context of QCD in extreme conditions, a research area that has naturally gained more interest with the growing availability of data on quark-gluon plasma from LHC and RHIC experiments.
There, not only high temperatures and high densities are present,
but also, when the plasma is created in non-central heavy ion collisions, very high magnetic fields (of the order of $10^{15}$ Tesla) \cite{Skokov:2009qp}.
For a review on strongly interacting matter in magnetic fields, see \cite{Kharzeev:2012ph} and references therein.

Many magnetic effects have been investigated in the Sakai-Sugimoto model, so, to avoid incompleteness, let us refer here to the review paper \cite{Bergman:2012na}
for a nice overview.

\section{Goal and strategy}

\paragraph{Basic argument for rho meson condensation in field theory}

In \cite{Chernodub:2010qx}, a possible magnetic instability of the QCD vacuum towards a phase where charged rho mesons are condensed is discussed. The basic argument for this rho meson condensation at some critical value of the magnetic field $B_c$, is that the charged rho meson combinations which have their spin aligned with the magnetic field $B$, have an effective mass squared
\begin{equation} \label{meff}
\mathit{m_{\rho,eff}^2}(B) = m_\rho^2 - eB
\end{equation}
which vanishes at
\begin{equation} \label{BcLandau}
eB_c = m_\rho^2 = 0.602 \text{ GeV}^2,
\end{equation}
based on the fact that the $n$-th energy level of a free, structureless spin-$s$ particle with mass $m$ in the presence of a background magnetic field $\vec B=B \vec e_3$ is given by the well-known Landau level quantization formula
\begin{align}\label{Elevels}
E^2 = m_\rho^2 + p_3^2 + (2n - 2 s_3 + 1) eB
\end{align}
with $p_3$ the particle's momentum in the direction of the magnetic field, and $s_3$ its spin projection on the same direction.  This leads to (\ref{meff}) for the lowest-energy rho meson $p_3=0$, $n=0$ with spin $s_3=1$.

The above argument holds in the context of the bosonic effective DSGS-model \cite{Djukanovic:2005ag} for rho meson quantum electrodynamics, used in \cite{Chernodub:2010qx}. Somewhat later, the rho meson condensation effect was also shown to emerge in the NJL-model \cite{Chernodub:2011mc}. It should be clear however that rho meson condensation is merely conjectured to occur in QCD based on these descriptions in effective QCD-models, not proven nor experimentally observed.
To date, the effect of rho meson condensation has been discussed in  \cite{Chernodub:2010qx,Chernodub:2011mc,Braguta:2011hq,Chernodub:2011gs,Chernodub:2012tf} using phenomenological and lattice approaches, in our work \cite{Callebaut:2011ab} using 
the Sakai-Sugimoto model, and in \cite{Ammon:2011je,Bu:2012mq,Cai:2013pda} using a bottom-up holographic approach. Its possible occurrence has been argued against in \cite{Hidaka:2012mz} -- followed by a rebuttal in \cite{Chernodub:2012zx} showing that the counterarguments of \cite{Hidaka:2012mz} should not apply.

\paragraph{Goal}

Our goal is to study the effective rho meson mass squared $\mathit{m_{\rho,eff}^2}(B)$ in a full-blown holographic top-down approach, using the Sakai-Sugimoto model.
In a simplified set-up, we were able to show in \cite{Callebaut:2011ab} that rho meson condensation does occur in this model. The $B$-dependence of the rho meson mass will be further investigated here, thereby
uplifting remaining approximations in \cite{Callebaut:2011ab}.
The influence of chiral magnetic catalysis on the differently charged constituents of the mesons is taken into account by considering the non-antipodal embedding. This will lead to a modification of the energy levels \eqref{Elevels}. We shall however continue to use the nomenclature
Landau levels. The instability is still present, at a somewhat higher value of $eB_c$  than the estimate \eqref{BcLandau}. We focus on the confinement phase of the model and set the number of flavours equal to two, $N_f=2$, necessary to describe charged mesons.

\paragraph{Outline}

We start with an outline of the set-up in Section \ref{setup}, including a short review of the Sakai-Sugimoto model. We fix the number of colours $N_c=3$ and the rest of the holographic parameters to numerical GeV units, in order to obtain results for $\mathit{m^2_{\rho,eff}}$ and $B_c$ in physical units, comparable to other -- phenomenological and lattice -- approaches.
In the same Section, the effect of the magnetic field on the probe branes' embedding is reviewed.

In Section \ref{stabilityfluctuations} we discuss the stability of the fluctuations.
For that purpose we plug a flavour gauge field ansatz containing a background ($\sim B$) and a fluctuation part ($\sim$ mesons) into the non-Abelian DBI-action governing the dynamics of the flavour gauge field living on the probe branes, and expand the action to second order in the fluctuations.
The eventual goal is to extract the effective rho meson mass from the 4-dimensional mass equation for the vector meson, the effective 4-dimensional action to be obtained from the DBI-action by integrating out the extra dimensions.

First, we have to choose a particular gauge to disentangle the scalar and vector fluctuations in the action, this is done in Section \ref{gaugefixing}.
Then we
discuss the stability with respect to scalar fluctuations, corresponding to the positions of the probe branes.
Next, we consider the vector fluctuations. This we already partly covered in our previous paper \cite{Callebaut:2011ab}, where we discussed the case of antipodal embedding and the case of non-antipodal embedding with the action approximated to second order in the total field strength $F$ and with the extra assumption of coinciding branes. Here, we extend on these analyses by considering the non-antipodal embedding with magnetically separated branes, both in the case of using the action expanded to second order in $F$ (Section \ref{F2approx}) and the full non-linear DBI-action in $F$ (Section \ref{4.4}).
Because the field strength $F$ in the DBI-action is accompanied with a factor proportional to the inverse of the 't Hooft coupling $\lambda$, which is large in the validity range of the gauge-gravity duality, the expansion to second order in $F$ is commonly used.  However, in the presence of large background fields, the higher order terms may become important (see Section \ref{ambiguities}). We therefore compare the outcome of using the $F^2$-approximated action versus the full DBI-action, from which we can conclude that the difference in $B_c$ is very small and the $F^2$-expansion was justified in our case after all.

In Section \ref{F2approx} the focus is on handling the magnetically separated branes. For non-coinciding branes, the symmetrized trace (STr) over flavour indices in the DBI-action no longer simplifies to a normal Tr. Instead, evaluating the STr (which can be done exactly to second order in the fluctuations) gives rise to complicated functions in the action (defined via integrals),
which depend on the background fields and are discontinuous in the holographic radius $u$. We pay some attention
to solving the eigenvalue equation for the rho meson eigenfunction with these functions present.
The evaluation of the STr is discussed in Section \ref{4.1.1}, with the used -- exact -- prescriptions outlined in the Appendix, including a sketch of their derivation.
In Section \ref{4.3.2}, for completeness, we briefly discuss the pions in the DBI-action. The Section ends with a comment on the validity of the use of the non-Abelian DBI-action for non-coincident branes.

In Section \ref{4.4} the focus is on handling the extra dependences on the magnetic field from considering the full DBI-action. The resulting effective 4-dimensional equation of motion (EOM) (to second order in the rho meson fields) has extra terms compared to the standard Proca EOM used in phenomenological descriptions of the rho meson in a background magnetic field, making it
harder to analyze. We solve the EOMs for the complete energy spectrum exactly
in Section \ref{4.4.3}, with the main result for the generalized Landau levels given in eq. (\ref{omegasquared}). The energy eigenstates are no longer spin eigenstates (as opposed to the Proca energy eigenstates), except for the condensing state.

We comment on the antipodal set-up with full DBI-action in Section \ref{commentantipodal} and summarize in Section \ref{summary}.

\section{Set-up} \label{setup}

\subsection{Review of the Sakai-Sugimoto model}

The Sakai-Sugimoto model \cite{Sakai:2004cn,Sakai:2005yt} is a holographic QCD-model,
involving $N_f$ pairs of $\text{D}8$-$\overline{\text{D}8}$ flavour probe branes placed in a
D4-brane background
\begin{eqnarray}\label{backgr}
ds^2 &=& g_{mn} dx^m dx^n \quad(m,n = 0 \cdots 9) \nonumber \\
 &=& \left(\frac{u}{R}\right)^{3/2} (\eta_{\mu\nu}dx^\mu dx^\nu + f(u)d\tau^2) + \left(\frac{R}{u}\right)^{3/2}
\left( \frac{du^2}{f(u)} + u^2 d\Omega_4^2 \right), \nonumber\\
&&e^\phi = g_s \left(\frac{u}{R}\right)^{3/4} \hspace{2mm}, \quad
F_4 = \frac{N_c}{V_4}\epsilon_4 \hspace{2mm}, \quad f(u) = 1-\frac{u_K^3}{u^3}\,\,,
\end{eqnarray}
\noindent where $d\Omega_4^2$, $\epsilon_4$ and $V_4=8\pi^2/3$ are,
respectively, the line element, the volume form and the volume of a unit
four-sphere,  while $R$ is a constant parameter related to the string
coupling constant $g_s$, the number of colours $N_c$ and the string
length $\ell_s$ through $R^3 = \pi g_s N_c \ell_s^3$.
This background has a natural cut-off at $u = u_K$ and is therefore dual to a confining QCD-like theory, living on the boundary at $u \rightarrow \infty$.
Imposing a smooth cut-off of space at $u = u_K$ uniquely determines
the period $\delta \tau$ of $\tau$:
\begin{eqnarray}
\delta \tau = \frac{4\pi}{3} \frac{R^{3/2}}{u_K^{1/2}} = 2\pi M_K^{-1}
\end{eqnarray}
with $M_K$ the inverse radius of the $\tau$-circle.

The parameters $R, g_s , \ell_s, M_K$, $u_K$ and 't Hooft coupling $\lambda = g^2_{YM} N_c$ are
related through the following equations:
\begin{equation} \label{relaties}
R^3 = \frac{1}{2} \frac{\lambda \ell_s^2}{M_K},  \quad g_s =
\frac{1}{2\pi}\frac{g^2_{YM}}{M_K \ell_s},  \quad u_K = \frac{2}{9} \lambda
M_K \ell_s^2.
\end{equation}
Since all physical results are independent of the choice of $\lambda \ell_s^2$, one can
moreover impose, without loss of generality, that
$\frac{2}{9} M_K^2 \ell_s^2 = \frac{1}{\lambda}$ \cite{Sakai:2005yt} which is the same as stating that
\begin{equation} \label{SS2(2.4)a}
u_K = \frac{1}{M_K}.
\end{equation}
Consequently, the remaining parameter relations reduce to
\begin{eqnarray} R^3 = \frac{9}{4} \frac{1}{M_K^3} \quad \mbox{and}\quad \frac{1}{g_s \ell_s^3} = \frac{4\pi}{9} N_c M_K^3 = \frac{4\pi}{3} M_K^3. \end{eqnarray}

The duality is valid in the limit of a large number of colours $N_c \rightarrow \infty$ and large but fixed 't Hooft coupling $\lambda = g^2_{YM} N_c \gg 1$  (with $g_{YM} \rightarrow 0$), which means one
probes the strong coupling regime of the 4-dimensional dual field theory in the 't Hooft limit. The backreaction of the $N_f \ll N_c$ flavour degrees of freedom on the D4-brane geometry is ignored. This is the so-called probe approximation \cite{Karch:2002sh} (or quenched approximation in QCD language).
Furthermore, bare quark masses are zero, so this model is in the chiral limit\footnote{To overcome this, in \cite{Bergman:2007pm} the bifundamental `tachyon'-field connecting D8- and $\overline{\text{D8}}$-branes is taken into account. Other possible mechanisms to include bare quark masses can be found in \cite{hashimoto}. We did not consider these options here for reasons of simplicity.
}.

\begin{figure}[h!]
  \hfill
  \begin{minipage}[t]{.45\textwidth}
    \begin{center}
      \scalebox{0.7}{
  \includegraphics{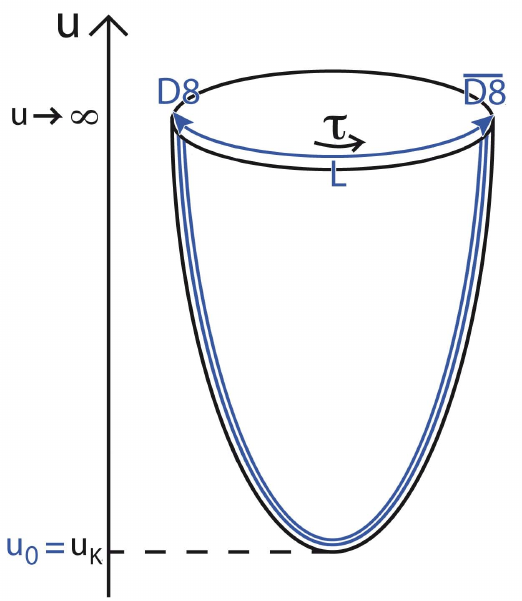}}
    \end{center}
  \end{minipage}
  \hfill
  \begin{minipage}[t]{.45\textwidth}
    \begin{center}
      \scalebox{0.7}{
  \includegraphics{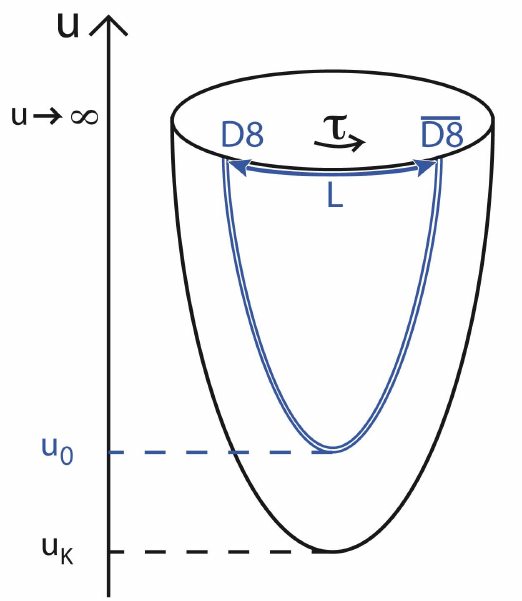}}
    \end{center}
  \end{minipage}
      \caption{The Sakai-Sugimoto model: antipodal ($u_0=u_K$) and non-antipodal ($u_0 > u_K$) embedding.}
	\label{SS}
  \hfill
\end{figure}

On the stack of $N_f$ coinciding $\text{D}8$-$\overline{\text{D}8}$ flavour pairs, there lives
a $U(N_f)_L \times U(N_f)_R$ gauge theory for the flavour gauge field $A_m(x^\mu,u)$($m=0,1,2,3,u$)  describing massless excitations of open strings attached to the branes.
This gauge theory is interpreted as corresponding to the global chiral symmetry in the dual QCD-like theory. The cigar-shape of the ($u,\tau$) subspace of the D4-brane background enforces a $\cup$-shaped embedding of the flavour branes, encoded in the embedding function $u(\tau)$. This particular form of the embedding represents the spontaneous
breaking of chiral symmetry $U(N_f)_L \times U(N_f)_R \rightarrow U(N_f)$ as the merging of the D8-branes and $\overline{\text{D}8}$-branes at $u = u_0$. The asymptotic separation $L$ (at $u\rightarrow \infty$) between D8- and $\overline{\mbox{D8}}$-branes, indicated in Figure \ref{SS}, is related to $u_0$ as
\begin{align}
L &= 2 \int_{u_0}^\infty \frac{du}{u'} \qquad (\text{with}~u' = du/d\tau)  \nonumber \\
   &= 2 \int_{u_0}^\infty du  \left(\frac{R}{u}\right)^{3/2} f(u)^{-1} \sqrt{\frac{u_0^8 f(u_0)}{u^8 f(u) - u_0^8 f(u_0)}}.
   \label{Lconf}
\end{align}
In the original set-up of \cite{Sakai:2004cn,Sakai:2005yt} the embedding is antipodal: the flavour branes merge at the tip of the cigar, $u_0=u_K$.
In the more general non-antipodal embedding, $u_0 > u_K$,  the distance between $u_K$ and $u_0$ is interpreted to be related to the constituent quark mass as the energy stored in a string stretching from $u_0$ to $u_K$ \cite{Aharony:2006da}:
\begin{equation}\label{constmass}
    m_q=\frac{1}{2\pi \alpha'}\int_{u_K}^{u_0}\frac{du}{\sqrt{f(u)}}\,\,,
\end{equation}
with $2\pi\alpha'$ the inverse string tension, related to the string length through $\alpha' = \ell_s^2$. In the latter set-up, unlike in the $u_0=u_K$ case, it is possible \cite{Bergman:2008sg} to model the effect of chiral magnetic catalysis \cite{Shushpanov:1997sf} which says that a magnetic field boosts the chiral symmetry breaking and hence the constituent quark masses.
More precisely, the authors of \cite{Shushpanov:1997sf} discuss a low-energy theorem in the context of chiral perturbation theory, thereby finding that the chiral condensate grows (linearly) in terms of an increasing magnetic field, with the coefficient a function of the pion decay constant $f_\pi$.

In this work, we will interpret $m_q$ as an indicator for the chiral symmetry breaking order parameter, for lack of a chiral condensate in the used set-up. Let us however remark that possible alternatives to define chiral order parameters can be found in, for example, \cite{Aharony:2008an} or \cite{Bergman:2007pm,hashimoto,Dhar:2007bz}.

\subsection{Numerical fixing of the holographic parameters} \label{numfixing}
In this paper, for the purpose of presenting the end results in physical GeV units, we will fix the number of colours to three, $N_c=3$. We choose the number of flavours to be two, $N_f=2$, in order to be able to model electromagnetically charged mesons consisting of up- and down quarks. This means we are stretching the validity of the probe approximation, but we will nonetheless
ignore the backreaction.
With these choices, we are then able to fix the remaining free parameters in the model, $R, \lambda, \ell_s, M_K, u_0, g_s$ and $L$, by matching to the following QCD input parameters
\begin{eqnarray} m_q = 0.310 \mbox{ GeV}, \quad f_\pi = 0.093  \mbox{ GeV} \quad \mbox{and} \quad m_\rho = 0.776 \mbox{ GeV} \end{eqnarray}
for resp. the constituent quark mass $m_q$, the pion decay constant $f_\pi$ and the $\rho$ meson mass $m_\rho$, in absence of magnetic field.

The results of our numerical analysis are (for the underlying details we refer to \cite{Callebaut:2011ab})
\begin{equation} \label{values}
M_K \approx 0.7209 \mbox{  GeV}, \quad  \frac{u_0}{u_K} \approx 1.38 \quad
\mbox{and}\quad \kappa  = \frac{\lambda N_c}{216 \pi^3} \approx 0.006778.
\end{equation}

From these values we do extract a relatively large 't Hooft coupling, $\lambda\approx 15$, and (via (\ref{Lconf})) a value for the asymptotic flavour brane separation $L \approx  1.574 \mbox{  GeV}^{-1}$ that is approximately 2.8 times smaller than the maximum value of $L$, given by $L_{max} = \frac{\delta \tau}{2} = \frac{\pi}{M_K} \approx 4.358 \mbox{  GeV}^{-1}$.
Our estimate for the effective string tension between a quark and an antiquark becomes $\sigma \approx 0.19$ GeV$^2$, in excellent agreement with the pure $SU(3)$ lattice-QCD value $\sigma\approx 0.18$-$0.19$ GeV$^2$ \cite{Bali:1992ru,Sommer:1993ce}.

Using the above values for the parameters enables us
to present all our results in physical units, and in particular compare our result for the critical magnetic field for the onset of rho meson condensation to the values obtained in other (phenomenological or lattice) QCD approaches.

\subsection{Non-Abelian probe brane action}

The dynamics of the stack of $N_f$ coinciding $\text{D}8$-$\overline{\text{D}8}$ flavour branes in the 10-dimensional D4-brane background is determined by the dynamics of open strings with their endpoints attached to the branes. The spectrum of vibrational modes of these attaching strings contains a massless $U(N_f)$ flavour gauge field with 10  components, which can be decomposed in a $U(N_f)$ flavour gauge field $A_m(x^\mu,u)$ ($m=0,1,2,3,u)$ living on the world volume of the branes (we set $A_{\Omega_4}=0$ and $\partial_{\Omega_4} A_m = 0$) and a scalar field $\tau$ describing fluctuations of the branes along their transversal ($\tau$-)direction.
Before writing down the action for the flavour branes in terms of $A_m$ and $\tau$, a few comments are in order.

While the low energy effective action for a single brane is known to be the Dirac-Born-Infeld action \cite{Leigh:1989jq},
valid in the static gauge (i.e.~alignment of the world volume with space-time coordinates) and for slowly varying field strengths, the full non-Abelian generalization of it
for the description of a stack of coinciding branes is not.
Tseytlin proposed in \cite{Tseytlin:1997csa} to non-Abelianize the Dirac-Born-Infeld action by introducing a symmetrized trace STr. The action is still restricted to static gauge and the (in the non-Abelian case slightly ambiguous) slowly-varying field strengths approximation, ignoring derivative terms including $[F,F] \sim [D,D]F$ terms.
This action was shown to be valid up to fourth order in the field strength, with deviations starting to appear at order $F^6$ \cite{Hashimoto:1997gm,Sevrin:2001ha}.
For the probe flavour branes we are dealing with, it is given by the following, which we will further refer to as `the' (non-Abelian) DBI-action \cite{Tseytlin:1997csa,Myers:2003bw,Howe:2006rv}:
\begin{equation} \label{nonabelian}
S_{DBI} = -T_8 \int d^4x \hspace{1mm} 2 \int_{u_0}^{\infty}  du \int \epsilon_4 \hspace{1mm} e^{-\phi}\hspace{1mm} \text{STr}
\sqrt{-\det \left[g_{mn}^{D8} + (2\pi\alpha') iF_{mn} \right]},
\end{equation}
where $T_8 = 1/((2\pi)^8 \ell_s^9)$ is the D8-brane tension, the factor 2 in front of the $u$-integration makes sure that we integrate over both halves of the $\cup$-shaped D8-branes,   $\text{STr}$ is the
symmetrized trace which is defined as
\begin{eqnarray} \label{STrdef}
\text{STr}(F_1 \cdots F_n) = \frac{1}{n!} \text{Tr}(F_1 \cdots F_n + \mbox{all permutations}), \end{eqnarray}
$g_{mn}^{D8}$ is the induced metric on the D8-branes,
\[
g_{mn}^{D8} = g_{mn} + g_{\tau\tau} (D_m \tau)(D_n \tau),
\]
with covariant derivative $D_m\tau = \partial_m \tau + [A_m,\tau]$,  and
\[
F_{mn} = \partial_m A_n - \partial_n A_m + [A_m, A_n] = F_{mn}^a t^a
\]
the field strength with anti-Hermitian generators
\begin{eqnarray} t^a  =  -\frac{i}{2} (\mathbb{1}, \sigma_1, \sigma_2, \sigma_3), \quad
\text{Tr}(t^a t^b) = - \frac{\delta_{ab}}{2}, \quad [t^a,t^b] = \epsilon_{abc}t^c. \end{eqnarray}

\subsection{Effect of uniform magnetic field on the probe branes' embedding} \label{embsection}

To model a uniform magnetic field $\vec B=B \vec e_3$ in the dual field theory, $B = F_{12}^{em} = \partial_1 A_2^{em}$, we assume the background gauge field ansatz
($e$ being the electromagnetic coupling constant and $Q_{em}$ the electric charge matrix) \cite{Sakai:2005yt}
\begin{align} \label{background}
A_\mu = \overline A_\mu &= -i e Q_{em} A_\mu^{em} \quad \mbox{(all other gauge field components zero)} \nonumber\\
& =  -i x_1  \left(
\begin{array}{cc}  \frac{2}{3}eB & 0 \\ 0 &  -\frac{1}{3}eB \end{array} \right) \delta_{\mu2}
= \frac{ x_1 e B
\delta_{\mu2}}{3} \left(  -\frac{i\mathbb{1}_2}{2} \right)   +  x_1 e B
\delta_{\mu2} \left(-\frac{i\sigma_3}{2}  \right),
\end{align}
or
\begin{equation}
\overline A_2^3 = x_1 e B  \quad \mbox{   and   } \quad \overline A_2^0 =
\overline A_2^3 / 3;
\end{equation}
and
\begin{eqnarray}
\overline F_{12}= \partial_1 \overline A_2 =  -i \left(\begin{array}{cc} \frac{2}{3}eB  & 0 \\ 0 & -\frac{1}{3}eB \end{array} \right) = -i \left(\begin{array}{cc} \overline F_u  & 0 \\ 0 & \overline F_d \end{array} \right),  \label{Fbardef}
 \end{eqnarray}
where in the last line we defined the up- and down-components of the background field strength, $\overline F_u$ and $\overline F_d$.
In the rest of the paper we will denote $eB$ as $B$.

\begin{figure}[h!]
  \centering
  \scalebox{0.8}{
  \includegraphics{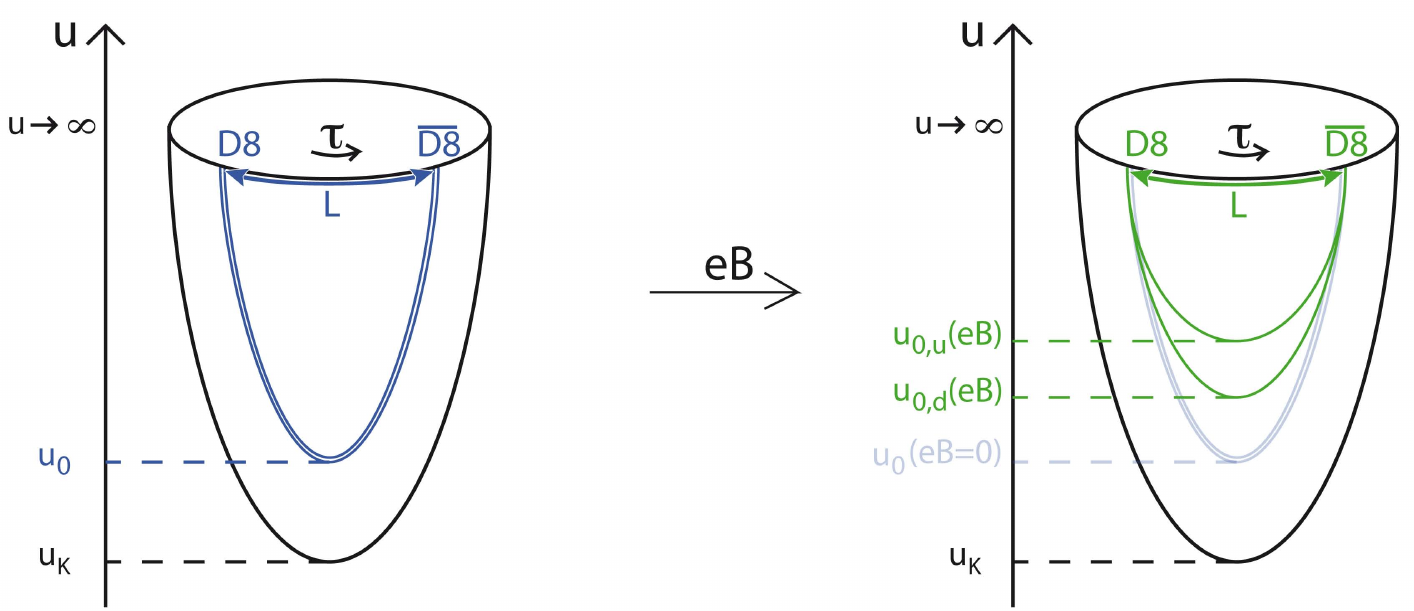}}
  \caption{The change in embedding of the flavour branes caused by the magnetic field $B$ models the chiral magnetic catalysis effect. The up-brane reacts the strongest to $B$, corresponding to a stronger chiral magnetic catalysis for the up-quarks than for the down-quarks.}\label{changedembedding}
\end{figure}

The embedding of the (8+1)-dimensional D8-branes in the 10-dimensional D4-brane background (\ref{backgr}) only requires the specification of one function, $\tau(u)$. This embedding function can be determined as a function of $B$ by first plugging the above gauge field ansatz into the DBI-action (\ref{nonabelian}), together with the metric ansatz
\begin{align}
g^{D8}
&= \left( \begin{array}{cc} g^{D8}_u & 0 \\ 0 & g^{D8}_d \end{array} \right),
\end{align}
to allow for a different response of up- and down-brane to the magnetic field. Subsequently one can solve for $u'=du/d\tau$ (for each flavour) by expressing conservation $\partial_\tau H = 0$ of the `Hamiltonian' $H = u' \frac{\delta \mathcal L^\tau}{\delta u'} - \mathcal L^\tau$ with $S_{DBI}\sim \int d\tau \mathcal L^\tau$  and assuming a $\cup$-shaped embedding, i.e.~$u' = 0$ at $u=u_{0}$.
The result for the $B$-dependent embedding is (for more details, see \cite{Callebaut:2011ab}):
\begin{eqnarray} \label{embmatrix}
\tau(u) = \overline \tau =\left( \begin{array}{cc} \overline \tau_u & 0 \\ 0 &  \overline \tau_d \end{array} \right)
\end{eqnarray}
with
\begin{equation} \label{tauembedding}
\partial_u \overline \tau_l = \sqrt{ \left(\frac{R}{u}\right)^{3} \frac{1}{f^2} \frac{u_{0,l}^8 f_{0,l} A_{0,l}}{u^8 f A_l - u_{0,l}^8 f_{0,l} A_{0,l}}}  \times \theta(u - u_{0,l}), \quad (l=u,d)
\end{equation}
where $f$ is short for $f(u) = 1-u_K^3/u^3$,  $A_0$ and $f_0$ stand for $A(u_0)$ and $f(u_0)$,
$\theta(u - u_{0,l})$ is the Heaviside stepfunction, and all the $B$-dependence is collected in the newly defined matrix $A$:
\begin{eqnarray} \label{A}
A=\left( \begin{array}{cc} A_u & 0 \\ 0 &  A_d \end{array} \right) = 1 - (2\pi\alpha')^2 \overline F_{12}^2  \left(\frac{R}{u}\right)^{3}, \quad
A_{l} = 1 + (2\pi\alpha')^2 \overline F_{l}^2  \left(\frac{R}{u}\right)^{3}, \quad (l=u,d).
\end{eqnarray}
The up- and down-brane are thus no longer coincident in the presence of $B$, as sketched in Figure \ref{changedembedding}.

The splitting of the branes represents the magnetically induced explicit breaking of
global chiral symmetry,
\begin{eqnarray}
U(2)_L \times U(2)_R \stackrel{B}{\rightarrow} (U(1)_L \times U(1)_R)^u \times (U(1)_L \times U(1)_R)^d,
\end{eqnarray}
caused by the up- and down-quarks' different coupling to the magnetic field.
This is also reflected in the fact that the non-Abelian DBI-action for the two D8-branes reduces to the sum of two Abelian actions (the STr reduces to an ordinary Tr because the embedding matrix (\ref{embmatrix}) is diagonal).

\begin{figure}[h!]
  \hfill
  \begin{minipage}[t]{.45\textwidth}
    \begin{center}
      \scalebox{0.65}{
  \includegraphics{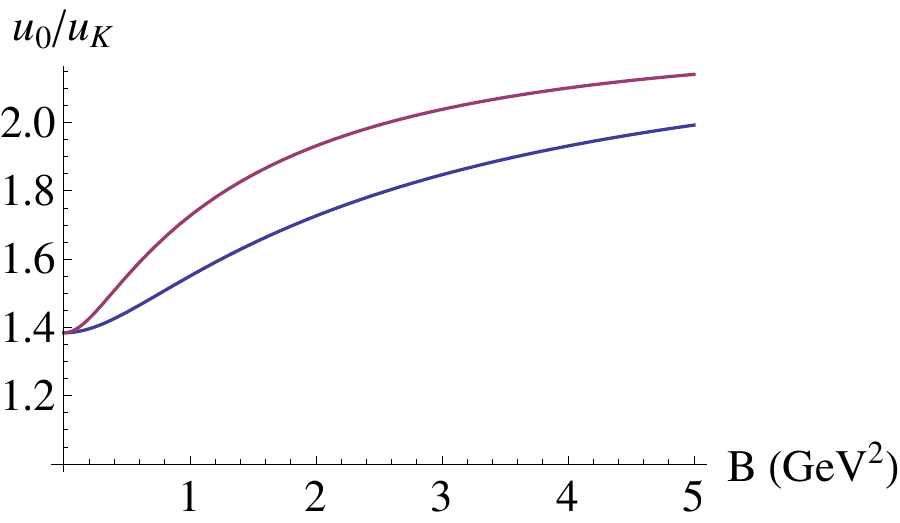}}
    \end{center}
  \end{minipage}
  \hfill
  \begin{minipage}[t]{.45\textwidth}
    \begin{center}
      \scalebox{0.65}{
  \includegraphics{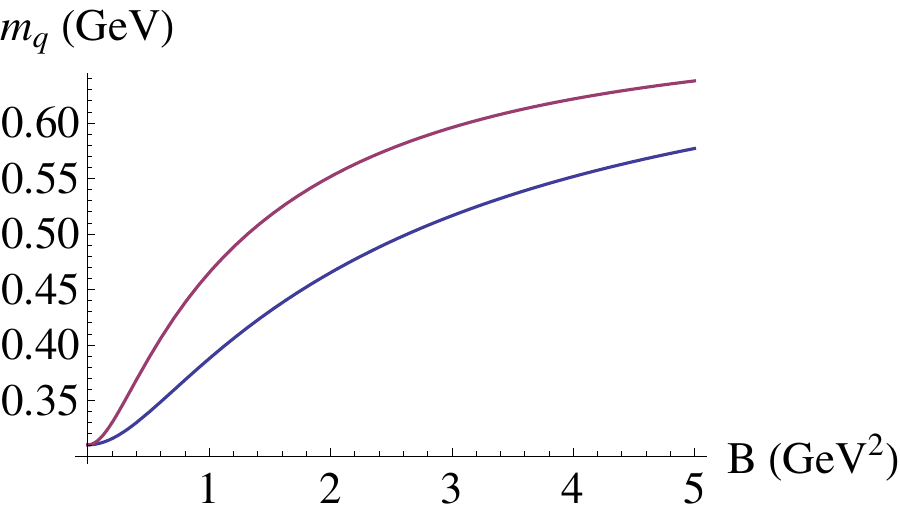}}
    \end{center}
  \end{minipage}
      \caption{(a) $\frac{u_0}{u_K}$ as a function of the magnetic field for the D8-brane corresponding to the up-quark (red), and the one corresponding to the down-quark (blue). (b) The constituent masses of the up-quark (red) and the down-quark (blue) as a function of the magnetic field. }
	\label{mqfig}
  \hfill
\end{figure}

The $B$-dependence of $u_{0,u}$ and $u_{0,d}$ is determined by keeping the asymptotic separation $L$ between D8- and $\overline{\mbox{D8}}$-branes, as a function of $B$ given by
\begin{align}
L     &= 2 \int_{u_0}^\infty du  \left(\frac{R}{u}\right)^{3/2} f^{-1} \sqrt{\frac{u_0^8 f_0  A_0}{u^8 f A - u_0^8 f_0  A_0}},  \label{LconfifvB}
\end{align}
fixed to its value at $B=0$. $L$ serves as the boundary condition on the branes' embedding\footnote{From the perspective of the asymptotic dual field theory, the flavour branes are infinitely extended, massive objects in the bulk, requiring an infinite amount of energy to move them. In this sense it is natural to keep $L$ fixed as a boundary condition to probe the effects of the bulk dynamics in the presence of the external field. The value of $L$ determines how much of the gluonic bulk dynamics is probed, ranging from all ($u_0=u_K$) for maximal $L$ to none ($u_0\rightarrow \infty$) for minimal $L$. In this interpretation, the choice of $L$ (which has no direct physical meaning in the dual field theory) corresponds to the choice of type of dual field theory, ranging from QCD-like to NJL-like in the limit of $L\rightarrow 0$ or $\tau$ non-compact. To avoid confusion, with ``NJL-like'' we refer to a model sharing some but not all features with NJL-models. For more details, see \cite{Antonyan:2006vw}.}, see also for example the work of Preis et~al.~\cite{Bergman:2008sg}.
The $B$-dependence of the constituent quark masses then follows directly from (\ref{constmass}), or in terms of the fixed parameters
\begin{equation}\label{masscont}
m_q(M_K,u_0,\kappa) =8\pi^2 M_K^2 \kappa \int_{1/M_K}^{u_0} du \frac{1}{\sqrt{1 - \frac{1}{(M_Ku)^3}}}.
\end{equation}
The results for $u_0(B)$ and $m_q(B)$ (for both flavours) are shown in Figure \ref{mqfig}.
The rising of the constituent masses $m_q$ with $B$ is consistent with the interpretation of the $B$-dependent  embedding as a modeling of the chiral magnetic catalysis effect (as already discussed in the Sakai-Sugimoto model in \cite{Bergman:2008sg}): as the value of $u_0$, where the branes merge, rises, the $\cup$-shaped embedding gets more strongly bent, diverging
more and more from the chirally invariant embedding of straight branes.
The up-brane reacts twice as strongly to the presence of $eB$, corresponding to a stronger chiral magnetic catalysis for the up-quarks than for the down-quarks.
In the special case of an antipodal embedding $u_0=u_K$ at $B=0$, turning on the magnetic field has no influence on the brane embedding: $f_0 = 0 \Rightarrow \partial \overline \tau = 0 \Rightarrow \overline \tau \sim \mathbb{1}$ so the branes remain antipodal and coincident for all values of the applied magnetic field. Hence choosing the extremal antipodal model corresponds to ignoring the magnetic catalysis.

It is interesting to notice that a similarly shaped plot as in Figure~\ref{mqfig}b was presented in \cite[Figure 12]{Argyres:2008sw} for the in \cite{Aharony:2008an} proposed chiral order parameter in terms of a background magnetic field (be it for the case of a non-compact $\tau$-direction).

\section{Stability analysis} \label{stabilityfluctuations}

To investigate the stability of the set-up with respect to gauge and scalar field fluctuations, let us first derive the form of the action to second order in the fluctuations by plugging the total gauge field ansatz
\begin{equation} \label{fieldansatz}
\left\{\begin{array}{ll} A_r = \overline A_r + \tilde A_r  \quad (r=\mu,u) \\ \tau = \overline \tau + \tilde \tau  \end{array} \right.
\end{equation}
with (see (\ref{background}) and (\ref{tauembedding}))
\begin{equation}
\left\{\begin{array}{ll}  \overline A_\mu = -i e Q_{em} x_1 B \delta_{\mu2}
\\ \partial_u\overline \tau =  \sqrt{\left(\frac{R}{u}\right)^{3} \frac{1}{f^2} \frac{u_0^8 f_0 A_0}{u^8 f A - u_0^8 f_0 A_0}} \times \theta(u-u_0)
\end{array} \right.,
\end{equation}
into the DBI-action (\ref{nonabelian}).
The background components of the field ansatz (\ref{fieldansatz}) describe the background magnetic field (in $\overline A_\mu$) and the ($B$-dependent) embedding of the branes (in $\partial_u \overline \tau$). The fluctuation components will be related to resp. vector and scalar mesons in the dual field theory.

We have to evaluate (for notational brevity we temporarily absorb the factor $(2\pi\alpha')$ into the field strength)
\begin{equation} \label{tensionabsorbed}
2 \int du \hspace{2mm} \text{STr} \sqrt{-\det (a_{rs})} = 2 \int du \hspace{2mm}  \text{STr} \sqrt{-\det (g_{rs}^{D8} + i F_{rs})},
\end{equation}
with
\begin{align}
 g_{rs}^{D8} &= g_{rs} + g_{\tau\tau} D_r \tau D_s \tau, \qquad \mbox{with $D_r \cdot = \partial_r + [\overline A_r,\cdot]$}
 \end{align}
and
\begin{align}
F_{rs} &= \partial_r A_s - \partial_s A_r + [A_r, A_s].
\end{align}
If the argument $a$ of the determinant (which runs over the Lorentz-indices) is written as
\[ a = \overline a + a^{(1)} + a^{(2)} + \cdots \]
with  $a^{(n)}$ being $n$-th order in the fluctuations $\tilde A$,
the determinant can be expanded to second order in the fluctuations as follows
\begin{align}
\sqrt{-\det a}|_{\tilde A^2} &= \sqrt{-\det \overline a}  \left\{ 1+\frac{1}{2} \text{tr} (\overline a^{-1} a^{(1)}) + \frac{1}{8} \left( \text{tr}(\overline a^{-1} a^{(1)}) \right)^2 - \frac{1}{4} \text{tr}\left( (\overline a^{-1} a^{(1)})^2 \right) + \frac{1}{2} \text{tr} (\overline a^{-1} a^{(2)}) \right\}.  \label{expansion}
\end{align}
We denote the trace in Lorentz-space with a small $\text{tr}$, and the trace in flavour space with a capital (S)Tr.
Splitting each component of $a$ in its symmetric and antisymmetric parts
\begin{equation} \label{a}
\left\{ \begin{array}{ll} \overline a^{-1} = \mathcal G + \mathcal B \\
 					    a^{(1)} = \overline a^{(1)} + \delta_1 F \\
					    a^{(2)} = \overline a^{(2)} + \delta_2 F \\
\end{array} \right.
\end{equation}
the expansion of the determinant (\ref{expansion}) to second order in the fluctuations becomes
\begin{align}
\sqrt{-\det a} |_{\tilde A^2}
&= \sqrt{-\det \overline a} + \sqrt{-\det \overline a} \times \nonumber\\
 & \hspace{-1cm}\left\{ \frac{1}{2}  \text{tr} ( \mathcal G \overline a^{(1)})  + \frac{1}{8}  \left(\text{tr} ( \mathcal G \overline a^{(1)})\right)^2  - \frac{1}{4}  \text{tr} ( \mathcal G \overline a^{(1)}\mathcal G \overline a^{(1)} + \mathcal B \overline a^{(1)}\mathcal B \overline a^{(1)}) + \frac{1}{2}  \text{tr} ( \mathcal G \overline a^{(2)}) \right.  \nonumber\\
& \hspace{-1cm} +\frac{1}{2}  \text{tr} ( \mathcal B \delta_1 F)  + \frac{1}{8}  \left(\text{tr} (\mathcal B \delta_1 F)\right)^2  - \frac{1}{4}  \text{tr} (\mathcal G \delta_1 F\mathcal G \delta_1 F + \mathcal B \delta_1 F\mathcal B \delta_1 F) + \frac{1}{2}  \text{tr} (\mathcal B \delta_2 F)   \nonumber \\
 &\hspace{-1cm} \left. +\frac{1}{4} \text{tr} ( \mathcal G \overline a^{(1)})\text{tr} ( \mathcal B \delta_1 F) - \frac{1}{2} \text{tr}(\mathcal G \overline a^{(1)} \mathcal B \delta_1 F) - \frac{1}{2} \text{tr}(
 \mathcal G \delta_1 F \mathcal B \overline a^{(1)}) \right\}. \label{derdelijnscalarplusvector}
\end{align}
For our field ansatz we have
\begin{align}
\overline a_{rs} &= g_{rs} + g_{\tau\tau} \partial_r \overline \tau \partial_s \overline \tau +  i \overline F_{rs}, \\
\overline a^{(1)}_{rs} &= g_{\tau\tau} \left(\partial_r \overline\tau \left([\tilde A_s,\overline\tau] + D_s \tilde \tau \right) + \left([\tilde A_r,\overline\tau] + D_r \tilde \tau\right) \partial_s \overline\tau \right), \\
\delta_1 F_{rs} &= i (D_r \tilde A_s - D_s \tilde A_r) \stackrel{notation}{=} i \tilde F_{rs} \label{notF1}\\
\overline a^{(2)}_{rs} &= g_{\tau\tau}\left([\tilde A_r,\overline\tau]+D_r \tilde \tau\right) \left([\tilde A_s,\overline\tau]+D_s \tilde \tau\right) + g_{\tau\tau} \left([\tilde A_r,\tilde \tau] \partial_s \overline \tau+  \partial_r \overline \tau [\tilde A_r,\tilde \tau] \right), \\
\delta_2 F_{rs} &= i [\tilde A_r, \tilde A_s].
\end{align}
The symmetric part $\mathcal G$ of $\overline a^{-1}$ is diagonal,
\begin{equation}
\mathcal G = \left( \begin{array}{ccccc}g_{00}^{-1} & & & & \\ & g_{11}^{-1} A^{-1} & & & \\ & & g_{22}^{-1} A^{-1} & & \\ & & & g_{33}^{-1} & \\ & & & & G_{uu}^{-1} \end{array} \right), \quad \mbox{ with } G_{uu} = g_{uu}+g_{\tau\tau} (\partial_u \overline \tau)^2
\end{equation}
and the antisymmetric part $\mathcal B$ has non-zero components
\begin{equation}
\mathcal B_{12}=-\mathcal B_{21} = i \overline F_{12} g_{11}^{-1} g_{22}^{-1} A^{-1}.
\end{equation}
As a check, the first order terms in (\ref{derdelijnscalarplusvector}) do vanish on-shell, that is upon using the embedding function (\ref{tauembedding}). The DBI-Lagrangian to second order in the fluctuations then reads
\begin{align}
&\text{STr} \hspace{1mm} e^{-\phi} \sqrt{-\det a} |_{\tilde A^2,\tilde \tau^2,\tilde A \tilde \tau}
= \mathcal L_1 + \mathcal L_2  + \mathcal L_3 + \mathcal L_4
\end{align}
with
\begin{align}
& \mathcal{L}_1 = \text{Tr} \hspace{1mm} e^{-\phi}  \sqrt{-\det \overline a} \nonumber \\
& \mathcal{L}_2 = \text{STr} \hspace{1mm} \overline x  \left\{
\frac{1}{2} \left( [\tilde A_u,\overline \tau] +D_u\tilde\tau \right)^2 G_{uu}^{-2}
+ \overline y [\tilde A_u,\tilde \tau] + \frac{1}{2} \left( [\tilde A_\mu,\overline \tau] +D_\mu \tilde\tau \right)^2 g_{\mu\mu}^{-1} A^{-1}|_{\mu=1,2} G_{uu}^{-1} \right\} \nonumber \\
& \mathcal{L}_3 = \text{STr} \hspace{1mm} \overline x  \left\{ - \overline F_{12} g_{11}^{-1} g_{22}^{-1} A^{-1} [\tilde A_1,\tilde A_2]
- \frac{1}{4} g_{\mu\mu}^{-1} g_{\nu\nu}^{-1} A^{-2}|_{\mu,\nu=1,2}\tilde  F_{\mu\nu}^2 - \frac{1}{2} g_{\mu\mu}^{-1} A^{-1}|_{\mu=1,2} G_{uu}^{-1}\tilde F_{\mu u}^2\right\}  \nonumber \\
&\mathcal{L}_4=  \text{STr} \hspace{1mm} \overline x  \left\{
-\overline z \left( \left( [\tilde A_u,\overline \tau]+ D_u \tilde\tau \right) \tilde F_{12} + \left( [\tilde A_1,\overline \tau]+ D_1 \tilde\tau \right) \tilde F_{2u} - \left( [\tilde A_2,\overline \tau]+ D_2 \tilde\tau \right) \tilde F_{1u} \right)
\right\}, \label{STR}
\end{align}
where
\begin{equation}  \label{xbardef}
\overline x =  e^{-\phi} \sqrt{- \det \overline a} =  e^{-\phi} g_{11}^2 \sqrt{G_{uu}} g_{S_4}^2 \sqrt{A}, \qquad \overline y = G_{uu}^{-1}g_{\tau\tau}\partial_u \overline \tau, \qquad \overline z =  \overline y \overline F_{12}g_{11}^{-1} g_{22}^{-1} A^{-1}
\end{equation}
are functions of the background fields $\partial_u \overline \tau$ and $\overline F_{12}$, so functions of $u$ only, and diagonal in flavour space.
The notation for the factors $g_{\mu\mu}^{-1} A^{-1}|_{\mu=1,2}$ coming from $\mathcal G$ means that $g_{\mu\mu}^{-1}$ is accompanied with a factor $A^{-1} = \frac{1}{1 - (2\pi\alpha')^2 \overline F_{12}^2  R^3/u^3}$ only for $\mu=1,2$.

\subsection{Gauge fixing} \label{gaugefixing}

\subsubsection{STr-evaluation} \label{4.1.1}

The action (\ref{STR}) contains mixing terms between the scalar and gauge fluctuations in $\mathcal{L}_2$ and $\mathcal{L}_4$. We will disentangle these couplings here by choosing a particular gauge.
First we work out $\mathcal{L}_2$ a bit further by evaluating the STr (\ref{STrdef}).
According to its definition in \cite{Myers:2003bw} the STr takes a symmetric average over all orderings of $F_{mn}$, $D_m \tau$ and $\tau$ appearing in the non-Abelian Taylor expansions of the fields in the action. In particular, commutators,  such as $[A_m,A_n]$ in $F_{mn}$ or $[A_m,\tau]$ in $D_m \tau$, are handled as one matrix.
The
STr-expressions we encounter in (\ref{STR}) can be classified into two types: expressions of the form STr$(\mathcal H(\partial_u \overline \tau) \mathcal G(\overline F_{12}) \tilde X)$ and STr$(\mathcal H(\partial_u \overline \tau) \mathcal G(\overline F_{12}) \tilde X^2)$. Here $\mathcal H$, resp.~$\mathcal G$ are even functions of the diagonal background field
\[
\overline \tau = \overline \tau^0 \sigma^0 + \overline \tau^3 \sigma^3,
\]
resp.~
\[
\overline F_{12}=F^0 \sigma^0 + F^3 \sigma^3 = -\frac{i}{2} \frac{B}{3} \sigma^0 -\frac{i}{2} B \, \sigma^3,
\]
and $\tilde X=\tilde X^a t^a$ is some fluctuation -- in the present case
fully general fluctuations $D_m \tilde\tau$ and off-diagonal fluctuations $[\tilde A_m,\overline \tau]$.
For expressions of these types the STr can be evaluated exactly \cite{Hashimoto:1997gm,Denef:2000rj} as elaborated on in the Appendix \ref{appendix}. Using the prescriptions presented and rederived there, we arrive at the following form for $\mathcal{L}_2$:
\begin{align}
\mathcal L_{2} = &\sum_{a=1}^2 \left\{ \gamma(u) \frac{1}{2} \left( [\tilde A_u,\overline \tau]^a +\partial_u\tilde\tau^a \right)^2 + \alpha(u) \frac{1}{2} \left( [\tilde A_\mu,\overline \tau]^a +D_\mu \tilde\tau^a \right)^2 + \beta(u) \sum_{\mu=1}^2 \frac{1}{2} \left( [\tilde A_\mu,\overline \tau]^a +D_\mu \tilde\tau^a \right)^2 \right\} \nonumber\\
&+  \text{Tr} \left( \overline x \overline y  [\tilde A_u,\tilde \tau]  \right) + \sum_{l=u,d} \left\{  \gamma_l(u) \frac{1}{2} \left(\partial_u\tilde\tau^l \right)^2 +  \alpha_l(u) \frac{1}{2} \left(D_\mu \tilde\tau^l \right)^2  +  \beta_l(u) \sum_{\mu=1}^2 \frac{1}{2} \left(D_\mu \tilde\tau^l \right)^2 \right\} \label{74}
\end{align}
with
\begin{align}
\gamma(u) &= -\frac{1}{2} I(\overline x G_{uu}^{-2}), \quad
\alpha(u) = -\frac{1}{2} I(\overline x g_{11}^{-1} G_{uu}^{-1}), \quad
\beta(u) = -\frac{1}{2} I(\overline x g_{11}^{-1} G_{uu}^{-1} \frac{1-A}{A}),  \label{gab} \\
\gamma_l(u) &= -\frac{1}{2} I_l(\overline x G_{uu}^{-2}), \quad
\alpha_l(u) = -\frac{1}{2} I_l(\overline x g_{11}^{-1} G_{uu}^{-1}), \quad
\beta_l(u) = -\frac{1}{2} I_l(\overline x g_{11}^{-1} G_{uu}^{-1} \frac{1-A}{A})
\end{align}
containing what we will refer to as `$I$-functions' and `$I_l$-functions', defined in (\ref{IfctionsGeneral}) and (\ref{IlfctionsGeneral}), e.g.
\begin{align*}
I(\overline x G_{uu}^{-2}) &=  e^{-\phi} g_{11}^2 g_{S_4}^2 I\left(G_{uu}^{-3/2}(\partial \overline \tau) A^{1/2}(\overline F_{12})\right) \nonumber \\
&= \frac{e^{-\phi} g_{11}^2 g_{S_4}^2}{2} \int_0^1 d\alpha \left\{G_{uu}^{-3/2}(\partial \overline \tau^0 + \alpha \partial \overline \tau^3) A^{1/2}(F^0 + \alpha F^3) + G_{uu}^{-3/2}(\partial \overline \tau^0 - \alpha \partial \overline \tau^3) A^{1/2}(F^0 - \alpha F^3)\right\},  \\
I_u(\overline x G_{uu}^{-2}) &= e^{-\phi} g_{11}^2 g_{S_4}^2 G_{uu}^{-3/2}(\partial \overline \tau^0 +  \partial \overline \tau^3) A^{1/2}(F^0 + F^3),
\end{align*}
with $\partial \overline \tau$ short for $\partial_u \overline \tau$
and (with $\tilde \tau = \tilde \tau^a t^a$)
\[
\tilde \tau^l = \frac{\tilde \tau^0 \pm \tilde \tau^3}{\sqrt 2}.
\]
Having used $g_{\mu\nu} = g_{11} \eta_{\mu\nu}$
and absorbing $\eta_{\mu\nu}$ in the notation of the squares, $(\partial_\mu \tilde\tau^a)^2 = \partial_\mu \tilde\tau^a \partial_\nu \tilde\tau^a \eta^{\mu\nu} =  \partial_\mu \tilde\tau^a \partial^\mu \tilde\tau^a$, all the products over $\mu$ in the above Lagrangian (and in all expressions following unless stated otherwise) are contracted Minkowski products.

The difficulty in evaluating the STr, although we restrict to second order in the fluctuations, comes from the presence of the background fields $\partial \overline \tau$ (appearing in the induced metric on the flavour branes through $G_{uu} = g_{uu} + g_{\tau\tau} (\partial_u \overline \tau)^2$) and $\overline F_{12}$ (appearing in $A$ as defined in (\ref{A})), which have to be ordered\footnote{There is some ambiguity here in the sense that the background scalar field $\partial_u \overline \tau$ itself depends on the background gauge field $\overline F_{12}$, so there is also the option to order the matrices $\overline F_{12}$ within $\partial_u \overline \tau$, as opposed to ordering $\partial_u \overline \tau$ as independent. We however opted for the latter, which seems more logical to us.} within the STr. The functions containing the background fields have to be Taylor expanded before the ordering and subsequently resummed. This gives rise to complicated $I$-functions as
 in (\ref
 {gab}), which in general have to be calculated numerically.

\subsubsection{Choosing a 't Hooft gauge}

We consider a 't Hooft gauge-fixing function \cite{'tHooft:1971rn} in the non-Abelian directions -- assuming the Einstein convention that double $SU(2)$-indices $b,c=1,2,3$  are summed over --
\begin{equation} \label{thooftgauge}
G^a = \frac{1}{\sqrt \xi} \left( \alpha(u) D_\mu \tilde A_\mu^a + \gamma(u) D_u \tilde A_u^a+ \sum_{i=1,2} \beta(u) D_i \tilde A_i^a \right)+ 2 i \sqrt \xi \epsilon_{abc} \tilde \tau^b \overline \tau^c \quad (a=1,2)
\end{equation}
such that the gauge-fixed Lagrangian
\begin{align}
&\mathcal L_{2} -\frac{1}{2} (G^a)^2 = \sum_{a=1}^2 \left\{ \gamma(u) \frac{1}{2}\left[ \left( [\tilde A_u,\overline \tau]^a \right)^2 +\left(\partial_u\tilde\tau^a \right)^2 \right] + \alpha(u) \frac{1}{2} \left[ \left( [\tilde A_\mu,\overline \tau]^a \right)^2 + \left(D_\mu \tilde\tau^a \right)^2 \right] \right. \nonumber\\
& \left.\quad + \beta(u) \sum_{\mu=1}^2 \frac{1}{2} \left[ \left( [\tilde A_\mu,\overline \tau]^a\right)^2 + \left(D_\mu \tilde\tau^a \right)^2 \right] -\frac{1}{2 \xi} \left[ \text{$(D\tilde A)^2$ terms} \right]
+\frac{1}{2} (\sqrt \xi \tilde \tau^a)^2 (2\overline \tau^3)^2
 +2 i \tilde A_u^a \epsilon_{abc} \tilde \tau^b \partial_u(\gamma(u)  \overline \tau^c) \right\} \nonumber\\
& \quad +  \text{Tr} \left( \overline x \overline y  [\tilde A_u,\tilde \tau]  \right) + \sum_{l=u,d} \left\{ \cdots \right\}
\label{82}
\end{align}
will be free of mixing terms for a sensible choice of the gauge parameter $\xi$. The Lagrangian
$\mathcal L$ is replaced by $\mathcal L -\frac{1}{2} (G^a)^2$ by virtue of the Faddeev-Popov trick: the partition function of a system with action $S = \int dx \mathcal L$ fulfilling the gauge-fixing constraints $G^a(A,\tau)=0$ is written
as
\begin{align}
Z = \int \mathcal D A \mathcal D \tau \hspace{1mm} e^{i \int dx \mathcal L(A,\tau)}
 \sim \int \mathcal D A \mathcal D \tau  \hspace{1mm} e^{i \int dx \mathcal L(A,\tau)} \delta\left[G(A,\tau)\right] \Delta_{G(A,\tau)}
\end{align}
with proportionality constant the volume of the
gauge group, $\delta \left[G(A,\tau)\right] = \Pi_{x,a}\left( \delta \left[G^a(A(x),\tau(x))\right] \right)$
and $\Delta_{G(A,\tau)}$ the associated Jacobian,
or alternatively -- through introducing the gauge-fixing as $\delta(G^a(A(x),\tau(x)) - \omega^a(x))$ and integrating over $\omega^a$ having a Gaussian distribution around zero -- as
\[
Z \sim \int \mathcal D A \mathcal D \tau e^{i \int dx \left[ \mathcal L(A,\tau) - \frac{1}{2} \left(G^a(A,\tau)\right)^2 \right]} \Delta_{G(A,\tau)}.
\]
Now we rescale the charged scalar fluctuations $\tilde \tau^{a=1,2} \rightarrow \frac{\tilde \tau^{a=1,2}}{\sqrt{\xi}}$
and choose the so-called `unitary' gauge
\begin{equation} \label{xiunitary}
\xi\rightarrow \infty. \end{equation}
This boils down to deleting all dynamical terms for the fluctuations  $\tilde \tau^{a=1,2}$ and we are left with
\begin{align}
\mathcal L_{2} -\frac{1}{2} (G^a)^2  &= \sum_{a=1}^2 \left\{ \gamma(u) \frac{1}{2}\left( [\tilde A_u,\overline \tau]^a \right)^2 + \alpha(u) \frac{1}{2} \left( [\tilde A_\mu,\overline \tau]^a \right)^2 + \beta(u) \sum_{\mu=1}^2 \frac{1}{2} \left( [\tilde A_\mu,\overline \tau]^a\right)^2
+\frac{1}{2} (\tilde \tau^a)^2 (2\overline \tau^3)^2 \right\} \nonumber \\ &+ \sum_{l=u,d} \left\{ \cdots \right\}.
\end{align}
With the above gauge choice we can see the Higgs mechanism at work that is associated with the magnetic field pulling the up- and down-brane apart:  the charged scalar fluctuations $\tilde \tau^{1,2}$ now serve as Goldstone bosons that are eaten by the gauge bosons $\tilde A_m^{1,2}$, acquiring a mass $\sim (\overline \tau^3)^2$, where $\overline \tau^3$ is essentially the vacuum expectation value of
the diagonal component $\tau^{3}$ of the $\tau$-field. The remaining fluctuations $\tilde \tau^{0,3}$ are the Higgs bosons.

\subsubsection{Fixing the remaining gauge freedom}

In the unitary gauge, $\mathcal L_4$, containing the only remaining mixing terms between gauge and scalar fluctuations, reads
\begin{align}
\mathcal L_4 &= \frac{1}{2} \left\{ I(\overline x \overline z) \sum_{a=1}^2 \left[ [\tilde A_u,\overline \tau]^a \tilde F_{12}^a + [\tilde A_1,\overline \tau]^a \tilde F_{2u}^a -  [\tilde A_2,\overline \tau]^a \tilde F_{1u}^a \right]  + \sum_{l=u,d} I_l(\overline x \overline z) \left[ D_u \tilde\tau^l \tilde F_{12}^l + D_1 \tilde\tau^l \tilde F_{2u}^l - D_2 \tilde\tau^l \tilde F_{1u}^l \right] \right\} \nonumber\\
&=  \frac{1}{2} I(\overline x \overline z) \sum_{a=1}^2 \left( -[\tilde A_1,\overline \tau]^a \partial_u \tilde A_2^a + [\tilde A_2,\overline \tau]^a \partial_u \tilde A_1^a \right)
\end{align}
where we used partial integration. The neutral part vanishes due to the gauge choice
\begin{equation} \label{a03gauge}
A_u^3 = A_u^0 = 0,
\end{equation}
hereby using the remaining gauge freedom in the $a=0,3$ directions, as the 't Hooft gauge (\ref{thooftgauge}) only fixes the gauge for $a=1,2$.

In the chosen gauge (\ref{thooftgauge}), (\ref{xiunitary}), (\ref{a03gauge}), the Lagrangian is free of $\tilde A_m \tilde \tau$ couplings:
\begin{align}
& \text{STr} \hspace{1mm} e^{-\phi} \sqrt{-\det a} |_{\tilde A^2,\tilde \tau^2}  = \overline{\mathcal L} + \mathcal L_{Higgs}  + \mathcal L_{scalar} + \mathcal L_{vector} + \mathcal L_{vector-mixing} \end{align}
with
\begin{align}
& \overline{\mathcal{L}} = \text{Tr} \hspace{1mm} e^{-\phi}  \sqrt{-\det \overline a} \nonumber \\
&\mathcal{L}_{Higgs}= \sum_{a=1}^2 \left\{ \gamma(u) \frac{1}{2}\left( [\tilde A_u,\overline \tau]^a \right)^2 + \alpha(u) \frac{1}{2} \left( [\tilde A_\mu,\overline \tau]^a \right)^2 + \beta(u) \sum_{\mu=1}^2 \frac{1}{2} \left( [\tilde A_\mu,\overline \tau]^a\right)^2
-\frac{1}{2} (\tilde \tau^a)^2 (\overline \tau^3)^2 \right\} \nonumber\\
&\mathcal{L}_{scalar}= \sum_{l=u,d} \left\{  \gamma_l(u) \frac{1}{2} \left(\partial_u\tilde\tau^l \right)^2 +  \alpha_l(u) \frac{1}{2} \left(D_\mu \tilde\tau^l \right)^2  +  \beta_l(u) \sum_{\mu=1}^2 \frac{1}{2} \left(D_\mu \tilde\tau^l \right)^2 \right\} \nonumber\\
&\mathcal{L}_{vector}= \text{STr} \hspace{1mm} \overline x \left\{
- \overline F_{12} g_{11}^{-2} A^{-1} [\tilde A_1,\tilde A_2]
- \frac{1}{4} g_{11}^{-2} \tilde F_{\mu\nu}^2 \hspace{1mm} A^{-2}|_{\mu,\nu=1,2}  - \frac{1}{2} g_{11}^{-1} G_{uu}^{-1} \tilde F_{\mu u}^2 \hspace{1mm} A^{-1}|_{\mu=1,2}  \right\} \nonumber \\
&\mathcal{L}_{vector-mixing} = \frac{1}{2} \left\{ I(\overline x \overline z) \sum_{a=1}^2 \left( -[\tilde A_1,\overline \tau]^a \partial_u \tilde A_2^a + [\tilde A_2,\overline \tau]^a \partial_u \tilde A_1^a \right) \right\}. \label{STRgauged}
\end{align}

\subsection{Stability in scalar sector} \label{4.2}

In this Section we discuss the scalar part of the DBI-Lagrangian (\ref{STRgauged}),
\begin{align}
\mathcal L_{scalar} &=  \text{STr} \hspace{1mm} e^{-\phi} \sqrt{-\det a} |_{\tilde \tau^2} \nonumber\\
&= \sum_{l=u,d} \left\{  \gamma_l(u) \frac{1}{2} \left(\partial_u\tilde\tau^l \right)^2 +  \alpha_l(u) \frac{1}{2} \left(D_\mu \tilde\tau^l \right)^2  +  \beta_l(u) \sum_{\mu=1}^2 \frac{1}{2} \left(D_\mu \tilde\tau^l \right)^2 \right\}.
\end{align}
With the purpose of checking the stability of the $B$-dependent configuration with respect to scalar fluctuations, it is important
to keep track of the correct signs in the action.
First of all, we therefore replace $(\tilde \tau^l)^2 \rightarrow -4(\tilde \tau^l)^2$ such that the
fluctuations $\tilde \tau^l = \frac{\tilde \tau^0 \pm \tilde \tau^3}{\sqrt 2}$ are now written in terms of the real components of the scalar fluctuation $\tilde \tau = \tilde \tau^a \sigma^a$
(where in (\ref{74}) it was implicitly assumed in evaluating the STr that $\tilde \tau = \tilde \tau^a t^a = -i \tilde \tau^a \sigma^a /2$ with imaginary components $\tilde \tau^a$).
Slightly redefining $\mathcal L_{scalar}$ to incorporate the sign of the full action,
\begin{equation*}
S_{DBI} |_{\tilde \tau^2} = -T_8 \int d^4x \hspace{1mm} 2 \int_{u_0}^{\infty} du \int \epsilon_4 \hspace{1mm} e^{-\phi}\hspace{1mm} \textrm{STr}
\sqrt{-\det a} |_{\tilde \tau^2} = T_8 \int d^4x \hspace{1mm} 2 \int_{u_0}^{\infty} du \int \epsilon_4 \mathcal L_{scalar}
\end{equation*}
we then end up with
\begin{align}
& \mathcal L_{scalar}
&= - \sum_{l=u,d} \left\{  I_l(\overline x G_{uu}^{-2}) \left(\partial_u\tilde\tau^l \right)^2 +  I_l(\overline x g_{\mu\mu}^{-1} G_{uu}^{-1}) \left(D_\mu \tilde\tau^l \right)^2  +  I_l(\overline x g_{\mu\mu}^{-1} G_{uu}^{-1} \frac{1-A}{A})  \sum_{\mu=1}^2 \left(D_\mu \tilde\tau^l \right)^2 \right\}
\end{align}
with the convention $(\partial_\mu \tilde\tau^l)^2 = \partial_\mu \tilde\tau^l \partial_\nu \tilde\tau^l \eta^{\mu\nu}$.

The Hamiltonian associated with the Lagrangian is given by
\begin{align}
\mathcal H &= \frac{\delta \mathcal L_{scalar}}{\delta \partial_0 \tau^l} \partial_0 \tau^l - \mathcal L_{scalar} \nonumber\\
&= \sum_{l=u,d} \left\{  I_l(\overline x G_{uu}^{-2}) \left(\partial_u\tilde\tau^l \right)^2 +  I_l(\overline x g_{\mu\mu}^{-1} G_{uu}^{-1}) \left(\left(\partial_0 \tilde\tau^l \right)^2 + \left(\partial_3 \tilde\tau^l \right)^2\right) +  I_l(\overline x g_{\mu\mu}^{-1} G_{uu}^{-1} A^{-1})  \sum_{i=1}^2 \left(D_i \tilde\tau^l \right)^2 \right\}
\end{align}
where we switched notation again to normal squares $(\partial_\mu\tau^l)^2 = \partial_\mu\tau^l \partial_\mu\tau^l$.
For the embedding to be stable towards scalar $\tilde \tau^l$-fluctuations, the associated energy density has to obey
\begin{equation}
\mathcal E = \int_{u_{0,d}}^\infty \hspace{1mm} \mathcal H \geq 0,
\end{equation}
which will be the case if each of the $I_l$-functions is positive.

\begin{figure}[h!]
  \centering
  \scalebox{0.8}{
  \includegraphics{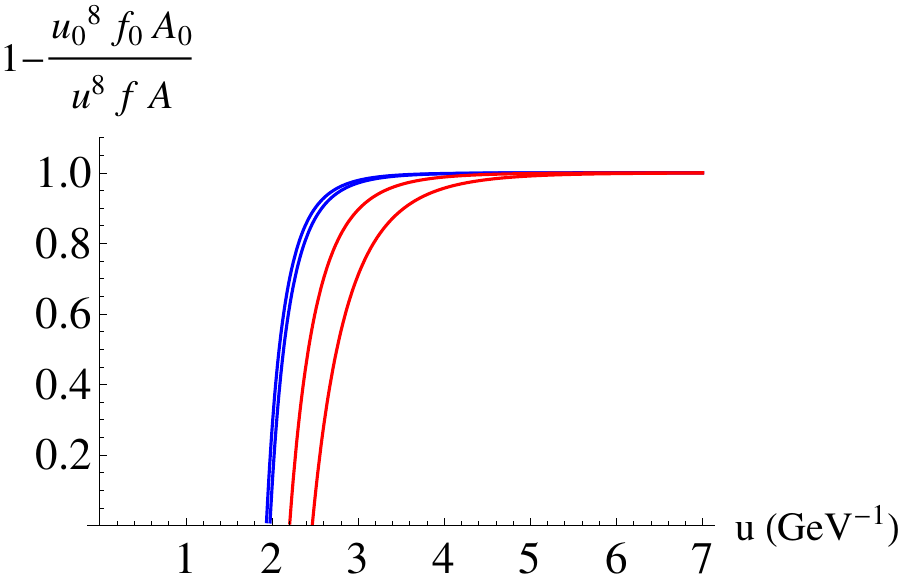}}
  \caption{The function $1 - \frac{u_{0,l}^8 f_{0,l} A_{0,l}}{u^8 f A_l}$ as a function of $u$ for $l=u,d$ for $B=0.2$ GeV$^2$ in blue and $B=1.2$ GeV$^2$ in red. Up distinguishable from down through $u_{0,u} > u_{0,d}$.}\label{function}
\end{figure}

Let us discuss the two background functions that appear in the $I_l$-functions, $A(\overline F_{12})$ and $G_{uu}(\partial \overline \tau)$.
Using (\ref{tauembedding}), the $uu$-component of the induced metric on the D8-branes as a function of the embedding $\partial \overline \tau$ reads
\begin{align}
G_{uu}(\partial \overline \tau^0 \pm \partial \overline \tau^3) = G_{uu}(\partial \overline \tau_l)  &= g_{uu} + g_{\tau\tau} (\partial_u \overline \tau_l)^2
=  \left(\frac{R}{u}\right)^{3/2} \frac{1}{f} \frac{1}{1 - \frac{u_{0,l}^8 f_{0,l} A_{0,l}}{u^8 f A_l}}, \quad (l=u,d)
\end{align}
with $u \geq u_{0,l}$ implicitly understood,
and, from (\ref{A}),
\begin{align}
A(F^0 \pm F^3) = A_l = 1 + (2\pi\alpha')^2 \overline F_{l}^2  \left(\frac{R}{u}\right)^{3}, \quad (l=u,d)
\end{align}
with the plus (minus) sign corresponding to $l=u$ ($l=d$).
$A_l$ is an increasing function of $B$, equal to 1 for $B=0$, and a decreasing function of $u$, equal to 1 for $u=\infty$ so
\[ A_l \geq 1 \quad (\text{for all $B$ and $u$}). \]
The function $1 - \frac{u_{0,l}^8 f_{0,l} A_{0,l}}{u^8 f A_l}$
is a monotonically increasing function of $u$ going from 0 at $u_{0,l}$ to 1 at $u\rightarrow \infty$ for any fixed value of $B$, see Figure \ref{function}.
Then,
\begin{align}
I_l(\overline x G_{uu}^{-2}) &= e^{-\phi} g_{11}^2 g^2_{S_4} \underbrace{I_l(G_{uu}^{-3/2} A^{1/2})}_{G_{uu}^{-3/2}(\partial \overline \tau_l) A_l^{1/2}(\overline F_l)} \nonumber\\
&\sim \underbrace{\vphantom{\left(1 - \frac{u_{0,l}^8 f_{0,l} A_{0,}l}{u^8 f A_l}\right)^{3/2}}\left(\frac{u}{R}\right)^{3/2} u^4}_{\left(\frac{u_0}{R}\right)^{3/2} u_0^4 \cdots \infty} \quad \underbrace{\vphantom{\left(1 - \frac{u_{0,l}^8 f_{0,l} A_{0,}l}{u^8 f A_l}\right)^{3/2}}\hspace{1cm}f^{3/2}\hspace{1cm}}_{\left( 1 - \frac{u_K}{u_0}^3\right)^{3/2} \cdots 1} \quad \underbrace{\hspace{0.2cm}\left(1 - \frac{u_{0,l}^8 f_{0,l} A_{0,}l}{u^8 f A_l}\right)^{3/2}\hspace{0.2cm}}_{0 \cdots 1 \hspace{1mm} \text{for any fixed value of $B$}}~~~\underbrace{\vphantom{\left(1 - \frac{u_{0,l}^8 f_{0,l} A_{0,}l}{u^8 f A_l}\right)^{3/2}}A_l^{1/2}}_{\geq 1} \nonumber\\
& \geq 0,
\end{align}
\begin{align}
I_l(\overline x g_{11}^{-1} G_{uu}^{-1}) &= e^{-\phi} g_{11}^2 g^2_{S_4} g_{11}^{-1} \underbrace{I_l( G_{uu}^{-1/2} A^{1/2})}_{G_{uu}^{-1/2}(\partial \overline \tau_l) A_l^{1/2}(\overline F_l)}  \nonumber\\
&\sim \left(\frac{u}{R}\right)^{-3/4} u^4 \left(\frac{u}{R}\right)^{-3/2} \left(\frac{R}{u}\right)^{-3/4} f^{1/2} \left(1 - \frac{u_{0,l}^8 f_{0,l} A_{0,l}}{u^8 f A_l}\right)^{1/2} A_l^{1/2} \nonumber\\
&\sim \underbrace{\vphantom{\left(1 - \frac{u_{0,l}^8 f_{0,l} A_{0,l}}{u^8 f A_l}\right)^{1/2}}\hspace{0.5cm} u^{5/2}f^{1/2}\hspace{0.5cm}}_{\sqrt{u_0^5 - u_K^3 u_0^2} \cdots \infty} \quad \underbrace{\vphantom{\left(1 - \frac{u_{0,l}^8 f_{0,l} A_{0,l}}{u^8 f A_l}\right)^{1/2}}\left(1 - \frac{u_{0,l}^8 f_{0,l} A_{0,l}}{u^8 f A_l}\right)^{1/2}}_{0 \cdots 1 \hspace{1mm}  \text{for any fixed value of $B$}}~~~\underbrace{\vphantom{\left(1 - \frac{u_{0,l}^8 f_{0,l} A_{0,l}}{u^8 f A_l}\right)^{1/2}}A_l^{1/2}}_{\geq 1}\nonumber \\
& \geq 0,
\end{align}
and for the same reasons
\begin{align}
I_l(\overline x g_{11}^{-1} G_{uu}^{-1} A^{-1}) &= e^{-\phi} g_{11}^2 g^2_{S_4} g_{11}^{-1} \underbrace{I_l( G_{uu}^{-1/2} A^{-1/2})}_{G_{uu}^{-1/2}(\partial \overline \tau_l) A_l^{-1/2}(\overline F_l)}  \geq 0.
\end{align}
This concludes the proof of stability of the flavour branes' embedding as depicted in Figure \ref{changedembedding} with respect to diagonal $\tilde \tau$-fluctuations. Note that the off-diagonal $\tilde \tau$-components have disappeared through the gauge fixing in Section \ref{gaugefixing} -- except for an irrelevant mass term for the undynamical $\tilde \tau^{1,2}$ in $\mathcal L_{Higgs}$. A similar mechanism in the context of the holographic description of heavy-light mesons can be found in \cite{Erdmenger:2007vj}.

Let us briefly expand on the physical interpretation of the discussion of stability in the scalar sector. While in the seminal work of \cite{Sakai:2004cn} (the $x^\mu$-dependent parts of) the scalar modes $\tilde \tau$ were identified with scalar mesons in the dual field theory, this interpretation was revisited in \cite{Imoto:2010ef}, where it is argued that the $\tilde \tau$-fluctuations are to be regarded as artifacts of the SSM\footnote{We would like to thank S.~Sugimoto for private communication about this.}. The reason is that they transform under a $\mathbb Z_2$-symmetry of the geometric configuration (strictly speaking in the antipodal set-up), which is redundant in the sense that it is not shared with QCD. This is similar to the gauge field components $A_{\Omega_4}$ not having a counterpart in the dual QCD-like field theory, as they transform under the $SO(5)$ isometry of the four-sphere in the background (\ref{backgr}).   Any concern about the interpretation of the off-diagonal $\tilde \tau$-components disappearing in the holographic Higgs mechanism coupled to the gauge fixing, is hence resolved: the `eaten' fluctuations do not correspond to physical QCD-particles. The above discussion of the stability is not to be interpreted in terms of mesons in the dual field theory, but rather establishes that the geometrical configuration we will employ further is stable against small perturbations.

\subsection{Vector sector in \texorpdfstring{$(2\pi\alpha')^2 F^2$-approximation}{F-squared approximation}} \label{F2approx} 

Consider the vector part of the DBI-Lagrangian (\ref{STRgauged}),
\begin{align}
\mathcal L &= \mathcal L_{Higgs} + \mathcal L_{vector} = \text{STr} \hspace{1mm} e^{-\phi} \sqrt{-\det a} |_{\tilde A^2} \nonumber\\
&=  \sum_{a=1}^2 \left\{ \gamma(u) \frac{1}{2}\left( [\tilde A_u,\overline \tau]^a \right)^2 + \alpha(u) \frac{1}{2} \left( [\tilde A_\mu,\overline \tau]^a \right)^2 + \beta(u) \sum_{\mu=1}^2 \frac{1}{2} \left( [\tilde A_\mu,\overline \tau]^a\right)^2  \right\} \nonumber\\
&+ \text{STr} \hspace{1mm} \overline x \left\{
- \overline F_{12} g_{11}^{-2} A^{-1} [\tilde A_1,\tilde A_2]
- \frac{1}{4} g_{11}^{-2} \tilde F_{\mu\nu}^2 \hspace{1mm} A^{-2}|_{\mu,\nu=1,2}  - \frac{1}{2} g_{11}^{-1} G_{uu}^{-1} \tilde F_{\mu u}^2 \hspace{1mm} A^{-1}|_{\mu=1,2}  \right\}.
\end{align}
We have anticipated the vanishing of $\mathcal L_{vector-mixing}$ upon filling in the gauge field expansion in terms of vector mesons, which we will come back to shortly.
Let us reinstate the factors $(2\pi\alpha')$ that we absorbed into the field strengths for notational convenience,  and further approximate\footnote{We assume here that the expansion in $1/\lambda$ is justified because $\lambda \approx 15$ is still large for the parameters that we fixed in Section \ref{numfixing}. We will elaborate on the validity of this expansion in the next Section.} the action to second order in $(2\pi\alpha')^2 \sim 1/\lambda^2$:
\begin{align}
\hspace{-1.2cm}
\mathcal L
&\sim u^{1/4}  (2\pi\alpha')^2 \sum_{a,b=1}^2 \left\{- \frac{1}{4} f_1 (\tilde F_{\mu\nu}^a)^2- \frac{1}{2} g_{11} f_2 (\tilde F_{\mu u}^a)^2  - \frac{1}{2} \frac{g_{11}}{ (2\pi\alpha')^2} \left(f_2 -  \frac{1}{2} g_{11}^{-2}  (2\pi\alpha')^2  f_3 \right)(\tilde A_\mu^a)^2 (2\overline \tau^3)^2 \right.  \nonumber \\
& \left. \hspace{3cm}  + \sum_{\mu=1}^2 \left( -\frac{1}{2} g_{11}^{-1} f_3 (\tilde A_\mu^a)^2 (2\overline \tau^3)^2  - \frac{1}{2}(\sqrt{G_{uu}}\overline F_{\mu\nu})^3 \epsilon_{3ab}\tilde A_\mu^a \tilde A_\nu^b \right) \right. \nonumber\\
& \qquad \qquad \quad \left.  - \frac{1}{2} \frac{g_{11}^2}{ (2\pi\alpha')^2} \left(f_4 -  \frac{1}{2} g_{11}^{-2}  (2\pi\alpha')^2  f_5 \right)(\tilde A_u^a)^2 (2\overline \tau^3)^2
\right\} \label{rhopi}
\end{align}
with proportionality factor $-\frac{1}{2} g_s^{-1} R^{\frac{3}{4}+3}$ and
\begin{align}
f_1 =  I(G_{uu}^{1/2}), \quad f_2 =  I(G_{uu}^{-1/2}), \quad f_3 = I(G_{uu}^{-1/2} \overline F_{12}^2), \quad f_4 = I(G_{uu}^{-3/2}) \quad \text{and} \quad f_5 = I(G_{uu}^{-3/2}\overline F_{12}^2) \label{f1f2f3f4f5}
\end{align}
similar $I$-functions as encountered in Section \ref{4.1.1}, again arising from the evaluation of the STr using the prescriptions in Appendix \ref{appendix}.

Effective 4-dimensional meson fields are introduced via
the assumption that the flavour gauge field can be expanded in complete sets
$\left\{\psi_n(u)\right\}_{n\geq 1}$ and $\left\{\phi_n(u)\right\}_{n\geq 0}$ as follows \cite{Sakai:2004cn}
\begin{align}
A_\mu(x^\mu,u) &= \sum_{n \geq 1} B_\mu^{(n)}(x^\mu) \psi_n(u) = \rho_\mu(x^\mu) \psi(u) + \cdots  \label{Amuexpansion}\\
A_u(x^\mu,u) &= \sum_{n \geq 0} \phi_\mu^{(n)}(x^\mu) \phi_n(u) = \pi(x^\mu) \phi_0(u) + \cdots.  \label{Auexpansion}
\end{align}
The rho meson appears as the lowest mode of the infinite vector meson tower $B_\mu^{(n)}$, and the pion as the lowest mode of the infinite (pseudo)scalar meson tower $ \phi_\mu^{(n)}$.
We will only retain these lowest-lying mesons in the fluctuation towers, as -- with the purpose of discussing a possible tachyonic vector instability -- it makes sense that the least massive vector meson will likely be the first to condense.

One obtains an effective 4-dimensional action for the mesons by plugging the above fluctuation expansion for the gauge field into the 5-dimensional DBI-action governing the dynamics of the flavour gauge field, and subsequently integrating out the $u$-dependence.
Some terms can already be understood to vanish during the integration over the extra radial dimension $u$ by looking at the parity of $\psi(z) \equiv \psi(u(z))$ and $\phi_0(z) \equiv \phi_0(u(z))$, with $u(z) = u_0^3 + u_0 z^2$ a commonly used coordinate transformation to the coordinate $z=-\infty \cdots \infty$ following the brane from one asymptotic endpoint to the other. Both $\psi(z)$ and $\phi_0(z)$ are even functions \cite{Sakai:2004cn}, hence coupling terms between rho mesons and pions of the form $\sim D_\mu \tilde A_u^a \partial_u \tilde A_\mu^a \sim D_\mu \pi^a \rho_\mu^a \phi_0(u) \partial_u \psi$ originating from $(\tilde F_{\mu u}^a)^2$ will give rise to vanishing integrals $\int_{-\infty}^{\infty} dz  \{ \text{odd function of $z$} \} = 0$.  This means we can discuss the rho meson and the pion terms separately.
For the same reason the terms $\sim \tilde A_i \partial_u \tilde A_j$ (with $i,j=1,2$) coming from $\mathcal L_{vector-mixing}$ will not survive the $u$-integration.
Note that this simplification is a consequence of cutting the meson towers down to their lowest states.

\subsubsection{Rho meson mass and rho meson condensation} \label{rhomesonc}

\paragraph{Background dependent functions in the action}

Before continuing with the strategy outlined above to extract the 4-dimensional effective action for the rho mesons, we take a closer look at
the relevant functions $f_1$, $f_2$ and $f_3$ as defined in (\ref{f1f2f3f4f5}),
as well as
the definitions for $\overline \tau^3$ and $(G_{uu}^{1/2}\overline F_{12})^3$ in terms of up- and down-components of the background fields.

\begin{figure}[h!]
  \hfill
  \begin{minipage}[t]{.3\textwidth}
    \begin{center}
      \scalebox{0.5}{
  \includegraphics{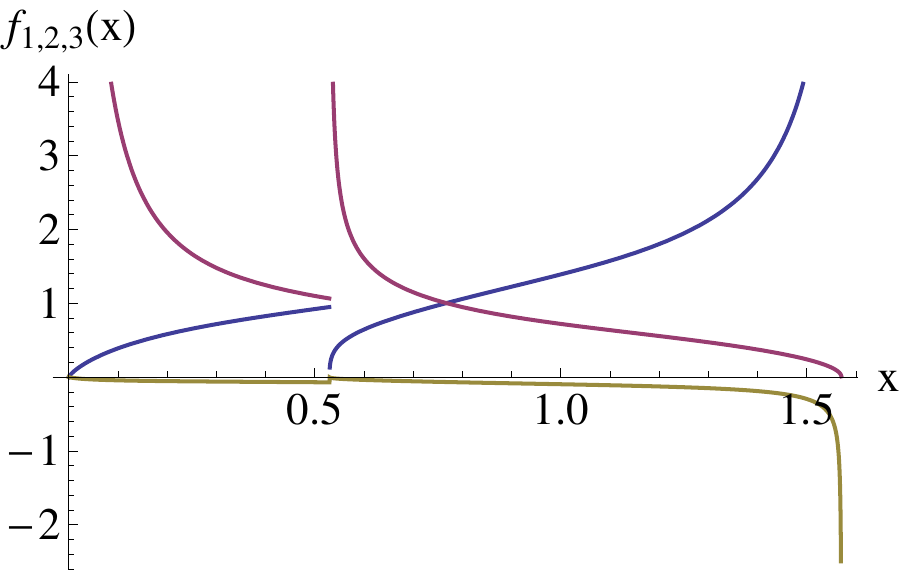}}
    \end{center}
  \end{minipage}
  \hfill
  \begin{minipage}[t]{.3\textwidth}
    \begin{center}
      \scalebox{0.5}{
  \includegraphics{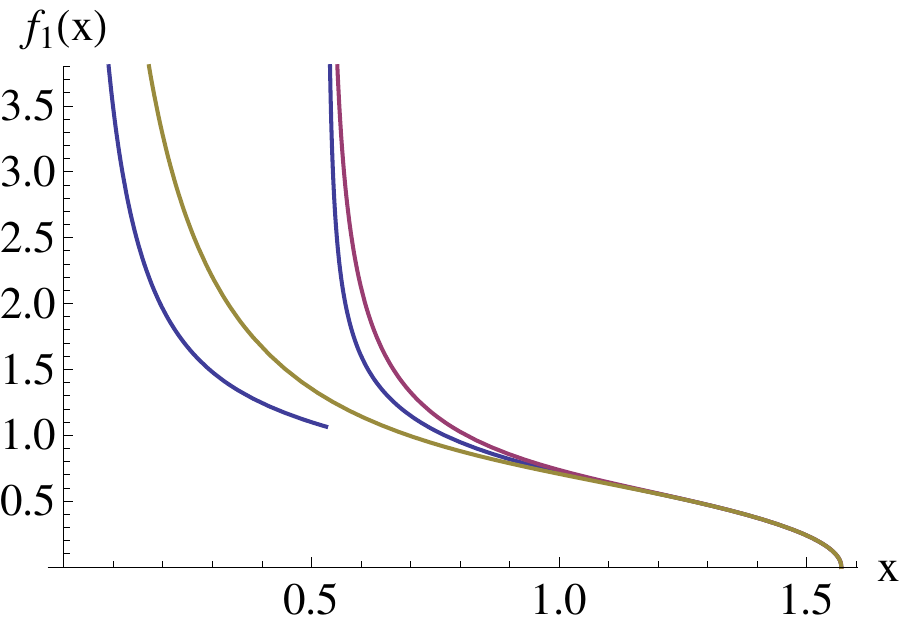}}
    \end{center}
  \end{minipage}
 \hfill
  \begin{minipage}[t]{.3\textwidth}
    \begin{center}
      \scalebox{0.5}{
  \includegraphics{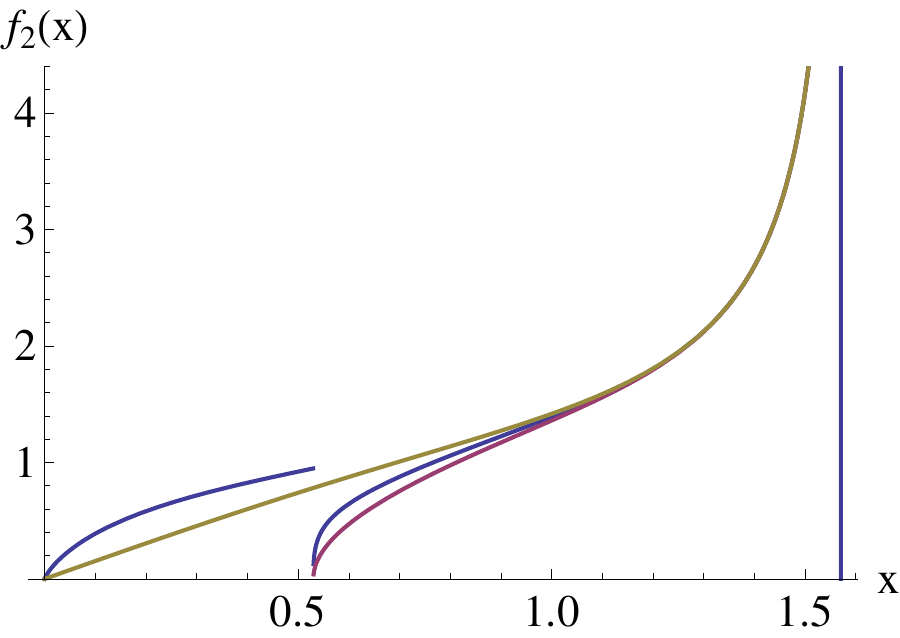}}
    \end{center}
  \end{minipage}
      \caption{(a) $f_1$ (red), $f_2$ (blue) and $f_3$ (yellow) plotted as functions of  $x$, related to $u$ through $u=u_{0,d}\cos^{-3/2}x$ mapping the infinite $u$-range to $x=0\cdots \pi/2$. (b) $f_1 = I(G_{uu}^{1/2})$ (blue) compared to $(G_{uu}^d)^{1/2}$ (yellow) and $(G_{uu}^u)^{1/2}$ (red), i.e.~the functions which would replace $f_1$ if there were a Tr instead of a STr in the action, reducing the non-Abelian to a sum of two Abelian actions. As required, $f_1 \rightarrow G_{uu}^{1/2}$ in the limit of coinciding branes at $u\rightarrow \infty$. (c) $f_2 = I(G_{uu}^{-1/2})$ (blue) compared to $(G_{uu}^d)^{-1/2}$ (yellow) and $(G_{uu}^u)^{-1/2}$ (red). All plots for $B=0.8$ GeV$^2$. }
	\label{f123fig}
  \hfill
\end{figure}

\begin{figure}[h!]
  \hfill
  \begin{minipage}[t]{.4\textwidth}
    \begin{center}
      \scalebox{0.5}{
  \includegraphics{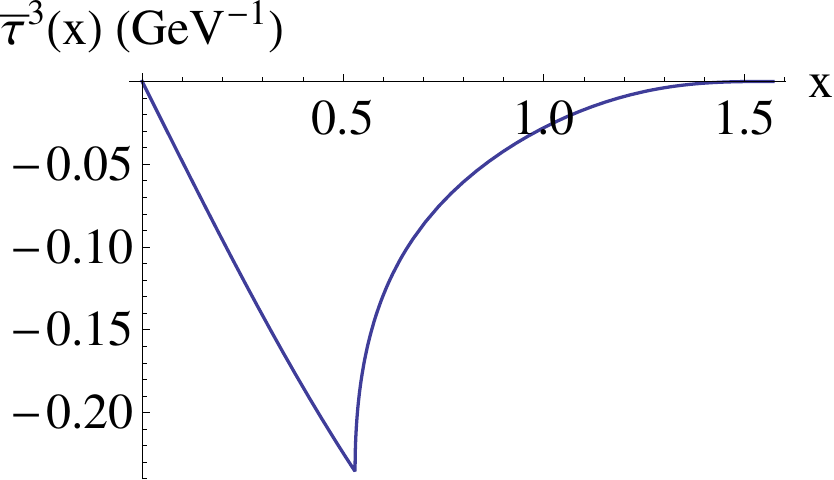}}
    \end{center}
  \end{minipage}
  \hfill
  \begin{minipage}[t]{.5\textwidth}
    \begin{center}
      \scalebox{0.5}{
  \includegraphics{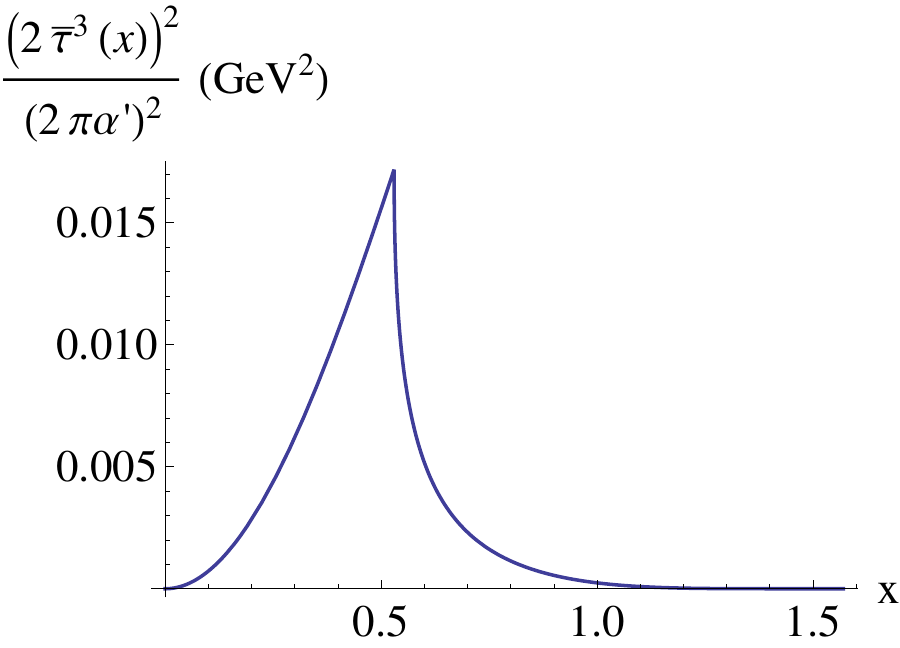}}
    \end{center}
  \end{minipage}
      \caption{The measure $\overline \tau^3(x)$ for the splitting of the branes and the resulting estimated contribution to the mass term for the flavour gauge field and indirectly the rho meson. The range $x=0\cdots \pi/2$ maps to $u=u_{0,d} \cdots \infty$ and we chose $B=0.8$ GeV$^2$. }
	\label{tau3fig}
  \hfill
\end{figure}

Using (\ref{Ifctions}) and (\ref{IfctionsGeneral}), we have
\begin{align}
f_{1} &=  I(G_{uu}^{1/2}) = \frac{1}{2(\partial \overline \tau_u-\partial \overline \tau_d)} \left( \sqrt{G_{uu}^u}\partial \overline \tau_u - \sqrt{G_{uu}^d}\partial \overline \tau_d + \frac{g_{uu}}{\sqrt{g_{\tau\tau}}} \ln \left[ \frac{\partial \overline \tau_u g_{\tau\tau} + \sqrt{g_{\tau\tau} G_{uu}^u}}{\partial \overline \tau_d g_{\tau\tau} + \sqrt{g_{\tau\tau} G_{uu}^d}} \right]\right) \\
f_{2} &= I(G_{uu}^{-1/2}) = \frac{1}{(\partial \overline \tau_u-\partial \overline \tau_d) \sqrt{g_{\tau\tau}}} \ln \left[ \frac{\partial \overline \tau_u g_{\tau\tau} + \sqrt{g_{\tau\tau} G_{uu}^u}}{\partial \overline \tau_d g_{\tau\tau} + \sqrt{g_{\tau\tau} G_{uu}^d}} \right]
\end{align}
\begin{align}
&f_3 = I(G_{uu}^{-1/2} \overline F_{12}^2) =
\frac{1}{2 (\partial \overline \tau_u - \partial \overline \tau_d)^3 g_{\tau\tau}^{3/2}} \left\{ (\overline F_d-\overline F_u)
\left[ \sqrt{g_{\tau\tau} G_{uu}^d} (\partial \overline \tau_d \overline F_d + 3 \partial \overline \tau_d \overline F_u - 4 \partial \overline \tau_u \overline F_d) \right. \right. \nonumber\\
& \qquad \qquad \qquad \qquad \qquad \qquad \qquad \qquad \left. + \sqrt{g_{\tau\tau} G_{uu}^u} (\partial \overline \tau_u \overline F_u + 3 \partial \overline \tau_u \overline F_d - 4 \partial \overline \tau_d \overline F_u) \right] \nonumber\\
& \left. - \left( 2(\partial \overline \tau_u \overline F_d - \partial \overline \tau_d \overline F_u)^2 g_{\tau\tau} - (\overline F_d-\overline F_u)^2 g_{uu} \right) \left[ \ln g_{\tau\tau} g_{uu} + \ln\left( \frac{\partial \overline \tau_u g_{\tau\tau} + \sqrt{g_{\tau\tau} G_{uu}^u}}{\partial \overline \tau_d g_{\tau\tau} - \sqrt{g_{\tau\tau} G_{uu}^d}} \right) \right] \right\},
\end{align}
with $\partial \overline \tau$ short for $\partial_u \overline \tau =$ (\ref{tauembedding}), $G_{uu}^l = G_{uu}(\partial_u \overline \tau_l)$ and $\overline F_u = \frac{2B}{3}$, $\overline F_d =- \frac{B}{3}$, as defined in (\ref{Fbardef}).
Because of the theta-functions $\theta(u-u_{0,l})$ contained in $\partial_u \overline \tau_l$, the contribution of $\partial_u \overline \tau_u$ only kicks in at $u>u_{0,u}$. Therefore these functions will all be discontinuous at $u=u_{0,u}$, as can be seen in the illustrative plot in Figure \ref{f123fig} for $B=0.8$ GeV$^2$. The dependence on $B$ is implicit through the embedding,
except for $f_3$ which also depends explicitly on $B$.
Further,
$\overline \tau^3$ gives a measure for the distance between up- and down-brane, defined as
\begin{align}\label{tau3math}
\overline \tau^3(u) = \int_\infty^u  \partial_u \overline \tau^3 du  =
\int_\infty^u  \frac{\partial_u \overline \tau_u - \partial_u \overline \tau_d}{2} du  = \int_\infty^{u_{0,u}}  \frac{\partial_u \overline \tau_u - \partial_u \overline \tau_d}{2} du + \int_{u_{0,u}}^{u}  \frac{- \partial_u \overline \tau_d}{2} du \nonumber
\end{align}
such that $\overline \tau^3$ fulfills the boundary condition that the flavour branes coincide at $u \rightarrow \infty$: $\overline \tau \sim \mathbb{1} \Rightarrow \overline \tau^3(\infty)=0$. In Figure \ref{tau3fig} the resulting discontinuous $\overline \tau^3$  is plotted for $B=0.8$ GeV$^2$, along with $(2\overline \tau^3)^2/(2\pi \alpha')^2$ which  contributes to the `$u$-dependent mass' of the 5-dimensional gauge field. The contribution is small -- although it is $(2\pi \alpha')^{-2}$-enhanced -- since the splitting itself is a small effect.
The last relevant background function in the action (\ref{rhopi}) for the discussion of the rho mesons is
\begin{align}
(G_{uu}^{1/2} \overline F_{12})^3 = \sqrt{G_{uu}^u} \overline F_u - \sqrt{G_{uu}^d} \overline F_d.
\end{align}

\paragraph{Eigenvalue problem} \label{paragraaf}

Upon substitution of the gauge field expansions (\ref{Amuexpansion}) and (\ref{Auexpansion}) into (\ref{rhopi}), the 5-dimensional DBI-Lagrangian to second order in the rho meson fluctuations (and second order in $(2\pi\alpha')$) reads
\begin{align}
\hspace{-1.2cm}
\mathcal L
&\sim u^{1/4} (2\pi\alpha')^2 \sum_{a,b=1}^2 \left\{- \frac{1}{4} f_1 (\mathcal F_{\mu\nu}^{a})^2 \psi^2 - \frac{1}{2} g_{11} f_2 (\rho_\mu^a)^2 (\partial_u \psi)^2  - \frac{1}{2} \frac{g_{11}}{(2\pi\alpha')^2} \tilde f_2 (\rho_\mu^a)^2 \psi^2 (2 \overline \tau^3)^2  \right.  \nonumber \\
& \left. \hspace{3cm}  + \sum_{\mu=1}^2 \left( -\frac{1}{2} g_{11}^{-1} f_3 (\rho_\mu^a)^2 \psi^2 (2 \overline \tau^3)^2     - \frac{1}{2}(\sqrt{G_{uu}}\overline F_{\mu\nu})^3 \epsilon_{3ab} \rho_\mu^a \rho_\nu^b \psi^2 \right) \right\} + \text{pions}, \label{70}
\end{align}
with $\mathcal F_{\mu\nu}^a = D_\mu \rho_\nu^a - D_\nu \rho_\mu^a$ and $\tilde f_2 = f_2 - \frac{1}{2}g_{11}^{-2}(2\pi\alpha')^2 f_3$.

Demanding that the first line of this Lagrangian reduces to the standard 4-dimensional form
\begin{align}
 \sum_{a=1,2} \left(  - \frac{1}{4} (\mathcal F_{\mu\nu}^a)^2 - \frac{1}{2} m_\rho^2 (\rho_\mu^a)^2  \right) \label{71}
\end{align}
after integrating out the $u$-dependences, leads to a normalization condition
\begin{equation} \label{normcondition}
\int_{u_{0,d}}^\infty du \hspace{1mm} u^{1/4} f_1 \psi^2 = 1
\end{equation}
and a mass term condition
\begin{equation} \label{masscondition}
\int_{u_{0,d}}^\infty du \hspace{1mm} \left\{ u^{1/4} g_{11} f_2 (\partial_u \psi)^2 +  u^{1/4} \frac{g_{11}}{(2\pi\alpha')^2} \tilde f_2  (2\overline \tau^3)^2\psi^2  \right\} = m_\rho^2
\end{equation}
on the $\psi(u)$ functions\footnote{We absorbed the total prefactor $\sqrt{V_4 T_8 g_s^{-1} R^{\frac{3}{4}+3} (2\pi\alpha')^2}$ into $\psi$ such that $\psi$ has a total mass dimension of $5/8$ instead of $2$ (without the prefactor).},
which
combine through partial integration
to an eigenvalue equation for $\psi(u)$:
\begin{equation} \label{eigeq}
u^{-1/4} f_1^{-1} \partial_u\left(u^{1/4} g_{11} f_2 \partial_u \psi \right) -  \frac{g_{11}}{(2\pi\alpha')^2} f_1^{-1}\tilde f_2 (2\overline \tau^3)^2 \psi  = -\Lambda \psi,
\end{equation}
with the eigenvalue $\Lambda = m_\rho^2$ the sought for rho meson mass squared.
We can separate the Higgs contribution to $m_\rho^2$ by defining
\begin{align} \label{Higgscontr}
\tilde m_\rho^2 &= \int_{u_{0,d}}^\infty du \hspace{1mm} u^{1/4} g_{11} f_2 (\partial_u \psi)^2 \quad \text{ and } \quad  m_{\rho,Higgs}^2 =  \int_{u_{0,d}}^\infty du \hspace{1mm}  u^{1/4} \frac{g_{11}}{(2\pi\alpha')^2} \tilde f_2  (2\overline \tau^3)^2\psi^2
\end{align}
such that
\begin{equation}
m_\rho^2 = \tilde m_\rho^2 + m_{\rho,Higgs}^2.
\end{equation}
Let us also mention that from (\ref{masscondition}) one can see that $m_\rho^2 >0$.

To solve the eigenvalue equation (\ref{eigeq}) numerically on a compact interval, we change to the coordinate $x=0\cdots x_{up} \cdots \frac{\pi}{2}$
related to $u=u_{0,d}\cdots u_{0,u} \cdots \infty$ by
\begin{equation}
u^3 = u_{0,d}^3 \cos^{-2} x.
\end{equation}
Rewritten as a function of $x$, the eigenvalue equation is invariant under $x\rightarrow-x$, so we can split up the eigenfunction set in even/odd $\psi_n(x)$'s, which correspond to odd/even parity mesons:
\begin{equation}
    \psi_n(0)=0\quad\text{or}\quad\partial_x\psi_n(0)=0.
\end{equation}

Asymptotically, the eigenvalue equation (\ref{eigeq}) reduces to $\partial_u \left( u^{5/2} \partial_u  \psi \right) = 0$, with the asymptotic solution $\psi(\infty) = c \frac{u^{-3/2}}{-3/2} + d$ only normalizable through (\ref{normcondition})  if $d=0$, i.e.~if
\begin{equation} \label{bdy}
\psi(u\rightarrow \infty) = 0 \quad \text{or} \quad \psi(x\rightarrow \pm \pi/2) = 0.
\end{equation}

The eigenvalue problem (\ref{eigeq}) for the (odd parity) rho meson with the appropriate boundary condition (\ref{bdy}) in the $x$-coordinate  is thus of the form
\begin{equation} \label{problem}
\cdots \partial_x^2\psi + \cdots \partial_x\psi + \cdots \psi = -\Lambda \psi \quad \text{with}\qquad \psi(\pm \pi/2)=0\,, \quad \partial_x\psi(0)=0.
\end{equation}
To solve it we employ a shooting method, which consists of temporarily replacing (\ref{problem}) with the well-defined initial value problem
\begin{equation} \label{diffeq}
    \cdots \partial_x^2\psi + \cdots \partial_x\psi + \cdots \psi = -\Lambda \psi \quad \text{with}\qquad \psi(0)=1\,, \quad \partial_x\psi(0)=0
\end{equation}
where $\Lambda$ is treated as a `shooting' parameter. We used the scaling freedom $\psi(x)\to h\psi(x)$ to impose that $\psi(0)=1$ (the value of $h$ will be fixed by the normalization condition in the end). For each value of $\Lambda$, (\ref{diffeq}) can be solved numerically for $\psi_\Lambda(x)$.
Next, solving the equation $\psi_\Lambda(\pi/2)=0$ finally determines the eigenvalue $\Lambda = m_\rho^2$.

\begin{figure}[h!]
  \hfill
  \begin{minipage}[t]{.3\textwidth}
    \begin{center}
      \scalebox{0.5}{
  \includegraphics{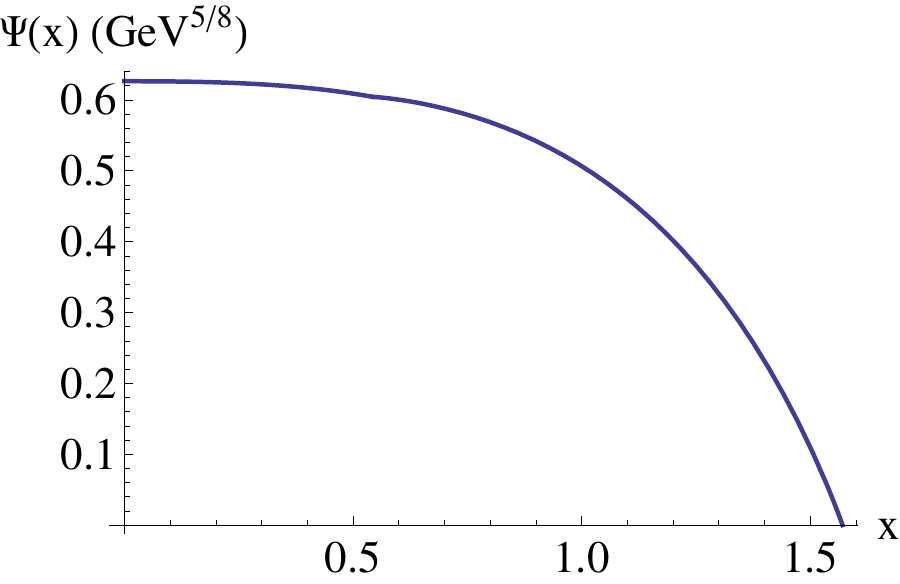}}
    \end{center}
  \end{minipage}
  \hfill
  \begin{minipage}[t]{.3\textwidth}
    \begin{center}
      \scalebox{0.5}{
  \includegraphics{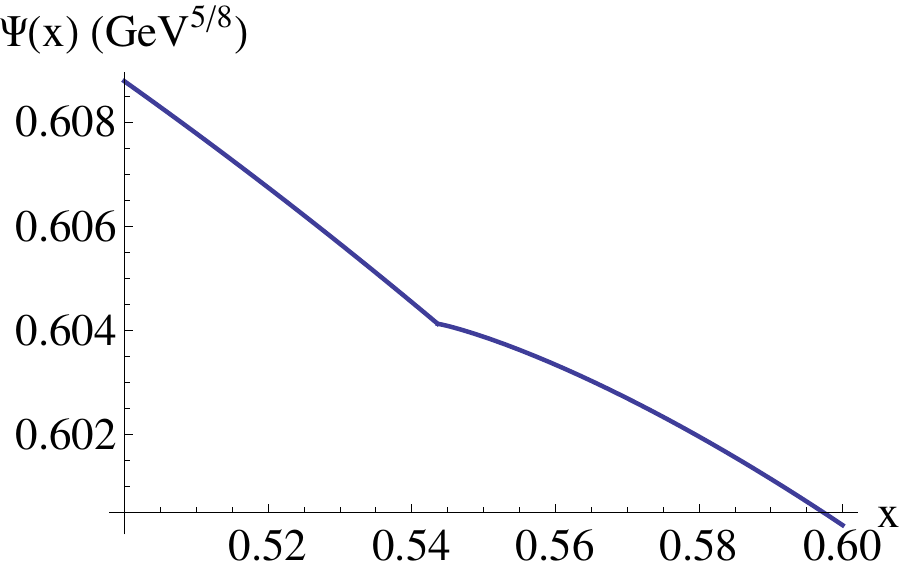}}
    \end{center}
  \end{minipage}
 \hfill
  \begin{minipage}[t]{.3\textwidth}
    \begin{center}
      \scalebox{0.5}{
  \includegraphics{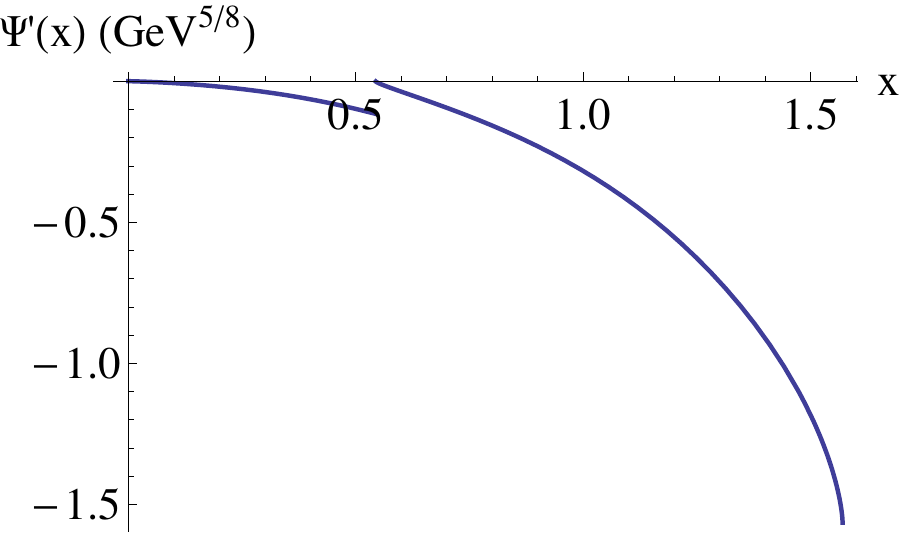}}
    \end{center}
  \end{minipage}
      \caption{Plots of the rho meson eigenfunction $\psi(x)$ and its derivative $\partial_x \psi(x)$, discontinuous at $x=x_{up} \approx 0.54$, for $B=0.9$ GeV$^2$, obtained numerically with a shooting method.}
	\label{psifigs}
  \hfill
\end{figure}

For completeness we add a few comments about the numerical method we used to solve the eigenvalue problem at hand (\ref{problem}), which in detail reads
\begin{align}
&\frac{9}{4}R^{-3/2}u_{0,d}^{-1/2} \frac{\cos^{11/6} x}{\sin x} f_{1}^{-1} \partial_x \left( f_{2}\frac{\cos^{1/2} x}{\sin x}\partial_x \psi \right)
- R^{-3/2} \frac{u_{0,d}^{3/2}}{(2\pi\alpha')^2} (\cos^{-1} x) \tilde f_{2} f_{1}^{-1} (2\overline \tau^3)^2 \psi = -m_\rho^2 \psi,
\end{align}
with $\psi(\pm \pi/2)=0$ and $\partial_x \psi(0)=0$. Near the origin $x \rightarrow 0$ the equation takes the form
\begin{align}
m_\rho^2 \psi + \partial_x^2 \psi - \ln x \hspace{1mm} \partial_x^2 \psi -\frac{1}{x}\partial_x \psi  &= 0,  \label{originbehaviour}
\end{align}
so we have to provide Mathematica with an ansatz for $\psi(x)$ at small $x$ to prevent the equation from blowing up there.
Demanding that $\partial_x \psi \sim x$ to avoid the last term in (\ref{originbehaviour}) from diverging, would still give $ \ln x \hspace{1mm} \partial_x^2 \psi \rightarrow -\infty$. Instead we demand that  $\partial_x^2 \psi \sim \frac{1}{\ln x}$ or $\psi(x\rightarrow 0) = 1 + x^2 \sum_{i=1}^n \frac{a_i}{\ln^i x}$ (in practice we have set $n=13$).
With this ansatz for $\psi \Rightarrow \partial_x \psi \sim \text{LogIntegral}(x) + c$, the term $\frac{1}{x}\partial_x \psi$ will only be finite if the integration constant $c=\partial_x \psi(0) = 0$ \footnote{This is consistent with vector mesons, but not with the initial condition on axial mesons (which we have not considered). We have not looked into it further to see if there is a way around this, in order to still be able to describe axial mesons in the presence of a magnetic field in this setting.}. Near $x = x_{up}$, or $y\rightarrow 0$ in the useful coordinate $y$ defined through $u^3 = u_{0,u}^3 \cos^{-2} y$, the differential equation's form
\begin{align}
m_\rho^2 \psi + \partial_y^2 \psi - \ln y \hspace{1mm} \partial_y^2 \psi -\frac{1}{y}\partial_y \psi  &= 0,  \label{xupbehaviour}
\end{align}
again needs to be fed with an ansatz for $\psi$ that keeps the equation finite, i.e.~$\psi(y\rightarrow 0) = \psi(x=x_{up}) + y^2 \sum_{i=1}^n \frac{a_i}{\ln^i y}$ with $\partial_y \psi(0) = 0$.
This means we can demand continuity of $\psi$ at $x=x_{up}$ but not of its derivative\footnote{It is known that the Schr\"odinger wave function can display kinks (thus jumps in its derivative), depending on the potential (singularities), see e.g.~\cite{levi}. This corresponds to the singular behaviour of some of the coefficient functions for $y\to0$ in \eqref{xupbehaviour}.}.
An example result of $\psi(x)$ and its derivative is shown in Figure \ref{psifigs}.

\paragraph{Effective 4-dimensional EOM and result for total eigenvalue}  \label{paragraaf2}

The effective 4-dimensional action becomes
\begin{align}
S_{4D}
&= \int d^4 x \sum_{a,b=1}^2 \left\{- \frac{1}{4}(\mathcal F_{\mu\nu}^{a})^2 -\frac{1}{2} m_\rho^2(B) (\rho_\mu^a)^2 + \sum_{\mu=1}^2 \left( -\frac{1}{2} m_{+}^2(B)  (\rho_\mu^a)^2 -\frac{1}{2} \epsilon_{3ab}\rho_\mu^a \rho_\nu^b \hspace{1mm} k(B) \overline F_{\mu\nu}^3 \right)\right\} \label{S4D}
\end{align}
with the normalized $\psi$, as determined in the previous paragraph, satisfying the normalization and mass conditions (\ref{normcondition}) and (\ref{masscondition}),
and the newly defined $m_+$ and $k$ to be calculated from
\begin{equation} \label{massplus}
\int_{u_{0,d}}^\infty du \hspace{1mm} u^{1/4} g_{11}^{-1} f_3 (2\overline \tau^3)^2 \psi^2 = m_{+}^2
\end{equation}
and
\begin{equation}
\int_{u_{0,d}}^\infty du \hspace{1mm} u^{1/4} (\sqrt{G_{uu}}\overline F_{12})^3 \psi^2 = k \hspace{1mm} \overline F_{12}^3
\end{equation}
with $\overline F_{12}^3=B$.
Here $m_+$ is an extra contribution to the mass of the transverse (w.r.t. the magnetic field $\vec B = B \vec e_3$) components of the charged rho meson, $\rho_{\mu=1,2}^{a=1,2}$, as a consequence of $B$ breaking Lorentz invariance. The parameter $k$ describes a non-minimal coupling of the charged rho meson to the magnetic field, related to the magnetic moment $\mu$ via $\mu = (1+k)e/(2m)$ so to the gyromagnetic ratio $g$ via $g=1+k$.

\begin{figure}[h!]
  \hfill
  \begin{minipage}[t]{.31\textwidth}
    \begin{center}
      \scalebox{0.55}{
  \includegraphics{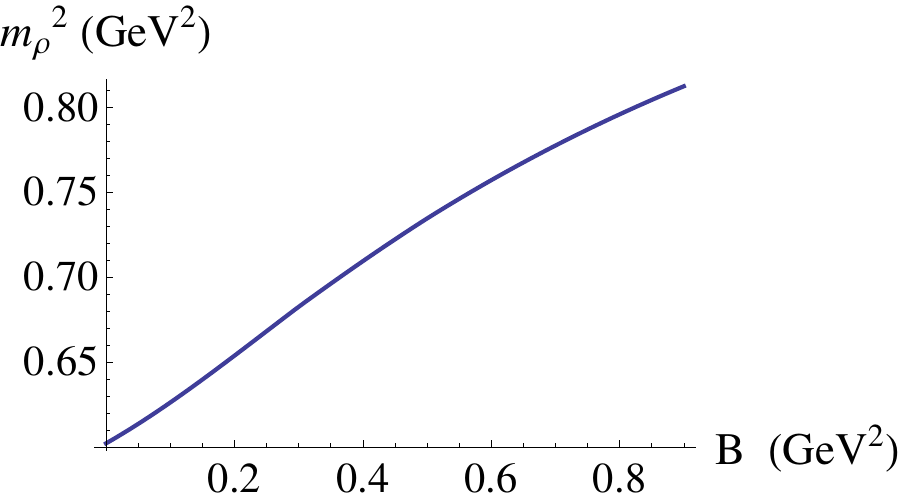}}
    \end{center}
  \end{minipage}
  \hfill
  \begin{minipage}[t]{.31\textwidth}
    \begin{center}
      \scalebox{0.55}{
  \includegraphics{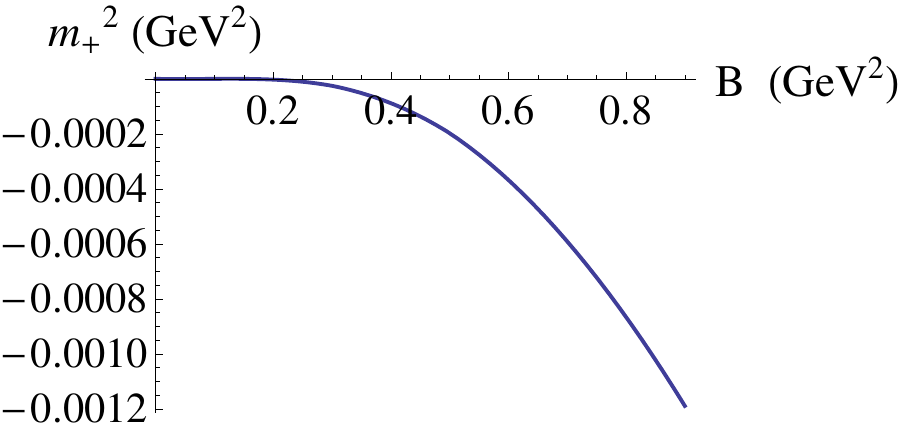}}
    \end{center}
  \end{minipage}
\hfill
  \begin{minipage}[t]{.31\textwidth}
    \begin{center}
      \scalebox{0.55}{
  \includegraphics{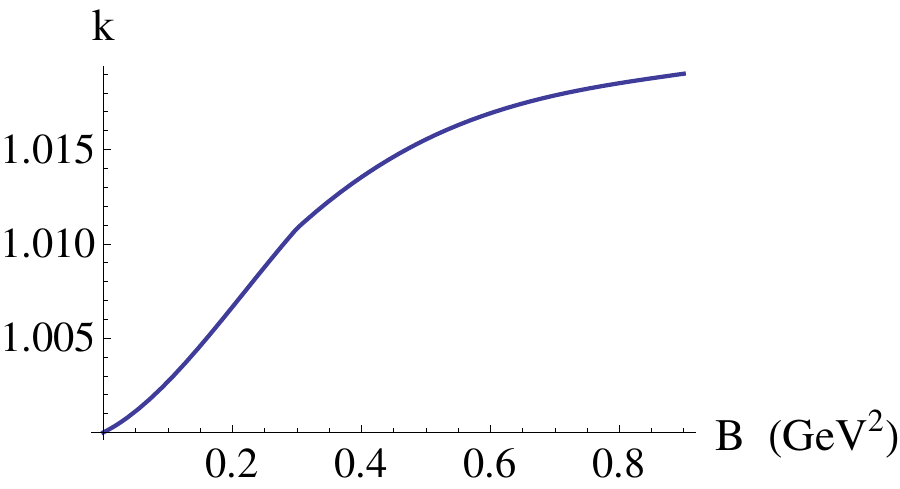}}
    \end{center}
  \end{minipage}
      \caption{Numerical results for $m_\rho^2(B)$, $m_+^2(B)$ and $k(B)$ in the $(2\pi\alpha')^2 F^2$-approximation of the DBI-action.
}
	\label{resultsfig}
  \hfill
\end{figure}

\begin{figure}[h!]
  \centering
  \scalebox{0.6}{
  \includegraphics{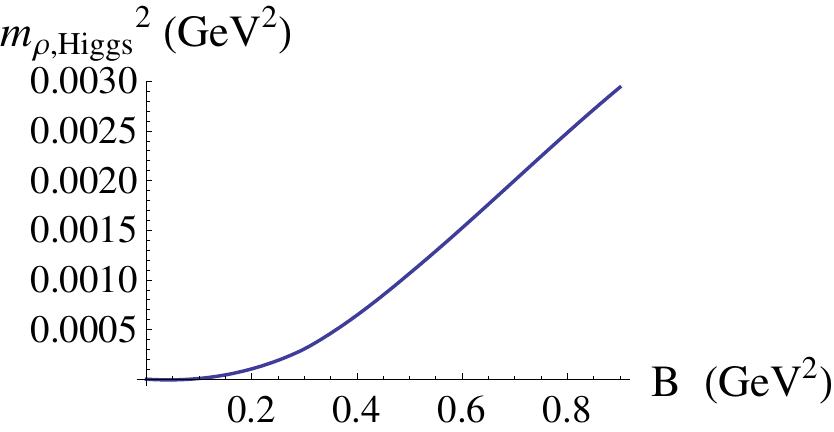}}
  \caption{The Higgs contribution $m_{\rho,Higgs}^2(B)$ to the rho meson mass squared $\mathit{m_{\rho}^2}(B)$, as defined in (\ref{Higgscontr}), in the $(2\pi\alpha')^2 F^2$-approximation of the DBI-action.
} \label{mHiggs2fig}
\end{figure}

The standard 4-dimensional action used to describe the coupling of charged rho mesons to an external magnetic field is the Proca action \cite{Obukhov:1984xb} (which is equivalent to the DSGS-action \cite{Djukanovic:2005ag} for self-consistent rho meson quantum electrodynamics to second order in the fields). The Proca action is equal to (\ref{S4D}) with $m_+ = 0$ and $m_\rho$ and $k(=1)$ independent of $B$: there is only explicit dependence of the action on $B$, which is to be traced back to the treatment of the rho mesons as point-like structureless particles.
Instead, in our approach, the effect of $B$ on the constituent quarks
is taken into account via the effect of $B$ on the embedding of the flavour probe branes\footnote{In the antipodal Sakai-Sugimoto model where the embedding is $B$-independent, one recovers exactly the Proca action \cite{Callebaut:2011ab}.}, leading to an implicit dependence on $B$ of both the mass $m_\rho^2(B)$ and the magnetic coupling $k(B)$. The effect of $B$ on the embedding is two-fold (see Section \ref{embsection} and in particular Figure \ref{changedembedding}): the branes move upwards in the holographic direction, corresponding to chiral magnetic catalysis, and the up- and down-brane get separated, corresponding to a stronger chiral magnetic catalysis
for the up-quark than for the down-quark.
Both effects translate into a mass generating effect for the rho meson, $m_\rho^2(B) \nearrow$, as can be seen in Figure \ref{resultsfig}. The chiral magnetic catalysis causes the rho meson to get heavier as its constituents do.
The split between the branes adds to the mass of the rho meson via a holographic Higgs mechanism: as the branes separate, the flavour gauge field strings between up and down branes (i.e.~representing charged quark-antiquark combinations $u\overline d$, $\overline u d$) get stretched. Because of their string tension this results in an extra Higgs mass term in the action for $\tilde A_\mu^{a=1,2}$ -- and thus for $\rho_\mu^{a=1,2}$ -- of the form $(A_\mu^a)^2 (\overline \tau^3)^2$, with $\overline \tau^3 \sim  \overline \tau_u - \overline \tau_d$, originating from $(D_\mu \tau)^2 \leadsto ([\tilde A_\mu,\overline \tau])^2$ in the start action.
Where in the absence of splitted branes, $\overline \tau^3=0$, the 4-dimensional mass $m_\rho$ as defined in going from (\ref{70}) to (\ref{71}) is purely effective, i.e.~only present after integrating out the fifth dimension $u$, the Higgs contributions to the mass stem from the stringy mass of the 5-dimensional gauge field itself.
A direct interpretation of the stringy mass contribution in effective QCD-terms we cannot offer.
Since the splitting of the branes is small though, the induced mass contribution is almost negligible, see Figure \ref{mHiggs2fig}.
Further, as can be seen in Figure \ref{resultsfig}, $m_+(B) \searrow$ as $f_3$ in (\ref{massplus}) is negative, so the mass of the transversal components of the charged rho mesons will already be slightly smaller than that of the longitudinal ones,
and $k(B) \nearrow$ is approximately equal to one, but not exactly, corresponding to a gyromagnetic ratio $g\approx 2$.


The 4-dimensional EOMs for the charged rho mesons $\rho_\mu^{a=1,2}$ are given by
\begin{align}\label{EOMeq}
D_\mu \mathcal F_{\mu\nu}^a - \epsilon_{a3b} \hspace{1mm} k(B) \hspace{1mm} \overline F_{\mu\nu}^3 \rho_\mu^b - M^2(B) \rho_\nu^a = 0,  \\
M^2(B) = m_\rho^2(B) + (\delta_{\nu 1}+\delta_{\nu 2}) m_{+}^2(B)
\end{align}
with $D_\mu = \partial_\mu +[\overline A_\mu,\cdot]$ and $\mathcal F_{\mu\nu}^a = D_\mu \rho_\nu^a - D_\nu \rho_\mu^a$.
They combine into the EOM
\begin{equation}  \label{rhoeq}
\text{D}_\mu(\text D_\mu \rho_\nu - \text D_\nu \rho_\mu) - i  \hspace{1mm}k(B) \hspace{1mm} \overline F_{\mu\nu}^3 \rho_\mu - M^2(B) \rho_\nu = 0
\end{equation}
with $\text D_\mu = \partial_\mu + i \overline A_\mu^3$ for the charged  combination $\rho_\mu = (\rho_\mu^1 + i \rho_\mu^2)/\sqrt 2$, and the complex conjugate of this equation for the other charged combination $\rho_\mu^* = (\rho_\mu^1 - i \rho_\mu^2)/\sqrt 2$.

Solving (\ref{rhoeq}) with $\rho_\nu \rightarrow e^{i(\vec p \cdot \vec x - E t)} \rho_\nu$ for the eigenvalues of the energy, one finds `modified Landau levels' that we will discuss in more detail in the next Section, where they will show up as a special case of the most general form of modified Landau levels that we encounter solving the 4-dimensional EOMs that come from the use of the full DBI-action.
Only in the case that $k=1$, $m_+=0$ and $m_\rho(B)=m_\rho(0)$ one retrieves the standard Landau levels for a free relativistic spin-$s$ particle  moving in the background of a constant magnetic field $\vec B=B \vec e_3$ (assuming $B>0$):
\begin{align}
E^2 = m_\rho^2 + p_3^2 + (2n - 2 s_3 + 1) B
\end{align}
with $n$ the Landau level number and $s_3$ the eigenvalue of the spin operator
\begin{equation}  \label{S3spin}
S_3 = \frac{1}{2} \left( \begin{array}{cc} 0 & \sigma_2-i \sigma_1 \\
\sigma_2+i \sigma_1 & 0 \end{array}\right)
\end{equation}
giving the projection of the spin of the particle onto the direction of the magnetic field.

\begin{figure}[h!]
  \centering
  \scalebox{0.7}{
  \includegraphics{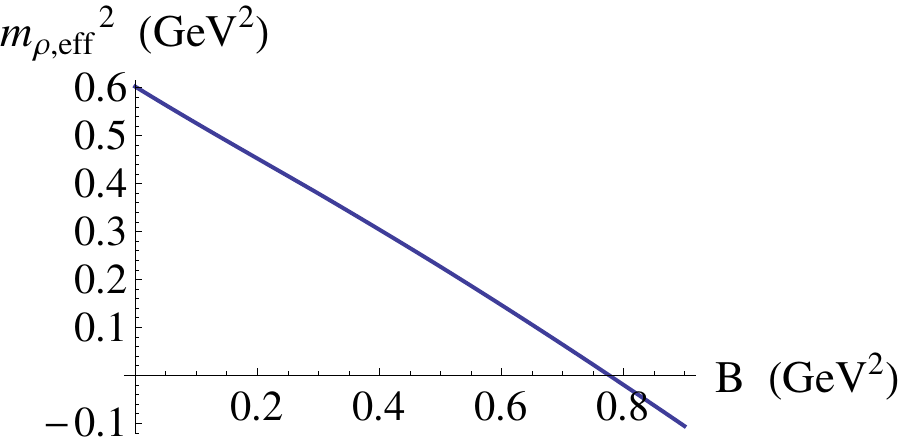}}
  \caption{The effective rho meson mass squared $\mathit{m_{\rho,eff}^2}(B)$  in the $(2\pi\alpha')^2 F^2$-approximation of the DBI-action.
} \label{mrhoeff2fig}
\end{figure}

While the modifications due to $k\neq1$, $m_+(B)\neq0$ and $m_\rho(B)$ are a bit subtle for higher levels, the energy of $s_3=1, p_3=0$ particles in the lowest Landau level $n=0$
is given by a straightforward generalization
\begin{equation} \label{}
E^2 = M^2(B) -  B  \hspace{1mm} k(B)
\end{equation}
of $E^2 = m_\rho^2 - B$.
We conclude that the combinations of charged rho mesons that have their spin aligned with the magnetic field, $s_3 = 1$,  i.e.
\begin{align}
\rho = \rho_1 + i \rho_2 \quad \mbox{ and }\quad  \rho^* =  \rho_1^* - i \rho_2^*, \label{fieldcom}
\end{align}
will have an effective mass squared
\begin{equation} \label{mrhoeff2}
\mathit{m_{\rho,eff}^2} = M^2(B) - B \hspace{1mm} k(B)
\end{equation}
going through zero at a critical magnetic field
\begin{equation} \label{Bc}
B_c \approx 0.78 \text{ GeV}^2,
\end{equation}
which marks the onset of rho meson condensation. Our result for $\mathit{m_{\rho,eff}^2}$ is shown in Figure \ref{mrhoeff2fig}.

The total action includes, next to the DBI-part, a Chern-Simons term. In general,  contributions from the Chern-Simons action are suppressed in the large
$\lambda$ expansion,
but in the presence of large background fields Chern-Simons effects can become important, similar to the higher order terms in the $(2\pi\alpha' \sim \frac{1}{\lambda})$-expansion of the DBI-action (see comments in the upcoming Section \ref{ambiguities}).
The intrinsic-parity-odd nature of the Chern-Simons action ensures that it will not contribute $\rho^2$-terms to the effective 4-dimensional action to second order in the fluctuations, but it will describe $\rho \pi B$ coupling terms between rho mesons and pions. However, as discussed in more detail in \cite{Callebaut:2011ab},
the antisymmetrization over spacetime indices in the Chern-Simons action
\begin{align}
S_{CS}
&\sim  \int \text{Tr} \left( \epsilon^{mnpqr} A_m F_{np} F_{qr} + \mathcal O(\tilde A^3) \right)
\end{align}
will make sure that the magnetic field $B = \overline F_{12}^3$ only induces couplings between longitudinal fluctuations ($\mu=0,3$), hence not affecting the
dynamics of transversal rho mesons (\ref{fieldcom}) and their condensation.

\subsubsection{Pion mass} \label{4.3.2}

We briefly discuss the pion part of the DBI-Lagrangian (\ref{rhopi}), which upon substitution of the gauge field expansion (\ref{Auexpansion}) and further approximation to second order in the pion fields reads
\begin{align}
\hspace{-1.2cm}
\mathcal L \sim u^{1/4}  (2\pi\alpha')^2 \sum_{a,b=1}^2 \left\{- \frac{1}{2} g_{11} f_2 (D_\mu \pi^a)^2 \phi_0^2  - \frac{1}{2} \frac{g_{11}^2}{ (2\pi\alpha')^2} \tilde f_4 (2\overline \tau^3)^2  (\pi^a)^2 \phi_0^2
\right\}
\end{align}
with $\tilde f_4 = f_4 -  \frac{1}{2} g_{11}^{-2}  (2\pi\alpha')^2  f_5$.
Ignoring in this Section the $1/\lambda$-suppressed $\rho \pi B$-contributions from the Chern-Simons action,
the effective 4-dimensional action for the pions becomes
\begin{align}
S_{4D} = \int d^4 x \sum_{a,b=1}^2 \left\{- \frac{1}{2}(D_\mu \pi^{a})^2 -\frac{1}{2} m_\pi^2(B) (\pi^a)^2 \right\}
\end{align}
with $\phi_0$ satisfying the normalization condition
\begin{equation} \label{normpion}
\int_{u_{0,d}}^\infty du \hspace{1mm} u^{1/4} g_{11} f_2 \phi_0^2 = 1
\end{equation}
and the pions no longer massless:
\begin{equation} \label{mpion}
\int_{u_{0,d}}^\infty du \hspace{1mm} u^{1/4} \frac{g_{11}^2}{ (2\pi\alpha')^2} \tilde f_4 (2\overline \tau^3)^2  \phi_0^2  = m_{\pi}^2.
\end{equation}
We can understand the emergence of this mass again as a consequence of the holographic Higgs mechanism. The magnetic field breaks chiral symmetry explicitly (albeit only slightly) by pulling the up- and down-brane apart. The previously massless pions, serving as Goldstone bosons associated with the spontaneous breaking of chiral symmetry, hence get a small mass, related to the distance $\overline \tau^3 \sim \overline \tau_u - \overline \tau_d$ between the branes.
Solving the effective 4-dimensional EOM for the charged pions with $\pi \rightarrow e^{i(\vec p \cdot \vec x - E t)} \pi$ for the eigenvalues of the energy, one finds `almost Landau levels' for a spinless particle
\begin{align}
E^2 = m_\pi^2(B)+ p_3^2 + (2n + 1) B
\end{align}
or an effective mass squared in the lowest Landau level
\begin{equation} \label{}
\mathit{m_{\pi,eff}^2} = m_\pi^2(B)  + B.
\end{equation}
The pion thus gets a mass in the presence of a magnetic field, although we are working in a model in the chiral limit (zero bare quark masses) and with no chiral condensate (at least not in the setting we used, without incorporating a tachyon field as was done in \cite{Bergman:2007pm}). This violates the GMOR-relation relating the  bare quark masses times chiral condensate to the mass of the pion. It was however already discussed in e.g.~\cite{Shushpanov:1997sf,Orlovsky:2013gha} that the GMOR-relation is no longer valid for charged pions in the presence of a magnetic field.

\begin{figure}[h!]
  \hfill
\hspace{-1cm}
  \begin{minipage}[t]{.3\textwidth}
    \begin{center}
      \scalebox{0.5}{
  \includegraphics{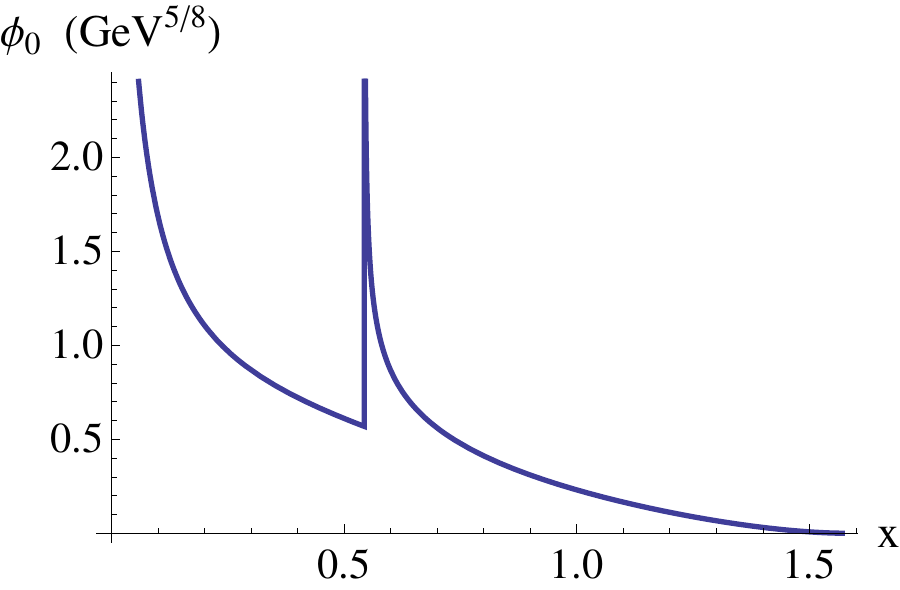}}
    \end{center}
  \end{minipage}
  \hfill
\hspace{-1cm}
  \begin{minipage}[t]{.35\textwidth}
    \begin{center}
      \scalebox{0.6}{
  \includegraphics{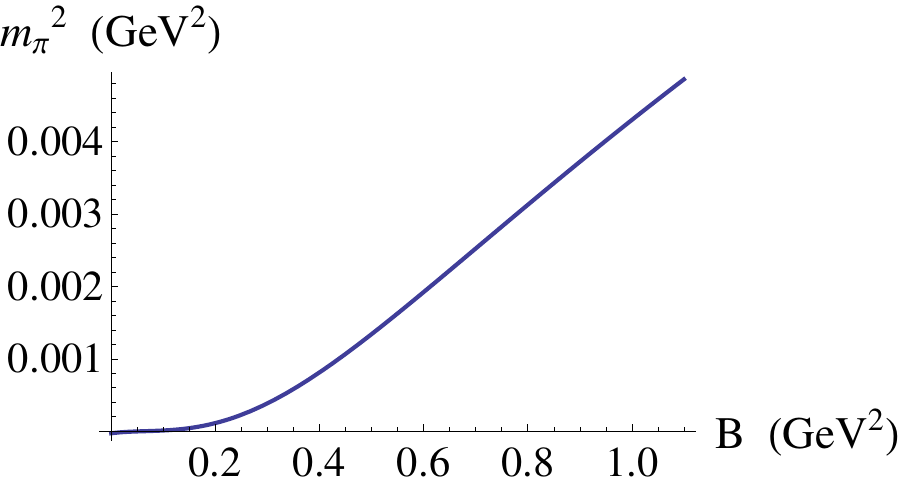}}
    \end{center}
  \end{minipage}
\hfill
\hspace{-1cm}
  \begin{minipage}[t]{.35\textwidth}
    \begin{center}
      \scalebox{0.6}{
  \includegraphics{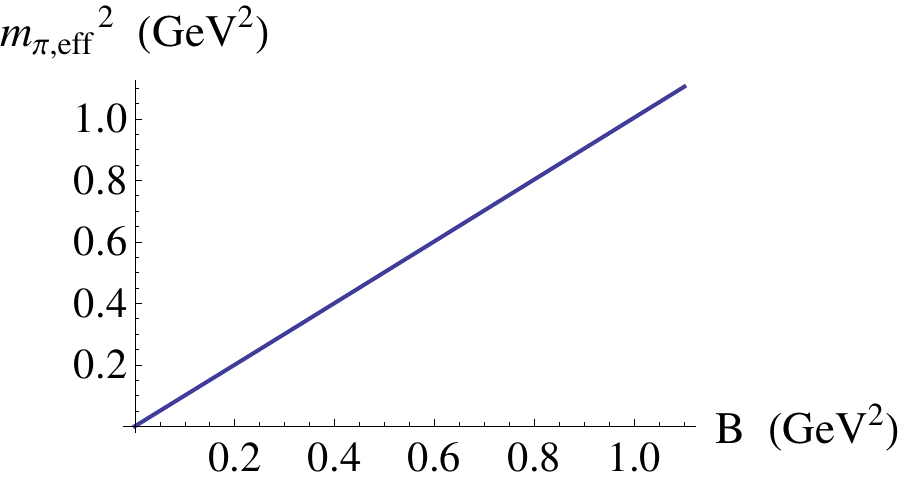}}
    \end{center}
  \end{minipage}
      \caption{Pion eigenfunction $\phi_0(x)$ (with $u=u_{0,d} \cos^{-3/2}x$) for $B=1.1$ GeV$^2$. Numerical result for $m_\pi^2(B)$ and the effective pion mass squared $\mathit{m_{\pi,eff}^2(B)}$ in the $(2\pi\alpha')^2 F^2$-approximation of the DBI-action.
}
	\label{pions}
  \hfill
\end{figure}

To calculate the mass $m_\pi$ in (\ref{mpion}),
we determine the form of the eigenfunction $\phi_0(u)$ analogously as in \cite{Sakai:2004cn}.
$\phi_0$ has to be orthogonal to all other $\phi_{n\geq 1}$ (the higher eigenfunctions that we left out in the expansion (\ref{Auexpansion})). The eigenfunctions $\phi_{n\geq 1}$ obey the same normalization condition (\ref{normpion}) as $\phi_0$, which upon comparison with the mass condition (\ref{Higgscontr}) for $\psi_{n \geq 1}$,
\begin{align}
&\int_{u_{0,d}}^\infty du \hspace{1mm} u^{1/4} g_{11} f_2 \phi_{n\geq 1}^2 = 1
\quad \text{and} \quad \int_{u_{0,d}}^\infty du \hspace{1mm} u^{1/4} g_{11} f_2 (\partial_u \psi_{n\geq1})^2  = \tilde m_\rho^2,
\end{align}
leads to
\begin{equation}
\phi_{n\geq 1} = \frac{\partial_u \psi_{n\geq 1}}{\sqrt{\tilde m_\rho^2}}.
\end{equation}
Then, orthogonality of $\phi_0$ and $\phi_{n\geq1} \sim \partial_u \psi_{n\geq1}$ is ensured
by proposing
\begin{equation} \label{phi0}
\phi_0 \sim u^{-1/4} g_{11}^{-1} f_2^{-1}
\end{equation}
(with normalization constant determined by the normalization condition (\ref{normpion})):
\begin{equation} \label{}
\int_{u_{0,d}}^\infty du \hspace{1mm} \phi_0 (u^{1/4} g_{11} f_2 \phi_{n\geq1})  \sim \int du  \hspace{1mm}\partial_u \psi_{n\geq1} = 0
\end{equation}
by virtue of the vanishing of $\psi_{n\geq1}$ at the boundary $u\rightarrow \infty$.
With $\phi_0$ given in (\ref{phi0}) we can determine the Higgs contribution to the mass $m_\pi$. In Figure \ref{pions} we plot the eigenfunction $\phi_0(u)$ (which is discontinuous due to the discontinuous nature of $f_2$), the mass $m_\pi$ and the total effective 4-dimensional mass $\mathit{m_{\pi,eff}}$.

We end this Section with a comment on the validity of the use of the non-Abelian DBI-action for non-coincident branes\footnote{We would like to thank  K.~Jensen for a private discussion about this.}.

In the context of heavy-light mesons, which we encounter here as magnetically induced through the splitting of the flavour branes, one often
studies the separated branes system by the use of
two (Abelian) DBI-actions plus a Nambu-Goto action for the classical, i.e.~macroscopic,  heavy-light meson string (e.g.~\cite{Erdmenger:2006bg}).
In \cite{Erdmenger:2007vj} however, one uses the non-Abelian DBI action for the description of heavy-light mesons, as we also did in this paper. They do remark
that as soon as the distance between the separated branes is larger than the fundamental string length $\ell_s$, the non-Abelian DBI-description is actually expected to break down.
So let us show here that in our case the separation between up- and down-brane and hence the length of the charged rho meson strings is not larger than $\ell_s$.

\begin{figure}[h!]
  \hfill
  \begin{minipage}[t]{.45\textwidth}
    \begin{center}
      \scalebox{0.7}{
  \includegraphics{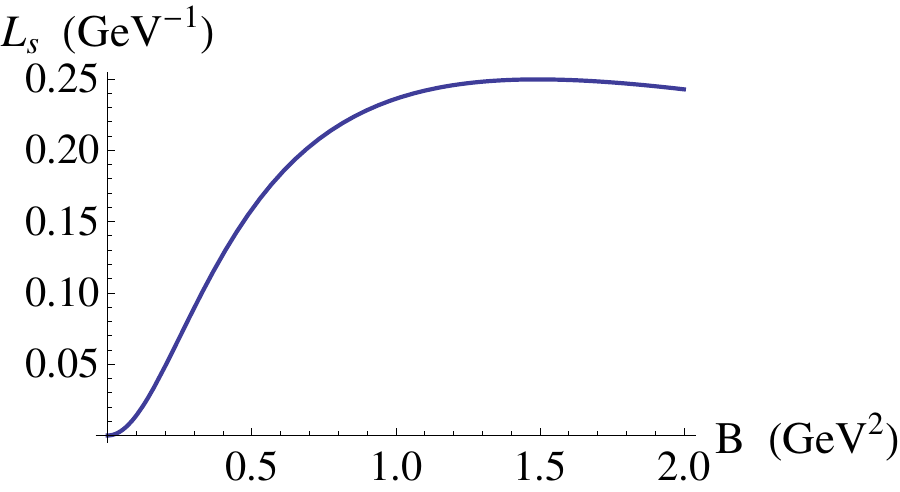}}
    \end{center}
  \end{minipage}
  \hfill
  \begin{minipage}[t]{.45\textwidth}
    \begin{center}
      \scalebox{0.7}{
  \includegraphics{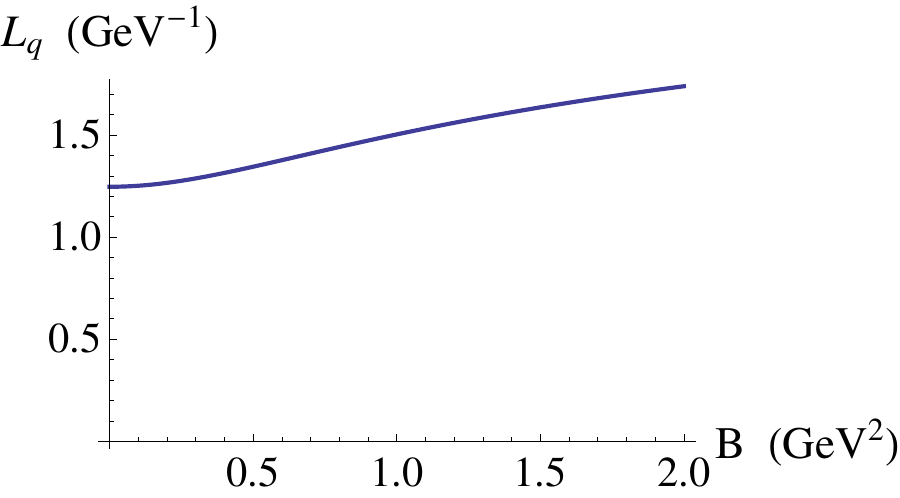}}
    \end{center}
  \end{minipage}
      \caption{The length $L_s(B)$ of a string at $\tau=0$ stretching between $u_{0,d}(B)$ and $u_{0,u}(B)$, and the length $L_q(B)$ of a down-quark string stretching between $u_K$ and $u_{0,d}(B)$.
}
	\label{Lsq}
  \hfill
\end{figure}

The total length of a string stretching in the $u$- and $\tau$-direction is given by
\[
L_s = \int ds = \int \sqrt{g_{\tau\tau} d\tau^2 + g_{uu}du^2}.
\]
Consider for example a string at $\tau=0$
 stretching from $u_{0,d}(B)$ to $u_{0,u}(B)$. It has a length
\begin{align*}
L_s &= \int ds = \int_{u_{0,d}(B)}^{u_{0,u}(B)} \sqrt{g_{uu}} du \nonumber\\
&= \int_{u_{0,d}(B)}^{u_{0,u}(B)} \left(\frac{R}{u}\right)^{3/4} f(u)^{-1/2} du  \nonumber\\
&= -\frac{R^{3/4}}{11 u_{0,d}^2 u_{0,u}^2 \sqrt{u_{0,d}^3-u_K^3} \sqrt{u_{0,u}^3-u_K^3}}4 (u_{0,d} u_{0,u})^{3/4} \nonumber\\
&\qquad \times
 \left\{ 11 u_{0,d}^3 u_{0,u}^{5/4} \sqrt{u_{0,u}^3-u_K^3}-6 u_{0,u}^{5/4} u_K^3 \sqrt{u_{0,u}^3-u_K^3}+u_{0,d}^{5/4} \sqrt{u_{0,d}^3-u_K^3} \left(-11 u_{0,u}^3+6 u_K^3\right) \right. \nonumber\\
&\qquad  \left.+6 u_K^3 \left(u_{0,u}^{5/4} \sqrt{u_{0,u}^3-u_K^3} \; {}_2 F_1\left[-\frac{11}{12},1,\frac{7}{12},\frac{u_{0,d}^3}{u_K^3}\right]-u_{0,d}^{5/4} \sqrt{u_{0,d}^3-u_K^3} \; {}_2 F_1\left[-\frac{11}{12},1,\frac{7}{12},\frac{u_{0,u}^3}{u_K^3}\right]\right)\right\},
\end{align*}
with the $B$-dependence of $u_{0,u}$ and $u_{0,d}$ implicit in the last line.
Similarly, the same string stretching between $u_0$ and $u_K$, corresponding to a constituent quark (i.e.~this one \emph{is} a macroscopic string, cfr. the use of the Nambu-Goto action to obtain the expression for the constituent quark mass (\ref{constmass})) has a length
\begin{align*}
L_q &= \int ds = \int_{u_K}^{u_0} \sqrt{g_{uu}} du \nonumber\\
&= \int_{u_K}^{u_0} \left(\frac{R}{u}\right)^{3/4} f(u)^{-1/2} du \nonumber\\
&= R^{3/4}\left(-\frac{4 \sqrt{\pi } u_K^{1/4} \Gamma \left[\frac{11}{12}\right]}{\Gamma \left[\frac{5}{12}\right]}+4 u_0^{1/4} \; {}_2 F_1 \left[-\frac{1}{12},\frac{1}{2},\frac{11}{12},\frac{u_K^3}{u_0^3}\right]\right).
\end{align*}
With our fixed holographic parameters, we have a numerical value for $\ell_s$ to compare these lengths to:
\begin{align*}
\ell_s=\sqrt{\alpha'}\approx 0.76  \hspace{1mm} \text{GeV$^{-1}$}.
\end{align*}
From the plots in Figure \ref{Lsq} of $L_s$ and $L_q$ as functions of $B$ up to 2 GeV$^2$, we read of estimations of the maximal $L_s \approx 0.25$ GeV$^{-1}$ and minimal $L_q \approx 1.25$ GeV$^{-1}$, from which we can conclude that
\[
L_s < \ell_s \quad \text{and} \quad L_q > \ell_s,
\]
consistent with using the classical Nambu-Goto action for the constituent quark string, but using the non-Abelian DBI-description for the charged rho meson string.

\subsection{Vector sector for full DBI-action} \label{4.4}

\subsubsection{Comments on the validity of the \texorpdfstring{$(2\pi\alpha')$-expansion}{inverse string tension expansion}}  \label{ambiguities}

In the previous Section \ref{F2approx} and the previous paper \cite{Callebaut:2011ab} we approximated the DBI-action to second order in $(2\pi\alpha') F$. The justification that we used for this expansion is roughly that $\alpha' \sim 1/\lambda$ with $\lambda \approx 15$ `large' in our fixed units.
The reader might worry that there is some ambiguity in the proportionality factor $\alpha' \sim 1/\lambda$ since the parameter $X=\lambda \ell_s^2$ can be chosen freely, as we did in (\ref{SS2(2.4)a}). The ambiguity should however disappear from all physical quantities and indeed will no longer be present in the full expansion parameter.
Let us take a closer look.

Expanding $\det (g_{mn}^{D8}+(2\pi\alpha') iF_{mn}) = \det g_{mk}^{D8} \times \det(\delta_{nk}+(g_{kr}^{D8})^{-1}(2\pi\alpha') i (\overline F_{rn}+ \delta_1 F_{rn}+\delta_2 F_{rn}))$ in the action (\ref{nonabelian}), the expansion parameter $(g_{11}^{D8})^{-1}(2\pi\alpha') i \overline F_{12}$ is supposed to be small compared to 1, with (\ref{Fbardef}):
\[
\left(\frac{u}{R}\right)^{-3/2} (2\pi \alpha') \left| \left(\begin{array}{cc} \frac{2}{3}eB  & 0 \\ 0 & -\frac{1}{3}eB \end{array} \right) \right| \ll 1.
\]
The same expansion parameter can be read off from the form of the matrix A as defined in (\ref{A}).
The most strict condition would then be
\[
\left(\frac{u_{0,d}(B=0)}{R}\right)^{-3/2} (2\pi \alpha') \frac{2}{3}eB \ll 1,
\]
or, in our fixed units,
\begin{align}
eB &\ll \frac{3}{2} \left(\frac{u_{0,d}(B=0)}{R}\right)^{3/2} (2\pi \alpha')^{-1}  \equiv  0.45 \text{ GeV}^2,  \label{Bmaxestimate}
\end{align}
with the appearing combination independent of our choice of $X$ since $u_0 \sim X$, $R^3 \sim X$ and $(2\pi \alpha') \sim X$.
The instability we found in the $F^2$-approximation sets in at $B_c \approx 0.8$ GeV$^2$ (see (\ref{Bc})), where the used approximation is thus not necessarily valid anymore.
On the other hand,
the above is the most strict condition we can impose, it is not so clear what the impact of the $u$-dependence is on this argument.
We will therefore use the full STr-action
 and compare with the $F^2$-approximation results to provide a conclusive answer to the question of the validity of the $(2\pi\alpha')$-expansion in our set-up.
It will turn out that using the full STr-action the instability is still present and the value of $B_c$ is only slightly higher.

In \cite{Bolognesi:2012gr} it is argued
that $\alpha'$-corrections can cause magnetically induced tachyonic instabilities of $W$-boson strings, stretching between separated D3-branes, to disappear when the inter-brane distance becomes larger than $2\pi \ell_s$. The Landau level spectrum for the $W$-boson is said to receive large $\alpha'$-corrections in general \cite{Bolognesi:2012gr,Ferrara:1993sq}.
The paper \cite{Lee:2010ay} also gives an example where consideration of the full non-Abelian DBI-action in all orders of $\alpha'$ -- be it using an adapted STr-prescription -- can change the physics, that is, the order of the there discussed phase transitions changes.

\subsubsection{Deriving the effective 4-dimensional equations of motion} \label{4.4.2}

Reconsider the vector part of the DBI-Lagrangian in unitary gauge (\ref{STRgauged}),
\begin{align}
\mathcal L &= \mathcal L_{Higgs} + \mathcal L_{vector} =   \text{STr} \hspace{1mm} e^{-\phi} \sqrt{-\det a} |_{\tilde A^2} \nonumber\\
&=  \sum_{a=1}^2 \left\{ \gamma(u) \frac{1}{2}\left( [\tilde A_u,\overline \tau]^a \right)^2 + \alpha(u) \frac{1}{2} \left( [\tilde A_\mu,\overline \tau]^a \right)^2 + \beta(u) \sum_{\mu=1}^2 \frac{1}{2} \left( [\tilde A_\mu,\overline \tau]^a\right)^2  \right\} \nonumber\\
&+ \text{STr} \hspace{1mm} \overline x \left\{
- \overline F_{12} g_{11}^{-2} A^{-1} [\tilde A_1,\tilde A_2]
- \frac{1}{4} g_{11}^{-2}\tilde F_{\mu\nu}^2 \hspace{1mm} A^{-2}|_{\mu,\nu=1,2}  - \frac{1}{2} g_{11}^{-1} G_{uu}^{-1} \tilde F_{\mu u}^2 \hspace{1mm} A^{-1}|_{\mu=1,2}  \right\}
\end{align}
where the notation $|_{\mu=1,2}$ as introduced in (\ref{STR}) can be written out as
\begin{align}
\tilde F_{\mu\nu}^2 A^{-2}|_{\mu,\nu=1,2} &= 2 A^{-1} (\tilde F_{i3}^2+\tilde F_{i0}^2) + 2 \tilde F_{03}^2 + 2 A^{-2} \tilde F_{12}^2 \qquad (i=1,2) \nonumber\\
&= \tilde F_{\mu\nu}^2 + 2 \frac{1-A}{A}(\tilde F_{i3}^2 + \tilde F_{i0}^2) + 2 \frac{1-A^2}{A^2} \tilde F_{12}^2 \nonumber\\
 \text{and} \quad \tilde F_{\mu u}^2 A^{-1}|_{\mu,\nu=1,2} &= \tilde F_{\mu u}^2 + \tilde F_{iu}^2 \frac{1-A}{A} \qquad (i=1,2).
\end{align}
Instead of approximating this action further to $(2\pi\alpha')^2 F^2$, we now keep all factors of $A =  1 - (2\pi\alpha')^2 \overline F_{12}^2  \frac{R^3}{u^3}$. Upon evaluating the STr we then obtain
\begin{align}
&
\mathcal L
\sim u^{1/4} (2\pi\alpha')^2 \sum_{a,b=1}^2 \left\{-(\sqrt{G_{uu}} \overline F_{12} A^{-1/2})^3 \epsilon_{3ab} \tilde A_1^a \tilde A_2^b - \frac{1}{4} f_1 (\tilde F_{\mu\nu}^a)^2  - \frac{1}{2} \sum_{i=1}^2 f_{1A} ((\tilde F_{i 3}^a)^2 + (\tilde F_{0 i}^a)^2) \right. \nonumber\\
& \left. - \frac{1}{2} f_{1B} (\tilde F_{12}^a)^2  - \frac{1}{2}  g_{11} f_2 (\tilde F_{\mu u}^a)^2 - \frac{1}{2} g_{11} \sum_{i=1}^2 f_{2A} (\tilde F_{i u}^a)^2
- \frac{1}{2}  g_{11} \frac{1}{T^2} f_2 (\tilde A_\mu^a)^2 (2\overline \tau^3)^2
- \frac{1}{2}  g_{11} \frac{1}{T^2} f_{2A} \sum_{i=1}^2 (\tilde A_i^a)^2 (2\overline \tau^3)^2 \right\},  \label{lagra}
\end{align}
where we defined the new $I$-functions
\begin{align}
f_1 &=  I(G_{uu}^{1/2}A^{1/2}), \quad f_{1A} = I(\sqrt{G_{uu}} \frac{1-A}{\sqrt{A}}), \quad f_{1B} = I(\sqrt{G_{uu}} \sqrt{A}\frac{1-A^2}{A^2}) \label{IfunctionsfullDBIa}  \\
f_2 &=  I(G_{uu}^{-1/2} A^{1/2}), \quad f_{2A} = I(G_{uu}^{-1/2} \frac{1-A}{\sqrt{A}}),  \label{IfunctionsfullDBI}
\end{align}
with $f_1$ and $f_2$ approaching their previous definition in (\ref{f1f2f3f4f5}) and $f_{1A}$, $f_{1B}$ and $f_{2A} \rightarrow 0$ for $A \rightarrow 1$ in the $(2\pi\alpha')^2$-approximation, as they should.

Extracting the effective 4-dimensional action from (\ref{lagra}) is completely analogous to the procedure described in Section \ref{F2approx}, so we will give a somewhat more schematic and short explanation here and refer to Section \ref{F2approx} for more details.

After plugging in the gauge field expansions (\ref{Amuexpansion})-(\ref{Auexpansion}) into the action in the approximation of only retaining the lowest modes of the meson towers, one can already notice the vanishing of $\int du \mathcal L_{vector-mixing}$ $= 0$ and of mixing terms between pions and rho mesons.
We will focus on the instability in the rho meson sector.

\paragraph{Background dependent functions in the action}

\begin{figure}[h!]
  \hfill
  \begin{minipage}[t]{.4\textwidth}
    \begin{center}
      \scalebox{0.55}{
  \includegraphics{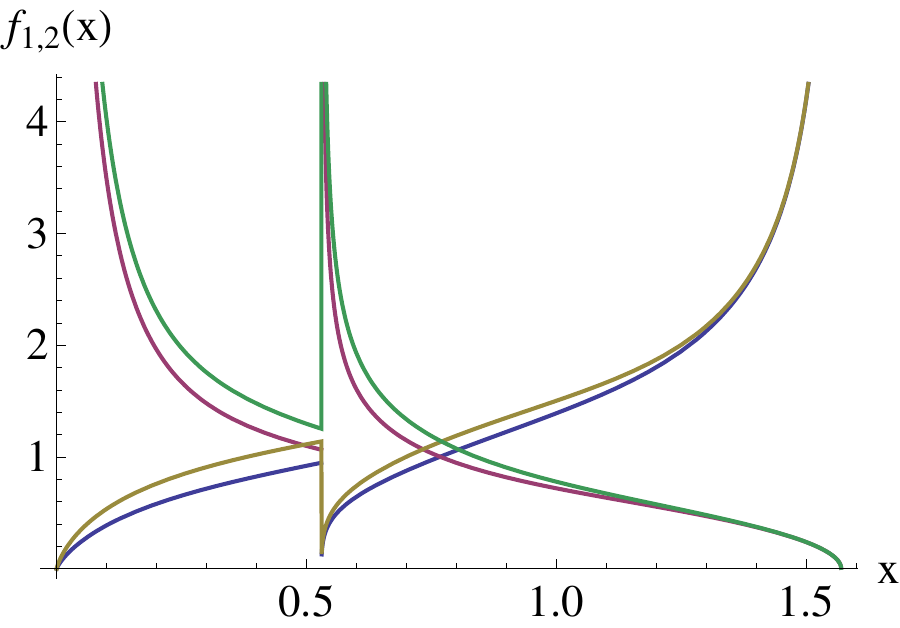}}
    \end{center}
  \end{minipage}
  \hfill
  \begin{minipage}[t]{.5\textwidth}
    \begin{center}
      \scalebox{0.55}{
  \includegraphics{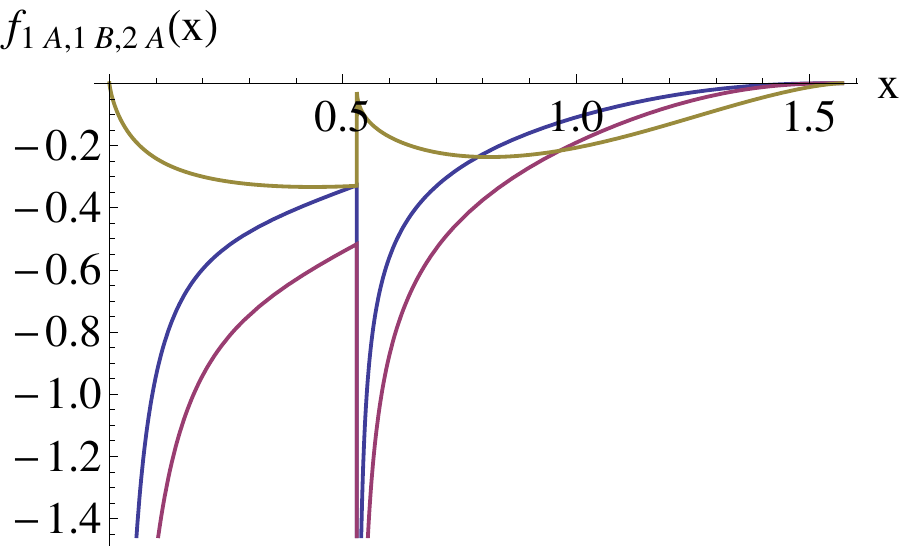}}
    \end{center}
  \end{minipage}
      \caption{(a) $f_1$ (green) and $f_2$ (yellow) compared to their $F^2$-approximated counterparts in red and blue resp. (b) $f_{1A}$ (blue), $f_{1B}$ (red) and $f_{2A}$ (yellow). For $B=0.8$ GeV$^2$ and $u=u_{0,d}\cos^{-3/2}x$.
}
	\label{fullDBIbackgroundfunctions}
  \hfill
\end{figure}

The generalized $I$-functions in (\ref{IfunctionsfullDBI}) have to be calculated numerically.
In Figure \ref{fullDBIbackgroundfunctions} we compare them to their approximated counterparts for some fixed values of the magnetic field.
The measure for the distance between up- and down-brane $\overline \tau^3(u)$ is still as defined in (\ref{tau3math}), and finally
\begin{align}
(G_{uu}^{1/2} \overline F_{12} A^{-1/2})^3 = \sqrt{G_{uu}^u} \overline F_u A_u^{-1/2} - \sqrt{G_{uu}^d} \overline F_d A_d^{-1/2}
\end{align}
with $G_{uu}^l = G_{uu}(\partial_u \overline \tau^l)$ (with flavour index $l=u,d)$, $\overline F_u = \frac{2B}{3}$ and $\overline F_d =- \frac{B}{3}$ (see (\ref{Fbardef})), and $A_l$ defined in (\ref{A}).

\paragraph{Eigenvalue problem}

The rho meson part of the DBI-Lagrangian to second order in fluctuations  (\ref{lagra})  after substituting (\ref{Amuexpansion}) reads
\begin{align}
\hspace{-1.2cm}
\mathcal L
&\sim u^{1/4} (2\pi\alpha')^2 \sum_{a,b=1}^2 \left\{- \frac{1}{4} f_1 (\mathcal F_{\mu\nu}^{a})^2 \psi^2 - \frac{1}{2} g_{11} f_2(\rho_\mu^a)^2 (\partial_u \psi)^2  - \frac{1}{2} \frac{g_{11}}{(2\pi\alpha')^2} f_2 (\rho_\mu^a)^2 \psi^2 (2\overline \tau^3)^2 \right.  \nonumber \\
& - \frac{1}{2} f_{1B} (F_{12}^a)^2
 + \sum_{\mu,\nu=1}^2 \left( -\frac{1}{2} \frac{g_{11}}{ (2\pi\alpha')^2} f_{2A} (\rho_\mu^a)^2 \psi^2 (2\overline \tau^3)^2 - \frac{1}{2}(\sqrt{G_{uu}}\overline F_{\mu\nu} A^{-1/2})^3 \epsilon_{3ab} \rho_\mu^a \rho_\nu^b \psi^2
\right. \nonumber\\
& \left. \left. \qquad \qquad
- \frac{1}{2} f_{1A} ((\mathcal F_{\mu 3}^a)^2 + (\mathcal F_{\mu 0}^a)^2) \psi^2
- \frac{1}{2} g_{11} f_{2A} (\rho_\mu^a)^2 (\partial_u \psi)^2 \right)
 \right\},  \label{}
\end{align}
which results in
the following effective 4-dimensional action
\begin{align}
\hspace{-1.2cm}
S_{4D}
&= \int d^4 x \sum_{a,b=1}^2 \left\{- \frac{1}{4}(\mathcal F_{\mu\nu}^{a})^2 -\frac{1}{2} m_\rho^2(B) (\rho_\mu^a)^2 - \frac{1}{2} b(B) (\mathcal F_{12}^a)^2 \right. \nonumber\\
& \left.+ \sum_{\mu,\nu=1}^2 \left(-\frac{1}{2} a(B) ((\mathcal F_{\mu 3}^a)^2 + (\mathcal F_{\mu 0}^a)^2)  -\frac{1}{2} m_{+}^2(B) (\rho_\mu^a)^2 -\frac{1}{2} \epsilon_{3ab}\rho_\mu^a \rho_\nu^b \hspace{1mm} k(B) \overline F_{\mu\nu}^3 \right)\right\}. \label{S4Dfull}
\end{align}
The function $\psi$ (rescaled to absorb all constant prefactors in the action) satisfies the normalization condition
\begin{equation} \label{normc}
\int_{u_{0,d}}^\infty du \hspace{1mm} u^{1/4} f_1 \psi^2 = 1
\end{equation}
and
\begin{equation} \label{massc}
\int_{u_{0,d}}^\infty du \hspace{1mm} \left\{ u^{1/4} g_{11} f_2 \partial_u \psi^2 + u^{1/4} \frac{g_{11}}{(2\pi\alpha')^2} f_2   (2\overline \tau^3)^2\psi^2  \right\} = m_\rho^2,
\end{equation}
combining into the eigenvalue equation
\begin{equation} \label{eigveq}
u^{-1/4} f_1^{-1} \partial_u\left(u^{1/4} g_{11} f_2 \partial_u \psi \right) -  \frac{g_{11}}{(2\pi\alpha')^2} f_1^{-1} f_2 (2\overline \tau^3)^2 \psi  = -m_\rho^2 \psi
\end{equation}
to be solved for its $B$-dependent eigenvalue $m_\rho^2$ and eigenfunction $\psi$. The $B$-dependent numbers $m_{+}, k, a$ and $b$ can subsequently be calculated with the obtained eigenfunctions from
\begin{equation} \label{}
\int_{u_{0,d}}^\infty du \hspace{1mm} \left\{ u^{1/4} g_{11} f_{2A} \partial_u \psi^2 +  u^{1/4} \frac{g_{11}}{(2\pi\alpha')^2} f_{2A} (2\overline \tau^3)^2\psi^2  \right\} = m_{+}^2,
\end{equation}
\begin{equation}
\int_{u_{0,d}}^\infty du \hspace{1mm} u^{1/4} (\sqrt{G_{uu}}\overline F_{12} A^{-1/2})^3 \psi^2 = k  \hspace{1mm}  \overline F_{12}^3
\end{equation}
and
\begin{equation} \label{aandb}
\int_{u_{0,d}}^\infty du \hspace{1mm} u^{1/4} f_{1A} \psi^2 = a, \quad \int_{u_{0,d}}^\infty du \hspace{1mm} u^{1/4} f_{1B} \psi^2 = b.
\end{equation}

\begin{figure}[h!]
  \hfill
\hspace{-0.5cm}
  \begin{minipage}[t]{.3\textwidth}
    \begin{center}
      \scalebox{0.55}{
  \includegraphics{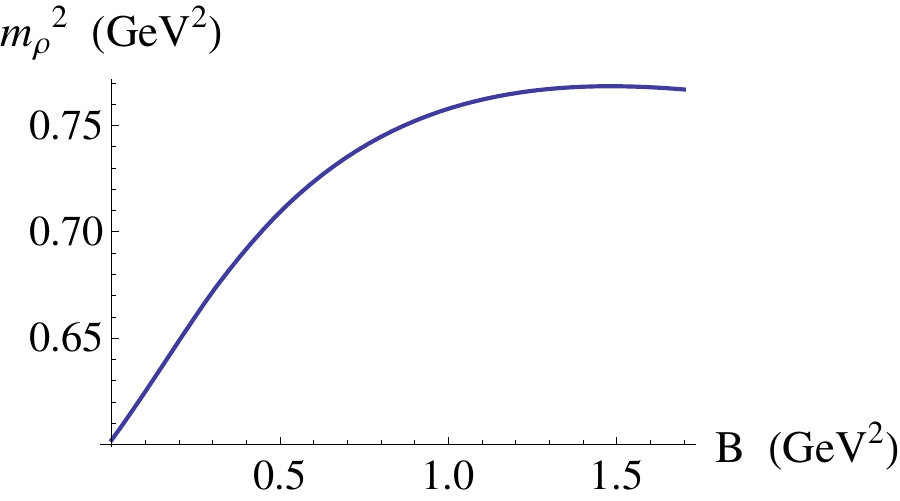}}
    \end{center}
  \end{minipage}
  \hfill
  \begin{minipage}[t]{.31\textwidth}
    \begin{center}
      \scalebox{0.55}{
  \includegraphics{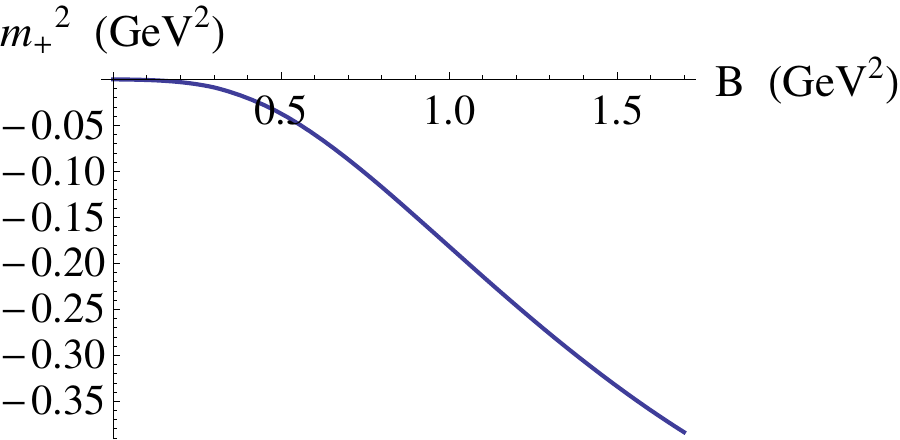}}
    \end{center}
  \end{minipage}
\hfill
  \begin{minipage}[t]{.31\textwidth}
    \begin{center}
      \scalebox{0.55}{
  \includegraphics{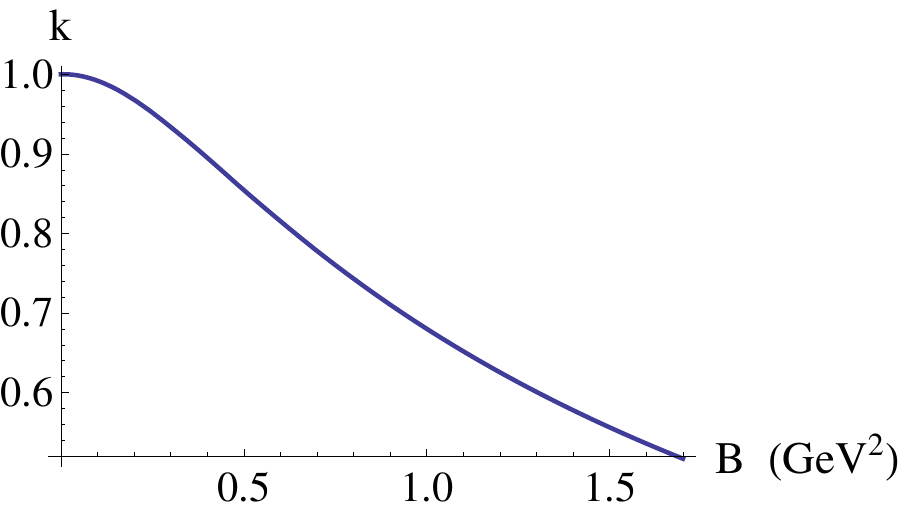}}
    \end{center}
  \end{minipage}
      \caption{Numerical results for $m_\rho^2(B)$, $m_+^2(B)$ and $k(B)$ from the full DBI-action. }
	\label{resultsfigDBI}
  \hfill
\end{figure}

\begin{figure}[h!]
  \hfill
  \begin{minipage}[t]{.5\textwidth}
    \begin{center}
      \scalebox{0.55}{
  \includegraphics{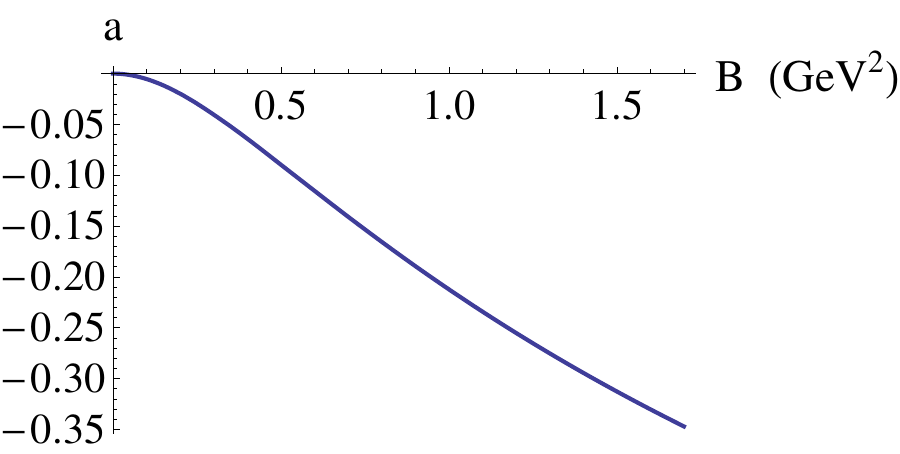}}
    \end{center}
  \end{minipage}
  \hfill
  \begin{minipage}[t]{.45\textwidth}
    \begin{center}
      \scalebox{0.55}{
  \includegraphics{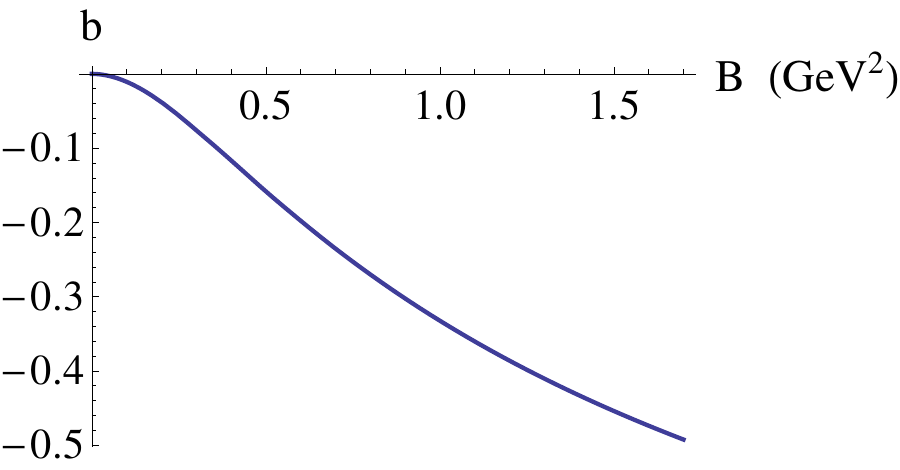}}
    \end{center}
  \end{minipage}
      \caption{Numerical results for $a(B)$ and $b(B)$.
}
	\label{resultsfigDBIpart2}
\end{figure}

The numerical results for $m_\rho^2$, $m_{+}^2$, $k$, $a$ and $b$ as functions of $B$, after having solved the eigenvalue problem with the techniques described in the second paragraph of \ref{paragraaf}, are shown in Figure \ref{resultsfigDBI}-\ref{resultsfigDBIpart2}. The discussion of the behaviour of $m_\rho^2(B)$ in the third paragraph of \ref{paragraaf2} is still applicable. The parameter $k$ specifying
the strength of the coupling to the magnetic field is again approximately equal to one, but now decreasing as a function of $B$ as opposed to
increasing in the $(2\pi\alpha')^2$-approximation.

\subsubsection{Solving the 4-dimensional equations of motion} \label{4.4.3}

The 4-dimensional EOMs for $\rho_\nu^a$ derived from the effective action (\ref{S4Dfull}) are given by
\begin{align}
D_\mu \mathcal F_{\mu\nu}^a - \epsilon_{a3b} k \overline F_{\mu\nu}^3 \rho_\mu^b - m_\rho^2 \rho_\nu^a &- \delta_{\nu i} (m_{+}^2 \rho_i^a + a (D_3 \mathcal F_{i3}^a - D_0 \mathcal F_{i0}^a) + b D_j \mathcal F_{ij}^a) + \delta_{\nu 3} a D_i \mathcal F_{i3}^a - \delta_{\nu 0} a D_i \mathcal F_{i0}^a = 0  \label{EOM}
\end{align}
with $D_\mu = \partial_\mu +[\overline A_\mu,\cdot]$
and $\mathcal F_{\mu\nu}^a = D_\mu \rho_\nu^a - D_\nu \rho_\mu^a$,
and where from now on we will not only keep assuming the Einstein convention that double $\mu,\nu$ indices are Minkowski sums over $\mu,\nu=0..3$ but also that double $i,j$ indices are sums over spatial indices $i,j=1,2$.
For notational clarity we will not explicitly write out the $B$-dependence of the parameters $m_\rho$, $m_+$, $k$, $a$ and $b$ in this Section, but assume it understood.

The equations $(a=1) \pm i (a=2)$ combine into the EOM for the charged rho meson $\rho_\mu = (\rho_\mu^1 + i \rho_\mu^2)/\sqrt 2$,
\begin{align}
& \text{D}_\mu \text F_{\mu\nu} - i k \overline F_{\mu\nu}^3 \rho_\mu - m_\rho^2 \rho_\nu - \delta_{\nu i} \left[ b \text D_j\text F_{ij} + a \left(\text D_3\text F_{i3}- \text D_0\text F_{i0} \right) + m_{+}^2 \rho_i \right] + \delta_{\nu 3} a \text{D}_j\text F_{j3} - \delta_{\nu 0} a \text{D}_j\text F_{j0} = 0,
\label{EOM4D}
\end{align}
with $\text D_\mu = \partial_\mu + i \overline A_\mu^3$ and $\text F_{\mu\nu} = \text D_\mu \rho_\nu - \text D_\nu \rho_\mu$ , and the complex conjugate of this equation for the other charged combination $\rho_\mu^* = (\rho_\mu^1 - i \rho_\mu^2)/\sqrt 2$.
Using $[\text D_\mu,\text D_\nu]=i \overline F_{\mu\nu}^3$, (\ref{EOM4D}) can be rewritten to the following EOMs for resp. $\nu=i$ and $\nu=3$:
\begin{itemize}
\item [\underline{$\nu=i$}]
\begin{align} \label{eom1}
& \hspace{-1cm}  (1+a) \text D_\mu^2 \rho_\nu
 - i (1+b+k) \overline F_{\mu\nu}^3 \rho_\mu - (1+a) \text D_\nu \text D_\mu \rho_\mu   - (m_\rho^2+m_{+}^2) \rho_\nu + (b-a) ( \text D_j^2 \rho_\nu - \text D_\nu \text D_j \rho_j)  = 0,\nonumber\\ & \end{align}
\item [\underline{$\nu=3$}]
\begin{align}
& \text D_\mu^2 \rho_\nu - \text D_\nu \text D_\mu \rho_\mu  - m_\rho^2 \rho_\nu + a  (\text D_j^2 \rho_\nu  - \text D_\nu \text D_j \rho_j )  = 0. \label{eom2}
\end{align}
\end{itemize}
These equations have to be complemented with a subsidiary condition, obtained
by acting with $\text D_\nu$  on the EOM (\ref{EOM4D}) and again using $[\text D_\mu,\text D_\nu]=i \overline F_{\mu\nu}^3$.
We find the generalized subsidiary condition (where by generalized we mean w.r.t. the Proca subsidiary condition $\text D_\nu \rho_\nu = 0$)
\begin{equation} \label{subsconditiongen}
\text D_\nu \rho_\nu = \frac{i}{m_\rho^2} (1+b-k)  \overline F_{\mu\nu}^3 \text D_\nu \rho_\mu  - \frac{m^2_{+}}{m_\rho^2}
\text D_i \rho_i,
\end{equation}
still relating $\text D_\nu \rho_\nu$ ($\nu=0..3)$ to transversal components $\rho_i$ $(i=1,2)$ only, such that the EOMs for the transverse rho mesons can be rewritten as independent from any longitudinal components.
Before doing so, let us remark that the above system of EOMs combined with the subsidiary condition reduces to its standard Proca form for $a$, $b$, $m_+ \rightarrow 0$, $k \rightarrow 1$ and no $B$-dependence in $m_\rho$ (or any of the previous parameters).
The non-zero and $B$-dependent $a$ and $b$ are present due to taking into account all powers in the field strength in the non-linear non-Abelian DBI-action, which is also partly the reason for
the $B$-dependence of $m_\rho$, $k$ and $m_+$, in addition to their implicit description of the response of the quark constituents to the magnetic field (cfr. the chiral magnetic catalysis and holographic Higgs mechanism for heavy-light mesons discussed earlier).

To determine the solutions of the EOMs we follow and generalize the procedure used in \cite{Mathews:1974ax}. In order to make comparisons with the original expressions in \cite{Mathews:1974ax} more clear, we temporarily change notation to
\begin{equation} \label{defphi}
\phi_\mu = \rho_\mu^* = (\rho_\mu^1 - i \rho_\mu^2)/\sqrt 2
\end{equation}
and
\begin{equation}
 i \pi_\mu = \text D_\mu^*  = \partial_\mu - i \overline A_\mu^3
\end{equation}
such that $\pi_\mu$ becomes $p_\mu - \overline A_\mu^3$ when substituting a plane wave ansatz $\phi_\mu \rightarrow e^{i \vec p \cdot \vec x - i E t} \phi_\mu$ into (\ref{eom1})-(\ref{eom2}), and in particular we can write $\pi_\nu^2 = -E^2 + \vec \pi^2$.
In this new notation the EOMs (\ref{eom1})-(\ref{eom2}) combined with (\ref{subsconditiongen}) can be recast in the form
\begin{align}
E^2 \phi_{\pm} =&  \left(\frac{m_\rho^2 + m_{+}^2}{1+a} + \mathcal B \vec \pi^2 \right) \phi_\pm + \frac{B}{2m^2}(1+b-k) \pi_\pm  (\pi_+ \phi_- - \pi_- \phi_+) \pm B \mathcal K  \phi_\pm - \frac{1}{2} \mathcal M \pi_\pm (\pi_+ \phi_- + \pi_- \phi_+)  \label{eom3}
\end{align}
with
\begin{equation}
\pi_\pm = \pi_1 \pm i \pi_2, \quad \phi_\pm = \phi_1 \pm i \phi_2,
\end{equation}
and
\begin{align}
E^2 \phi_3 =&  \left( m_\rho^2 + (1+a) \vec \pi^2 \right) \phi_3 + \frac{B}{2m^2}(1+b-k) \pi_3 (\pi_+ \phi_- - \pi_- \phi_+)  - \frac{1}{2} \left( a - \frac{m_{+}^2}{m_\rho^2} \right) \pi_3 (\pi_+ \phi_- + \pi_- \phi_+),  \label{eom4}
\end{align}
where we defined
\begin{align}
\mathcal B &= \frac{1+b}{1+a}, \quad \mathcal K = \frac{1+b+k}{1+a} \quad \text{and} \quad \mathcal M = \frac{b-a}{1+a}-\frac{m_{+}^2}{m_\rho^2}.
\end{align}
The main trick for solving the system is to notice that the operators $\pi_{\pm}$ obey the algebra of a simple harmonic oscillator, if one defines annihilation and creation operators $\hat a$ and $\hat a^\dagger$ as
\begin{align}
& \hat a = (2B)^{-1/2} \pi_+ \quad \text{and} \quad \hat a^\dagger = (2B)^{-1/2} \pi_-,
\end{align}
which obey
\begin{align}
& [\hat a,\hat a^\dagger]=1  \quad \text{and} \quad  [\hat a,\pi_3]=[\hat a^\dagger,\pi_3]=0.
\end{align}
The `number operator' $\hat N$ is then defined as
\begin{align}
 \hat N =\hat a^\dagger \hat a,
\end{align}
allowing us to rewrite the system (\ref{eom3})-(\ref{eom4}), using $\vec \pi^2 = p_3^2 + B(2 \hat N+1)$ and $\pi_+ \pi_- = 2B(1+\hat N)$,
to
\begin{align}
\hspace{-0.2cm}
\left\{ \begin{array}{ll}
(\omega^2 - \hat X_+) \phi_+ = A_\xi  \hat a^2 \phi_- \\
(\omega^2 - \hat X_-) \phi_- = -B_\xi (\hat a^\dagger)^2 \phi_+  \\
\left[ \omega_3^2 - (1+a)(2\hat N +1) \xi \right] \phi_3 = \xi^2 (1+b-k) a_3 (\hat  a \phi_- - \hat a^\dagger \phi_+) - \left( a - \frac{m_{+}^2}{m_\rho^2} \right) \xi a_3 (\hat  a \phi_- + \hat a^\dagger \phi_+),
\end{array} \right. \label{mathEOM}
\end{align}
with $\xi = \frac{B}{m_\rho^2}$ and
\begin{align}
\omega^2 &= \frac{E^2 - (m_\rho^2+m_+^2)/(1+a) - \mathcal B p_3^2}{m_\rho^2} \label{defomega}\\
\omega_3^2 &= \frac{E^2 - m_\rho^2 - (1+a) p_3^2}{m_\rho^2} \label{omega3} \\
\hat X_+ &= (2\hat N+1) \mathcal B \hspace{1mm} \xi - \frac{B_\xi}{2} + \mathcal K \hspace{1mm} \xi - (2 \hat N+1)\frac{B_\xi}{2} \\
\hat X_- &= (2\hat N+1) \mathcal B \hspace{1mm} \xi - \frac{A_\xi}{2} - \mathcal K \hspace{1mm} \xi + (2\hat N+1)\frac{A_\xi}{2} \\
A_\xi &= (1+b-k) \hspace{1mm} \xi^2 - \mathcal M \hspace{1mm} \xi \quad \text{and} \quad B_\xi = (1+b-k) \hspace{1mm} \xi^2 + \mathcal M \hspace{1mm} \xi,
\end{align}
and with $\pi_3$ replaced by its eigenvalue $p_3$ since it commutes with everything, or where convenient for the notation by the number $a_3 = (2B)^{-1/2} \pi_3$.
The system (\ref{mathEOM}) decouples completely in the special case where  $A_\xi = B_\xi = 0$ as well as $1+b-k = a - \frac{m_{+}^2}{m_\rho^2}=0$, which is for example the case for standard Proca parameters $a=b=m_+=0$ and $k=1$. In the latter situation the independent solutions for any $n$ are given by
\begin{align}
\phi_+ = |n-2\rangle, \quad \phi_-=\phi_3=0 \quad (n=2,3,\cdots) \nonumber\\
\phi_- = |n\rangle, \quad \phi_+=\phi_3=0 \quad (n=0,1,\cdots) \nonumber\\
\phi_3 = |n-1\rangle, \quad \phi_-=\phi_+=0 \quad (n=1,2,\cdots) \label{Proca}
\end{align}
with eigenvalue $\omega^2 = \omega_3^2 = (2n-1) \xi$.
Here we formally defined the `number eigenstates' $|n\rangle$ as
\begin{equation}
\hat N |n\rangle = n |n\rangle, \quad \hat a |0\rangle = 0, \quad |n\rangle = (n!)^{-1/2} (\hat a^\dagger)^n |0\rangle.
\end{equation}
In the rest of the discussion of possible solutions
below, we consider $A_\xi$ and $B_\xi$ different from zero.

\paragraph{Condensing solution}

Before decoupling the first two equations of (\ref{mathEOM}) to discuss the general form of the solution, let us first look at the one we are most interested in, the condensing solution:
\begin{equation} \label{solcondensing}
\phi_3 = \phi_+ = 0, \quad \phi_- = |0\rangle \quad (\Rightarrow \hat a \phi_- = 0),
\end{equation}
for which the EOM reduces to
\[
(\omega^2 - \hat X_-) |0\rangle = 0 \Rightarrow \omega^2 = \hat X_- (\hat N \rightarrow 0) = (\mathcal B \hspace{1mm} - \mathcal K) \xi = - \frac{k}{1+a} \xi
\]
with total eigenvalue
\begin{equation}
E^2 = \frac{m_\rho^2+m_{+}^2}{1+a} + \left(\frac{1+b}{1+a}\right) p_3^2 - \frac{k}{1+a} m_\rho^2 \xi,
\end{equation}
or, in the lowest state $p_3=0$ (and $\frac{1+b}{1+a}>0$ in the considered range of $B$):
\begin{equation}   \label{eigvcondensing}
\mathit{m_{\rho,eff}^2} = \frac{m_\rho^2+m_{+}^2}{1+a} - \frac{k}{1+a} m_\rho^2 \xi.
\end{equation}
This indeed reduces to its $(2\pi\alpha')^2$-approximated equivalent (\ref{mrhoeff2}),
$\mathit{m_{\rho,eff}^2} = m_\rho^2 + m_{+}^2  - k \xi m_\rho^2$,
for $a\rightarrow 0$.

\paragraph{Family of solutions}
We present the general discussion of the family of solutions of (\ref{mathEOM}).
One family of solutions is
\begin{equation} \label{sol1}
\phi_+ = \phi_- = 0, \quad \phi_3 = |n\rangle, \quad \omega_3^2 = (1+a)(2n+1)\xi, \quad n=0,1,2,\cdots,
\end{equation}
the other one
\begin{equation} \label{gensol}
\phi_- = |n+1\rangle, \quad \phi_+ = c_n |n-1\rangle, \quad \phi_3 = c_n' |n\rangle, \quad n=1,2,3,\cdots.
\end{equation}
The corresponding eigenvalue $\omega$ can be determined from decoupling the first two equations of  (\ref{mathEOM}) to
\begin{align}
\left\{ (\omega^2 - \hat X_-)(\omega^2 - \hat X_+) + (\hat N^2+3\hat N+2) A_\xi B_\xi - 2 (2 \mathcal B \hspace{1mm} \xi + A_\xi) (\omega^2 - \hat X_+) \right\} \phi_+ = 0   \label{16a}\\
\left\{ (\omega^2 - \hat X_-)(\omega^2 - \hat X_+) + (\hat N^2-\hat N) A_\xi B_\xi + 2 (2 \mathcal B \hspace{1mm} \xi - B_\xi) (\omega^2 - \hat X_-) \right\} \phi_- = 0. \label{16b}
\end{align}

Substitution of (\ref{gensol}) has the effect of replacing $\hat N$ in (\ref{16a}) by $(n-1)$ and in (\ref{16b}) by $(n+1)$. With these replacements, the curly-bracketed expressions in the two equations become identical, and either of them can be solved,
with the result for our generalized Landau levels finally given by
\begin{align}
& \omega^2 = (2n+1) \xi (\mathcal B -\frac{\mathcal M}{2}) + \frac{(1+b-k)}{2} \xi^2 \nonumber\\
& \pm \xi \sqrt{\mathcal M \left(\frac{(2n+1)^2}{4} + \mathcal K - 2 \mathcal B \right) + (\mathcal K - 2 \mathcal B)^2 - (1+b-k) (2n+1) \xi (\mathcal K - 2 \mathcal B + \frac{\mathcal M}{2}) + \frac{(1+b-k)^2}{4} \xi^2}.  \label{omegasquared}
\end{align}
This reduces to Mathews' solution for general $k\neq1$, eq. (19) in \cite{Mathews:1974ax}, for $a, b,  m_{+} \rightarrow 0$, i.e.~$\mathcal B \rightarrow 1,  \mathcal M \rightarrow 0, \mathcal K \rightarrow 1+k$:
\begin{align}
\omega^2(a, b,  m_{+} \rightarrow 0) = (2n+1) \xi  + \frac{1}{2}(1-k) \xi^2 \pm (1-k) \xi \sqrt{1 + (2n+1) \xi + \frac{1}{4} \xi^2},
\end{align}
and the modified Landau levels mentioned in Section \ref{rhomesonc} are given by (\ref{omegasquared}) with $a,b \rightarrow 0$. Given the value of $E^2$ from (\ref{omegasquared})
and the ansatz (\ref{gensol}) for $\phi_3$, the equation (\ref{omega3}) can be solved for $c_n'$. The constant $c_n$ can be determined from substituting the solution  (\ref{gensol}) and (\ref{omegasquared}) into either one of the first two equations of (\ref{mathEOM}).  \\
For completeness, we mention the last remaining possible solution
\begin{equation*}
\phi_- = |1\rangle, \quad \phi_+ = 0, \quad \phi_3 = c_0' |0\rangle
\end{equation*}
with $\omega^2 = \hat X_-(\hat N \rightarrow 1) = (3 \mathcal B - \mathcal K - \mathcal M) \xi + (1+b-k) \hspace{1mm} \xi^2$ and $c_0'$ to be determined from
$\omega_3^2 - (1+a) \xi c_0' =  (\xi^2 (1+b-k) - a \xi) a_3$.

In this whole discussion of the solutions of the EOMs for the rho meson, the key observation is that the energy eigenstates are so-called `number eigenstates', labeled by the Landau level number $n$. They are not necessarily spin eigenstates, as we will discuss next.

\paragraph{Discussion of the spin of the solutions}
Consider the eigenstates of the spin operator $S_3$ as defined in (\ref{S3spin}),
\begin{align*}
\phi_+ &= \phi_-= 0 \quad (s_3=0) \\
\phi_+ &= \phi_3=0 \quad (s_3=+1) \\
\phi_- &= \phi_3=0 \quad (s_3 = -1).
\end{align*}
It is clear that only the branch of solutions (\ref{sol1}) and the condensing solution (\ref{solcondensing}) are spin eigenstates, resp. with eigenvalues $s_3 = 0$ and $s_3 = +1$; the other branches of solutions for general $k\neq 1$  case
are not.  This is in contrast with the special $k=1$ Proca case (\ref{Proca}) where all Landau levels, including the excited states, are also spin eigenstates.

\begin{figure}[h!]
  \centering
  \scalebox{0.7}{
  \includegraphics{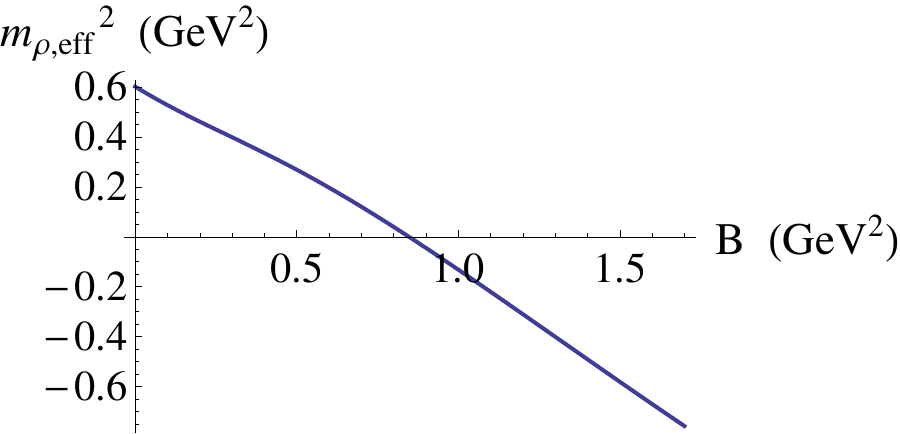}}
  \caption{The effective rho meson mass squared $\mathit{m_{\rho,eff}^2}(B)$ from the full DBI-action.
} \label{mrhoeff2figDBI}
\end{figure}

We conclude by summarizing that the condensing states are given by (\ref{solcondensing}) and its conjugate,
\[
\rho^*=\phi_- = \rho^*_1- i \rho^*_2
\quad \mbox{ and }\quad
\rho = \phi_-^* =  \rho_1 + i \rho_2
\]
--where we translated back to the previously used notation--
with energy eigenvalue $\mathit{m_{\rho,eff}^2}=$ (\ref{eigvcondensing}) and spin eigenvalue $s_3=+1$ corresponding to the spins being aligned with the magnetic field.
Our result for the effective rho meson mass squared $\mathit{m_{\rho,eff}^2}$, as  shown in Figure \ref{mrhoeff2figDBI}, again demonstrates the tachyonic instability, with the critical magnetic field for rho meson condensation given this time by
\[
B_c \approx 0.85 \text{ GeV}^2.
\]
The increase compared to the estimate for $B_c$  in (\ref{Bc}) using the $(2\pi\alpha')^2$-approximation is pretty small.
This indicates that the expansion to second order in $(2\pi\alpha')F$ was a valid approximation, despite the ambiguities mentioned in Section \ref{ambiguities}.

\subsection{Comment on the antipodal case} \label{commentantipodal}

For completeness, we consider the effect in the antipodal SSM, $u_0=u_K$, of including all higher order terms in the total field strength in the DBI-action. As mentioned before, the embedding of the flavour branes is independent of $B$ in this case, resulting in standard Landau levels and thus $\mathit{m_{\rho,eff}^2}(B) = m_\rho^2 - B$ if the action is approximated to second order in $(2\pi\alpha') F$.
In this set-up there is no constituent quark mass (\ref{constmass}) and no chiral magnetic catalysis.

To reproduce $m_\rho^2 = 0.602$ GeV$^2$ at zero magnetic field, along with $f_\pi = 0.093$ GeV for the pion decay constant, we have to use the holographic parameters fixed in \cite{Sakai:2005yt} to
\begin{equation} \label{antipodalvalues}
M_K \approx 0.949 \mbox{  GeV} \quad
\mbox{and}\quad \kappa  = \frac{\lambda N_c}{216 \pi^3} \approx 0.00745,
\end{equation}
instead of the values (\ref{values}) for $u_0>u_K$.
With these fixed parameters the estimate for the maximum value of the magnetic field for the  $(2\pi\alpha')$-expansion of the action to be valid, as discussed in Section \ref{ambiguities}, changes to
\begin{align}
eB \ll \frac{3}{2} \left(\frac{u_K}{R}\right)^{3/2} (2\pi \alpha')^{-1}  \equiv 0.31 \text{ GeV}^2,
\end{align}
which is even lower than the value 0.45 $\text{GeV}^2$ obtained for the non-antipodal case.

\begin{figure}[h!]
  \centering
  \scalebox{0.7}{
  \includegraphics{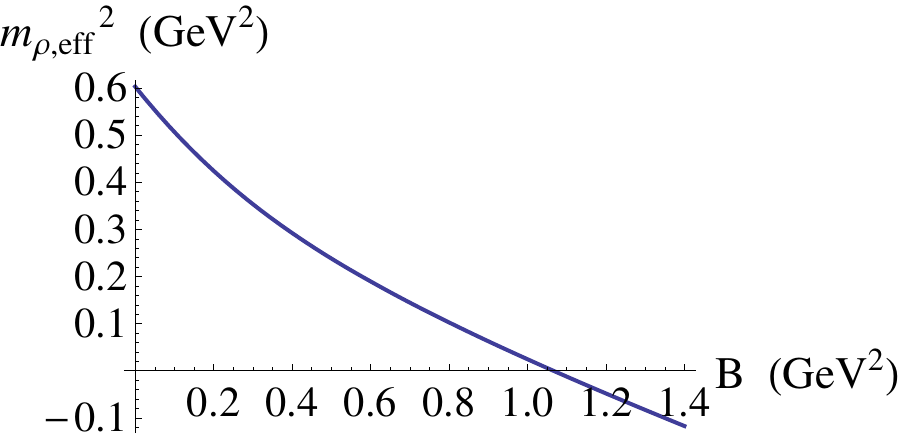}}
  \caption{
The effective rho meson mass squared $\mathit{m_{\rho,eff}^2}(B)$ from the full DBI-action for the antipodal embedded flavour branes.
} \label{antipodalfullDBI}
\end{figure}

As the flavour branes now remain coincident for any value of $B$, that is $\overline \tau \sim \mathbb{1} \Rightarrow \overline \tau^3 = 0$ and
$\partial_u \overline \tau = 0 \Rightarrow  G_{uu} = g_{uu}$,
we again obtain the effective 4-dimensional action (\ref{S4Dfull}), but with the integrals and equations (\ref{normc})-(\ref{aandb}) changed in the sense that
$u_{0,d} \rightarrow u_K$, $\overline \tau^3 \rightarrow 0$ and every $G_{uu} \rightarrow g_{uu}$, in particular in the $I$-functions $f_{1(A,B)}, f_{2(A)}$ defined in (\ref{IfunctionsfullDBIa})-(\ref{IfunctionsfullDBI}).
The eigenvalue equation can be recast in the form
\begin{equation*}
\frac{9}{4} \frac{u_K}{R^3} \cos^{4/3}x \left[ \partial_x^2 \psi + I(A^{1/2})^{-1} \partial_x  I(A^{1/2}) \partial_x \psi \right] = -m_\rho^2 \psi
\end{equation*}
with $u = u_K \cos^{-2/3}x$ this time and $I(A^{1/2})$ reducing to 1 for $B=0$.
With the numerical result for the eigenfunction $\psi$ and eigenvalue $m_\rho^2$, the total effective rho meson squared can be obtained using (\ref{eigvcondensing}),
\begin{equation*}
\mathit{m_{\rho,eff}^2} = \frac{m_\rho^2+m_{+}^2}{1+a} - \frac{k}{1+a} B.
\end{equation*}
The result is shown in Figure \ref{antipodalfullDBI}, where the corresponding critical magnetic field can be read off to be $B_c \approx 1.07$ GeV$^2$.

\section{Summary}\label{summary}

\begin{figure}[h!]
  \hfill
  \begin{minipage}[t]{.5\textwidth}
    \begin{center}
      \scalebox{0.7}{
  \includegraphics{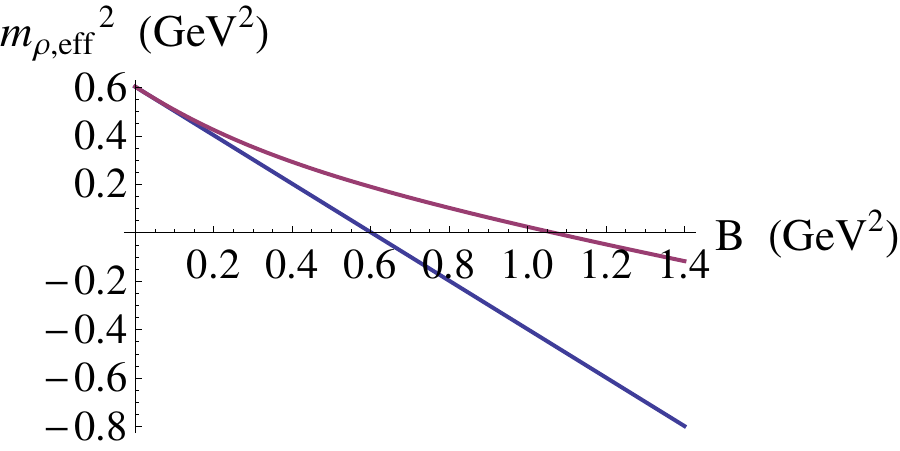}}
    \end{center}
  \end{minipage}
  \hfill
  \begin{minipage}[t]{.45\textwidth}
    \begin{center}
      \scalebox{0.7}{
  \includegraphics{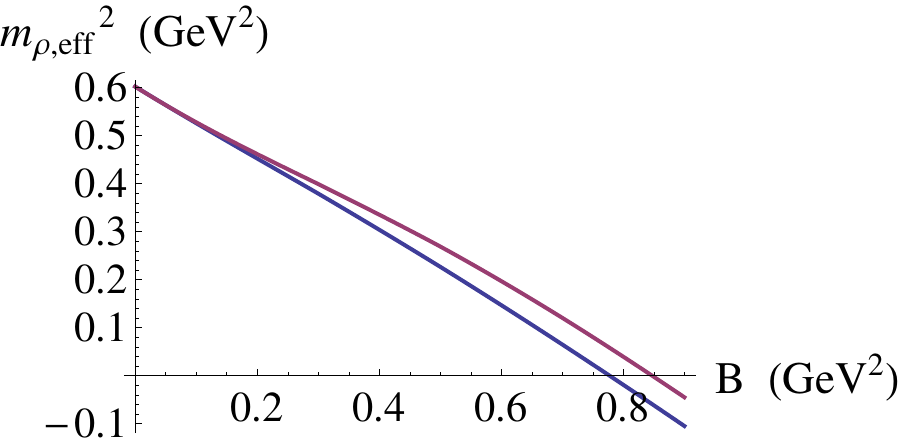}}
    \end{center}
  \end{minipage}
      \caption{The effective rho meson mass squared $\mathit{m_{\rho,eff}^2}(B)$ in the antipodal embedding (left) and the non-antipodal embedding (right), comparing the $(2\pi\alpha')^2 F^2$-approximated result in blue to the full DBI-result in red.
}\label{summaryfig}
\end{figure}

We studied a magnetically induced tachyonic instability in the charged rho meson sector, arising from the DBI-part of the two-flavour Sakai-Sugimoto model. We examined both the case of antipodal and the more general non-antipodal embedding, each in the $(2\pi\alpha')^2 F^2$-approximation of the action versus the full DBI-action, non-linear in the total field strength $F$.
The results for the effective rho meson mass squared $\mathit{m_{\rho,eff}^2}(B)$,   vanishing at the critical magnetic field $B_c$ and thereby signaling the onset of the tachyonic instability, are shown in Figure \ref{summaryfig} for each of the four set-ups.

The antipodal SSM reproduces exactly the standard 4-dimensional Proca picture and Landau levels of the effective QCD-model used in \cite{Chernodub:2010qx}, with $B_c = m_\rho^2 \approx 0.602$ GeV$^2$. The same picture was obtained in a holographic toy model involving an $SU(2)$ Einstein-Yang-Mills action for an $SU(2)$ bulk gauge field in a (4+1)-dimensional AdS-Schwarzschild black hole background \cite{Ammon:2011je},
and more recently for a 3-dimensional field theory in a (3+1)-dimensional DSGS-model generalized to AdS  \cite{Cai:2013pda}.
The non-antipodal SSM predicts a larger value of $B_c \approx 0.78 \text{ GeV}^2$  as a result of taking two mass-generating effects for the charged rho meson into account, i.e.~chiral magnetic catalysis for the rho meson constituents on one hand, and a stringy Higgs-contribution to the mass from stretching the rho meson string between the magnetically separated up- and down-brane. Both effects are a direct result from the $B$-dependence of the non-antipodal flavour branes' embedding, and hence absent in the antipodal set-up.
Considering the full DBI-action instead of approximating it to second order in the total field strength further increases the value of the magnetic field $B_c$ at the onset of rho meson condensation, more precisely to $B_c \approx 0.85$ GeV$^2$ in the non-antipodal case. The effect of taking the non-linear contributions in   $\overline F_{12}$
  into account seems to be stronger for the antipodal set of parameters compared to the non-antipodal one -- in both cases parameters are fixed to reproduce QCD parameters at zero magnetic field.
This leads us to conclude that the $F^2$-approximation is better justified for the considered problem in the non-antipodal embedding than in the antipodal one.
We are however very well aware of the fact that the full DBI-action is not the complete non-Abelian action for a system of $N_f$ branes -- a closed form of which is still to be found --, starting to show deviations at order $F^6$ \cite{Hashimoto:1997gm,Sevrin:2001ha}. We do not claim the DBI-result is necessarily  more correct than the $F^2$-result, yet we wanted to examine
the extent of the difference.
In conclusion, the SSM-predictions for $B_c$ are close to order 1 GeV$^2$, as  obtained in the NJL-model in \cite{Chernodub:2011mc} and on the lattice in \cite{Braguta:2011hq}.

A main motivation for these comparisons within the SSM was to investigate
what holography can add to the QCD-phenomenological picture of rho meson condensation, purposely
working in a top-down approach -- the downside of which are the technical complications. We for example elaborated on evaluating the STr exactly (to second order in fluctuations in the presence of an Abelian background field), the gauge fixing necessary to disentangle scalar and vector fluctuations, the contribution of the Chern-Simons action, the pion sector in the $F^2$-approximated DBI-part of the action, the Higgs mechanism associated with the magnetically induced heavy-light character of the charged rho mesons, numerically solving the eigenvalue equation for $m_\rho^2$ with a shooting method, and analytically solving the generalized effective 4-dimensional EOMs.
For the above reasons of complexity we have not yet been able to construct the new ground state in which the rho mesons are condensed.  This ground state is expected to be an Abrikosov lattice of rho meson vortices, as constructed in the DSGS-model in \cite{Chernodub:2011gs} and in a bottom-up holographic model in \cite{Bu:2012mq}.
The Abrikosov lattice forms an anisotropic and spatially inhomogeneous,  type II  superconducting ground state of the QCD vacuum in the presence of a strong magnetic field \cite{Chernodub:2012tf}, with the interesting property that the magnetic field creates the superconducting state instead of destroying it (cfr. Meissner effect). In \cite{Cai:2013pda}, the real part of the optical conductivity in the condensed phase is shown to contain a delta peak at the origin, consistent with a superconducting condensed state.
Another downside of the top-down approach and in particular the SSM is the abundance of extra fields in the bulk that do not have counterparts in the dual field theory. The mass scale $\sim M_K$ of these artifacts of the model is actually of the same order as the masses of the mesons. Nevertheless the SSM can present a nice record of QCD-effects and properties that can be modeled, suggesting the influence
of the redundant modes is not necessarily substantial.

We have been able to show that the SSM has a magnetically induced instability towards rho meson condensation,
consistent with the studies of this phenomenon
in phenomenological \cite{Chernodub:2010qx,Chernodub:2011mc}, lattice \cite{Braguta:2011hq} and bottom-up holographic \cite{Ammon:2011je,Cai:2013pda} approaches.
To come closer to the real-life quark-gluon plasma conditions where the presence of  magnetic fields of the order of $\sim 1$ GeV$^2$ might eventually be obtained,
it should be taken into account that there are also very high temperatures/densities present, and that the magnetic field is very localized both in space and time (see the more recent works cited under \cite{Skokov:2009qp}).
These features may in the end seriously influence the possible occurrence of
rho meson condensation.

\section*{Acknowledgements}
It is a pleasure to thank A.~Sevrin for helpful discussions. N.~C.~and D.~D.~are supported by the Research-Foundation Flanders.

\newpage
\appendix
\section{Appendix: STr-prescription} \label{appendix}

\paragraph{Prescription}
We write down the prescription for the evaluation of the symmetrized trace STr to second order in fluctuations in the presence of a constant Abelian background, as derived in \cite{Hashimoto:1997gm} and \cite{Denef:2000rj}.

For an even function $\mathcal H(F)$ of a diagonal background field $F = F^0 \sigma^0 +F^3 \sigma^3$ and fluctuation $\tilde X = \tilde X^a t^a$ (generator $t^a = -\frac{i}{2} \sigma^a$), one finds that
\begin{equation} \label{prescr}
\text{STr}\left(\mathcal H(F) \tilde X^2 \right) = -\frac{1}{2} \sum_{a=1}^2  (\tilde X^a)^2 \hspace{2mm} I(\mathcal H) -\frac{1}{2}\sum_{l=u,d} (\tilde X^l)^2 \hspace{2mm} I_l(\mathcal H)
\end{equation}
with
\begin{equation}  \label{Ifctions}
I(\mathcal H) = \frac{\int_0^1 d\alpha \mathcal H(F^0 + \alpha F^3) + \int_0^1 d\alpha \mathcal H(F^0 - \alpha F^3)}{2},
\end{equation}
\begin{equation}  \label{Ilfctions}
I_u(\mathcal H) = \mathcal H(F^0+F^3), \quad I_d(\mathcal H) = \mathcal H(F^0-F^3),
\end{equation}
\begin{equation}
\tilde X^u = \frac{\tilde X^0 + \tilde X^3}{\sqrt{2}}, \quad \tilde X^d = \frac{\tilde X^0 - \tilde X^3}{\sqrt{2}};
\end{equation}
and
\begin{equation} \label{typeI}
\text{STr}\left(\mathcal H(F) \tilde X \right) = \text{Tr}\left(\mathcal H(F) \tilde X \right).
\end{equation}

\paragraph{Generalized prescription} A straightforward generalization of the prescription when dealing with two Abelian background fields can be written down.

For even functions $\mathcal H(\partial \overline \tau)$ and $\mathcal G(F)$ of diagonal background fields $\partial \overline \tau = \partial \overline \tau^0 \sigma^0 + \partial \overline \tau^3 \sigma^3$  and $F = F^0 \sigma^0 + F^3 \sigma^3$, and fluctuation $\tilde X = \tilde X^a t^a$ (generator $t^a = -\frac{i}{2} \sigma^a$), it reads
\begin{equation}
\text{STr}\left(\mathcal H(\partial \overline \tau) \mathcal G(F) \tilde X^2 \right) = -\frac{1}{2} \sum_{a=1}^2 (\tilde X^a)^2 \hspace{2mm} I(\mathcal H \mathcal G) -\frac{1}{2}\sum_{l=u,d} (\tilde X^l)^2 \hspace{2mm} I_l(\mathcal H \mathcal G)
\end{equation}
with
\begin{equation} \label{IfctionsGeneral}
I(\mathcal H\mathcal G) = \frac{\int_0^1 d\alpha \mathcal H(\partial \overline \tau^0 + \alpha \partial \overline \tau^3) \mathcal G(F^0 + \alpha F^3) + \int_0^1 d\alpha \mathcal H(\partial \overline \tau^0 - \alpha \partial \overline \tau^3)\mathcal G(F^0 -\alpha F^3)}{2},
\end{equation}
\begin{equation} \label{IlfctionsGeneral}
I_u(\mathcal H\mathcal G) = \mathcal H(\partial \overline \tau^0+\partial  \overline \tau^3)\mathcal G(F^0 + F^3), \quad I_d(\mathcal H\mathcal G) = \mathcal H(\partial \overline \tau^0-\partial \overline \tau^3)\mathcal G(F^0 - F^3),
\end{equation}
\begin{equation}
\tilde X^u = \frac{\tilde X^0 + \tilde X^3}{\sqrt{2}}, \quad \tilde X^d = \frac{\tilde X^0 - \tilde X^3}{\sqrt{2}};
\end{equation}
and
\begin{equation} \label{typeIgeneral}
\text{STr}\left(\mathcal H(\partial \overline \tau) \mathcal G(F) \tilde X \right) = \text{Tr}\left(\mathcal H(\partial \overline \tau) \mathcal G(F) \tilde X \right).
\end{equation}

\subsection{Derivation of the prescription}
For completeness, let us schematically recapitulate
how the above prescription was obtained. In this derivation we will temporarily write $U(2)$-indices as lower instead of upper indices, to avoid notational clutter.

\begin{itemize}
\item Properties of the Pauli matrices ($a=1,2,3$):
\[ \text{Tr} (\sigma_a) = 0, \quad \text{Tr} (\sigma_a \sigma_b) = 2 \delta_{ab}, \quad \sigma_a \sigma_b = \delta_{ab} \mathbb{1} + i \epsilon_{abc} \sigma_c  \]
\[ \{\sigma_a,\sigma_b\} = 2 \delta_{ab} \mathbb{1}, \quad [\sigma_a,\sigma_b] = 2 i \epsilon_{abc} \sigma_c \]

\item $\text{STr}(\sigma_3^m \sigma_a \sigma_b)$:
\begin{align}
\text{STr}(\sigma_3^m \sigma_a \sigma_b) &= \frac{1}{(m+2)!} \sum_{\text{all permutations}} \text{Tr}(\sigma_3^m \sigma_a \sigma_b) \nonumber\\
&= \frac{1}{m+1} \sum_{k=0}^m \text{Tr}(\sigma_3^k \sigma_a \sigma_3^{m-k} \sigma_b) \nonumber\\
&= \left\{ \begin{array}{ll} 2 [\delta_{0a} \delta_{0b} + \delta_{3a} \delta_{3b} + \frac{\delta_{ab}}{m+1}|_{a,b=1,2}] \qquad \text{for $m$ even} \\
 2 [\delta_{0a} \delta_{3b} + \delta_{3a} \delta_{0b}]\qquad \text{for $m$ odd} \end{array} \right.
\end{align}
where now $a,b=0,1,2,3$ with $\sigma_0 = \mathbb{1}$,
and where we used
\begin{equation}
\sum_{k=0}^m \text{Tr}(\sigma_3^k \sigma_a \sigma_3^{m-k} \sigma_b) = \sum_{k=0}^m \text{Tr}((-1)^k \sigma_3^{m} \sigma_b \sigma_a).
\end{equation}
\item $\text{STr}(F^m \tilde X^2)$ with $m$ even, $F=F_0 \sigma_0 + F_3 \sigma_3$ and $\tilde X = \tilde X_a t_a$ with $t_a = -i \left(\frac{\mathbb{1}}{2},\frac{\sigma_a}{2}\right)$:
\begin{align}
\text{STr} (F^m \tilde X^2) &= F_3^m \text{STr}(\sigma_3^m \tilde X^2) + F_3^{m-1} F_0 \binom{m}{1} \text{STr}(\sigma_3^{m-1} \tilde X^2)  + F_3^{m-2} F_0^2 \binom{m}{2} \text{STr}(\sigma_3^{m-2} \tilde X^2) \nonumber \\& + \cdots + F_0^m \text{STr}(\tilde X^2) \nonumber\\
&= -\frac{1}{2}F_3^m [\tilde X_0^2 + \tilde X_3^2 + \sum_{a=1}^2 \frac{\tilde X_a^2}{m+1}] -\frac{1}{2} F_3^{m-1}F_0 \binom{m}{1} [\tilde X_0 \tilde X_3 + \tilde X_3 \tilde X_0] \nonumber\\ &\quad -\frac{1}{2} F_3^{m-2}F_0^2 \binom{m}{2} [\tilde X_0^2 + \tilde X_3^2 + \sum_{a=1}^2 \frac{\tilde X_a^2}{m-1}]  + \cdots -\frac{1}{2} F_0^m \sum_{a=0}^3 \tilde X_a^2 \nonumber\\
&= -\frac{1}{2} \sum_{a=1}^2 \tilde X_a^2 \left\{ \frac{F_3^m}{m+1} + \frac{F_3^{m-2}F_0^2}{m-1} \binom{m}{2} + \cdots + \frac{F_3^{2}F_0^{m-2}}{3} \binom{m}{2} + F_0^m \right\}  \nonumber\\
&\quad -\frac{1}{2} (\tilde X_0^2 + \tilde X_3^2) \left\{ F_3^m + F_3^{m-2}F_0^2 \binom{m}{2} + \cdots + F_0^m \right\} \nonumber\\
&\quad -\frac{1}{2} (2\tilde X_0 \tilde X_3) \left\{ F_3^{m-1} F_0 \binom{m}{1} + F_3^{m-3} F_0^3\binom{m}{3} + \cdots+ F_3 F_0^{m-1}\binom{m}{1}\right\}
\end{align}

\item $\text{STr}(\mathcal H(F) X^2)$ with $\mathcal H(F) = a_0  + a_1 F^2 + a_2 F^4 + \cdots + a_m F^{2m} + \cdots$ an even function of the background field $F$:

\begin{align}
\text{STr}(\mathcal H(F) \tilde X^2)
&= -\frac{1}{2} \sum_{a=1}^2 \tilde X_a^2 \left\{ a_0 + a_1 \left[ \frac{F_3^2}{3}+F_0^2\right] + a_2\left[ \frac{F_3^4}{5} + \binom{4}{2} \frac{F_3^2 F_0^2}{3} +  F_0^4\right] + \cdots \right\}  \nonumber\\
&\quad -\frac{1}{2} (\tilde X_0^2 + \tilde X_3^2) \left\{ a_0 + a_1 \left[ F_3^2+F_0^2\right] + a_2 \left[ F_3^4 + \binom{4}{2} F_3^2 F_0^2 + F_0^4\right] + \cdots \right\} \nonumber\\
&\quad -\frac{1}{2} (2\tilde X_0 \tilde X_3) \left\{ a_1 \left[ \binom{2}{1}F_0 F_3\right] + a_2 \left[ \binom{4}{1} F_0^3 F_3 + \binom{4}{1} F_0 F_3^3 \right]+ \cdots \right\} \nonumber\\
&= -\frac{1}{2} \sum_{a=1}^2 \tilde X_a^2 \left\{ \frac{\int_0^1 d\alpha \mathcal H(F_0 + \alpha F_3) + \int_0^1 d\alpha \mathcal H(F_0 - \alpha F_3)}{2}\right\} \nonumber\\
&\quad -\frac{1}{2} (\tilde X_0^2 + \tilde X_3^2) \left\{\frac{\mathcal H(F_0+F_3)+\mathcal H(F_0-F_3)}{2}\right\} \nonumber\\
&\quad -\frac{1}{2} (2\tilde X_0 \tilde X_3) \left\{\frac{\mathcal H(F_0+F_3)-\mathcal H(F_0-F_3)}{2}\right\}
\end{align}
which is the prescription (\ref{prescr}).

\end{itemize}

\end{document}